\documentclass[print]{aa}
\usepackage{ulem}
\usepackage{blindtext}
\usepackage{setspace}
\usepackage{geometry}
\usepackage{subfigure}
\usepackage{graphicx}
\usepackage{multirow}
\usepackage{longtable}
\usepackage{subfigure}
\usepackage{rotating}
\usepackage{txfonts}
\usepackage{lscape}
\usepackage{CJK}
\usepackage[figuresright]{rotating}
\usepackage{longtable}
\usepackage{pdflscape} 
\usepackage{longtable}
\usepackage{rotating}
\usepackage{amssymb}  
\usepackage{bbding}
\usepackage{longtable}
\usepackage[figuresright]{rotating}
\usepackage{graphicx}
\usepackage{epsfig}
\usepackage{epstopdf}
\usepackage{natbib}
\usepackage{CJK}
\usepackage{longtable}
\usepackage{amsmath}
\usepackage{multirow}
\usepackage{subfigure}
\usepackage{graphicx}
\usepackage{txfonts}
\usepackage{epsfig}
\usepackage{anyfontsize}
\usepackage{color,xcolor}
\usepackage[colorlinks,linkcolor=blue, anchorcolor=blue, citecolor=blue]{hyperref}
\usepackage{orcidlink}
\begin{document} 
\begin{CJK*}{UTF8}{gkai}

\title{Molecules in the peculiar age-defying source
 IRAS\,19312+1950}

\author{Jian-Jie Qiu (邱建杰)\orcidlink{0000-0002-9829-8655}
          \inst{1,2},
          Yong Zhang (张泳)\orcidlink{0000-0002-1086-7922}
          \inst{1,2,3}
          \thanks{e-mail: zhangyong5@mail.sysu.edu.cn},
          Jun-ichi Nakashima(中岛淳一)\orcidlink{0000-0003-3324-9462}
          \inst{1,2}
          \thanks{e-mail: nakashima.junichi@gmail.sysu.edu.cn},
          Jiang-Shui Zhang (张江水)\orcidlink{0000-0002-5161-8180}
          \inst{4}
          \thanks{e-mail: jszhang@gzhu.edu.cn},
          Nico Koning
          \inst{5},
          Xin-Di Tang (汤新弟)\orcidlink{0000-0002-4154-4309}
          \inst{6},
          Yao-Ting Yan (闫耀庭)\orcidlink{0000-0001-5574-0549}
          \inst{7},
          Huan-Xue Feng (冯焕雪)
          \inst{1}
          }

\institute{School of Physics and Astronomy, Sun Yat-sen University, Zhuhai, 519082, PR China
               \and
   CSST Science Center for the Guangdong-Hongkong-Macau Greater Bay Area, Sun Yat-Sen University, Zhuhai, PR China
   \and
    Laboratory for Space Research, The University of Hong Kong, Hong Kong, PR China
    \and
           Center For Astrophysics, Guangzhou University, Guangzhou, 510006, PR China
               \and
         Department of Physics and Astronomy, University of Calgary, 2500 University Drive NW, Calgary, AB, T2N 1N4, Canada 
               \and
           Xinjiang Astronomical Observatory, Chinese Academy of Sciences, Urumqi, 830011, PR China 
               \and
           Max-Planck-Institut f\"{u}r Radioastronomie, Auf dem H\"{u}gel 69, Bonn, 53121, Germany\\
             }


 
\abstract
   {
   IRAS\,19312+1950 is an isolated infrared source that exhibits a characteristic quasi-point-symmetric morphology in the near- and mid-infrared images and is also very bright in molecular radio lines. 
   Because of its unique observational characteristics, various observational studies have been conducted and several hypotheses have been proposed regarding its origin, which is still unclear.
   So far, it has been suggested that it could be a peculiar evolved star, a young stellar object, or even a red nova remnant.
   Regardless of which type of object it is ultimately classified as, IRAS\,19312+1950 is exceptionally bright in the infrared and molecular radio lines and therefore will undoubtedly be crucial as a prototype of this kind of object having a peculiar nature or unusual evolutionary phase.
   }
   {
   This study aims to reveal the molecular composition of the central part of IRAS\,19312+1950 by performing an unbiased molecular radio line survey and discussing the origin of the object from a molecular chemical point of view.
   }
   {
   We carried out a spectral line survey with the Institut de Radioastronomie Millim$\'{e}$trique (IRAM) 30 m telescope towards the center of IRAS\,19312+1950  
   in the 3 and 1.3 mm windows with frequency coverage of 83.9--91.8 and 218.2--226.0 GHz, respectively.
   }
   {
   In total, 28 transition lines of 22 molecular species and those isotopologues are detected towards IRAS\,19312+1950, some of which exhibit a broad and a narrow components. 
   Seventeen thermal lines and 1 maser line are newly detected. 
   The molecular species of C$^{17}$O, $^{30}$SiO, HN$^{13}$C, HC$^{18}$O$^{+}$, H$_{2}$CO, 
   and $c$-C$_{3}$H$_{2}$ are detected for the first time in this object.
   We calculated the optical depths of the transition lines of $^{13}$CO, C$^{18}$O, HCN, H$^{13}$CN, 
   and C$_{2}$H, and determined the rotational temperatures, column densities, 
   and fractional abundances of the detected molecules.
   We got the isotopic ratios of $^{12}$C/$^{13}$C, $^{14}$N/$^{15}$N, $^{16}$O/$^{18}$O, $^{16}$O/$^{17}$O, $^{18}$O/$^{17}$O, $^{28}$Si/$^{30}$Si, and $^{29}$Si/$^{30}$Si in IRAS\,19312+1950 
   and the values were compared to those of evolved stars, red novae, young stellar objects, and the Sun.
   The intensities of the molecular radio lines of IRAS\,19312+1950 were compared with those of different categories of objects, 
   indicating that the spectral pattern of the broad-line region is similar to that 
   of a red nova or a low-mass young stellar object, while the narrow-line region behaves 
   like an envelope of the asymptotic giant branch star.
   }
   { 
   Our results, in combination with previous studies, favor the hypothesis that IRAS\,19312+1950 might be a red nova remnant, 
   in which the progenitors that merged to become a red nova may have contained at least two evolved stars with oxygen-rich and carbon-rich chemistry, respectively. 
   }
   
\keywords{Individual objects: IRAS\,19312+1950 --- ISM: molecules --- circumstellar matter --- Line: identification --- Surveys}
\titlerunning{Molecules in IRAS\,19312+1950}
\authorrunning{J.-J. Qiu et al.}
\maketitle
%
\section{Introduction}

IRAS\,19312+1950 (hereafter, I19312) is a bright, isolated mid-infrared source located in the Galactic plane with many antithetical features which cannot be readily explained within the standard scheme of any known celestial object; its identity is still unknown.
The heliocentric distance to I19312 derived from the trigonometric parallax method using the VLBI Exploration of Radio Astrometry (VERA) is about 3.8 kpc \citep{Imai11}, 
which is consistent with the distances of 2.2$^{+5.2}_{-0.9}$ and 2.4$^{+3.1}_{-0.9}$ kpc 
derived from the archive data of Gaia DR2 and DR3 \citep[corresponding parallaxes are 0.45 $\pm$ 0.31 and 0.42 $\pm$ 0.24 mas, respectively,][]{Gaia18,Gaia21}.
The bolometric luminosity of I19312 is about 1--3 $\times$ 10$^{4}$ $L_{\odot}$ \citep{Cooper13,Cordiner16,Nakashima16} assuming a distance of 3.1--3.8 kpc. 
Until now, 15 molecular species consisting of the main isotopic elements, including both carbon- and oxygen-bearing molecules have been detected towards I19312: 
CN, CS, CO, OH, SO, SiO, SO$_{2}$, HCN, HNC, HCO$^{+}$, N$_{2}$H$^{+}$, NH$_{3}$, H$_{2}$CS, HC$_{3}$N, and CH$_{3}$OH 
(of these, OH, SiO, H$_{2}$O, and CH$_{3}$OH exhibit maser emission) 
\citep{Nakashima00,Nakashima03,Nakashima04a, Nakashima04b, Deguchi04, Nakashima05, Nakashima07, Nakashima08, Nakashima11, Nakashima15, Nakashima16, Cordiner16}.
The thermal rotational lines show "broad" and "narrow" features in the line profiles. 
These features are attributed to the different morpho-kinematic components lying in the nebulosity of I19312. 
Assuming that the broad feature originates from a spherically expanding molecular shell, 
\cite{Deguchi04} derived a mass-loss rate of 2.6 $\times$ 10$^{-4}$ $M_{\odot}$\,yr$^{-1}$ with large-velocity-gradient (LVG) calculations. 
This value of the mass-loss rate is considerably larger than those of the typical values of the asymptotic giant branch (AGB) stars 
\citep[$\dot{M}$ = 10$^{-8}$--10$^{-5}$ $M_{\odot}$\,yr$^{-1}$; see, e.g.][]{Hofner18}, 
implying the existence of a remarkable gas ejection from the central part of I19312. 

I19312 first came to attention when \cite{Nakashima00} detected the SiO maser emission. 
They conducted a systematic survey of 43 GHz SiO maser emission towards the samples of cold mid-infrared point sources, 
which are selected from the {\it Infrared Astronomical Satellite} (IRAS) point source catalog in terms of the mid-infrared colors. 
The effective temperatures of the surveyed samples are less than about 250 K, and I19312 is one of the reddest objects in the samples; 
i.e. the IRAS colors of I19312 are log($F_{25}/F_{12}$) = 0.5 and log($F_{60}/F_{25}$) = 0.7. 
These IRAS colors are consistent with those of post-AGB stars or proto-planetary nebulae (PPNe).  
SiO maser emission is usually detected only from low- and intermediate-mass evolved stars in the AGB phase or red supergiants (RSGs), 
which show active mass-loss, and the detection rate from stars in the post-AGB phase and beyond is suddenly going down \citep{Habing96}. 
Based on its red IRAS colors, near-infrared morphology (specifically, the quasi point-symmetric structure was revealed in 2MASS $J$, $H$, and $K$ band images), 
and the detection of SiO masers, \cite{Nakashima00} initially suggested that I19312 could be a PPN candidate similar to OH\,231.8+4.2 (Rotten Egg Nebula), 
which is a rare PPN with SiO maser emission.

Three OH maser lines at the rest frequencies of 1612, 1665, and 1667 MHz have been also detected towards I19312, with the 1612 MHz line being the strongest. 
This characteristic of OH masers is consistent with that of evolved stars \citep{Nakashima11}. 
Interferometric observations with the Very Large Baseline Array \citep[VLBA,][]{Nakashima11} in the 22 GHz H$_{2}$O maser line have revealed a double-peak profile with a velocity range of about 35 km\,s$^{-1}$. 
The VLBA images suggested that this double peak profile is most likely attributed to a bipolar molecular flow, 
suggesting that H$_{2}$O maser properties are also consistent with the evolved star interpretation. 
In fact, it is known that there exists a class of low- and intermediate-mass evolved stars at the AGB or post-AGB phase, 
which often show a double-peaked profile in the H$_2$O maser lines \citep[see, e.g.][]{Yung13, Tafoya20}; 
this class of objects is called ``water fountain''. 
The characteristic line profiles of water fountains are known to be formed by a bipolar molecular jet.

However, subsequent studies have revealed various puzzling aspects of I19312 for an evolved star, 
and the interpretations of this object became uncertain. 
For example, CH$_{3}$OH lines have been detected towards I19312 \citep{Deguchi04, Nakashima15} but have never been detected in evolved stars \citep{Cernicharo11,McGuire22}, 
making I19312 an unusual object from a chemical point of view if the true identity is really an evolved star.

When SiO, H$_{2}$O, and OH maser are detected in an evolved star, 
the star is usually considered to have oxygen-rich chemistry, 
in which the amount of oxygen atoms present exceeds that of carbon atoms \citep{Habing96, Hofner18}. 
\cite{Deguchi04}, however, revealed that not only molecules specific to the oxygen-rich chemistry but also a non-negligible number of molecular species specific to carbon-rich chemistry are found in I19312.

Based on the high total mass of 500--700 $M_{\odot}$ derived from the infrared SED modeling ($\sim$1--100 $\mu$m) and its IR spectral features, 
\cite{Cordiner16} suggested that I19312 might be a massive young stellar object (YSO) embedded in a dusty large collapsing circumstellar molecular envelope. 
However, we note that their infrared SED modeling is carried out on the basis of theoretical YSO models \citep{Robitaille06}. 
Based on radio interferometric observations of the $^{12}$C line, 
\citet{Nakashima05} found that the broad line region can be attributed to a spherical envelope 
with an expanding velocity of $\sim$30 km\,s$^{-1}$.
In addition, among all the known YSOs, SiO and OH masers have only been detected in the massive YSOs with ionized regions (i.e. H\,{\sc ii} regions). 
If I19312 is a massive YSO with such a bright luminosity of $\sim$2 $\times$ 10$^{4}$ $L_{\odot}$ and ionized regions \citep{Cordiner16}, 
free-free radio emission and the Br-$\gamma$ line should be detectable \citep{Shang04,Beck10} but so far, they are not.

\cite{Nakashima11,Nakashima15,Nakashima16} mentioned the possibility of I19312 as a "red nova remnant", 
which is typically created by the merger of two main-sequence stars or stars going to the main sequence. 
It is known that after this process, a single star with a cold molecular envelope may form which could potentially become a maser source.
Recent studies suggest that even if the progenitors include non-main-sequence stars, 
such as AGB and post-AGB stars, a remnant with cold matter could be similarly created by the mergers. 
For example, it is known that V838\,Mon, a prototypical red nova that erupted in 2002 \citep{Ivanova13}, exhibits SiO maser emission \citep{Deguchi05,Ortiz-Leon20}. 
The complex envelope morphology \citep{Nakashima11}, relatively large gas mass \citep{Nakashima16}, 
and characteristic molecular chemistry of I19312 \citep{Nakashima15} could be explained by the stellar merger hypothesis.
Importantly, \cite{Kaminski17} detected CH$_{3}$OH lines towards CK\,Vul (Nova\,1670), which is another well-studied red nova remnant. 
They also found lines both of carbon- and oxygen-bearing molecules, showing the complex chemistry of CK\,Vul, which is similar to that of I19312.
Although the infrared bolometric luminosity of I19312 ($\sim$10$^{4}$~$L_{\odot}$ with a distance of 3.8 kpc) is clearly too bright for a typical YSO (i.e. non-massive YSO with the ionized region), 
the theoretical calculations \citep{Tylenda06} suggested that the red nova remnant could explain the luminosity up to about 10$^{6}$~$L_{\odot}$ when two stars with masses of about 0.3 and 8 $M_{\odot}$ are merged. 
Additionally, we note that the infrared bolometric luminosity of I19312 is comparable to that of He\,3-1475 ($\sim$22,000 $L_{\odot}$), 
which is a PPN with a bipolar molecular jet \citep{Riera95}. 
\cite{Borkowski01} suggested that the high luminosity of He\,3-1475 is due to a mass transfer and/or a merger event that happened in an interacting binary system lying at the center of the envelope. 
In the case of He\,3-1475, SiO maser emission has not been detected, but since it is a PPN, many other properties are shared with I19312.

Although not much is known about stellar merger rates, they may be a ubiquitous phenomenon in the universe since more than 50 percent of the stars in the sky are part of a binary or multiple stellar systems. 
If we assume that the rate of stellar mergers is high, 
the characteristic chemical composition of red nova remnants may affect not only the chemistry of the interstellar material but also its chemical evolution. 
Although some theoretical works have suggested that nova ejecta is the major contributor of $^{13}$C, $^{15}$N, and $^{17}$O elements in the Galaxy \citep{Jose98, Romano03, Li16}, red novae would be equally important. 
We need to emphasize that if the true nature of I19312 is a red nova remnant, 
this object will be the brightest sample in radio and infrared wavelengths, 
making it an unparalleled laboratory for the observational study of the details of red novae and red nova remnants.
From this perspective, the motivation to determine if I19312 is a red nova remnant is relatively high, but there is one significant difficulty; there is no record of a red nova explosion of I19312. 
If the moment of the merger event is not recorded, identification of red nova remnants is not an easy task because the observational properties of the red nova remnants are similar to those of evolved stars. 
In fact, one of the known red nova remnants (CK\,Vul) has long been incorrectly identified as a planetary nebula (PN). 
In the case of CK\,Vul, the merger event was documented in classical European astronomical records in the 1600s \citep[][and references therein]{Kaminski20}. 
In fact, all of the red nova remnants found so far have had the moment of merger recorded, which provides conclusive evidence that they are red nova remnants.
We checked the old astronomical records, to the best of our ability, in China as well as in Europe, but were unable to find any record of a nova explosion occurring at the position of I19312 \citep[and references therein]{Xi55,Xi65,Clark82}. 
Given the incompleteness of the nova's record, which is several hundred years ago, presumably, there could be a non-negligible number of red nova remnants contaminating known PN and evolved star samples. 
In this regard, it is also important to establish a method for identifying red nova remnants for which the moment of the merger has not been recorded.

Molecular gases are key components in understanding star formation in interstellar space and the material cycle in the universe \citep{Tielens13}. 
So far, more than 240 kinds of molecular species have been detected in space \citep[and references therein]{McGuire18, McGuire22}.
Each of these species traces different aspects of the physical and chemical state of the celestial objects and the relative abundances of these species vary with their astrophysical environment \citep[e.g.][]{Cernicharo11,Aladro15,Jorgensen20}. 
The molecular rotational transition lines at millimeter wavelength are an excellent probe to investigate the physical and chemical properties of molecular gas components of the circumstellar envelope (CSE) and the interstellar medium.
Although \cite{Deguchi04} did a selected-line survey towards I19312, there has been no large-scale unbiased survey so far.  
An unbiased line survey enables us to detect multiple transition lines of a particular molecule and also the lines of different molecular species, 
providing an unbiased view of the molecular chemical environment of the target source.
Here, we report the results of an unbiased molecular line survey towards I19312 by using the IRAM 30 m telescope at the 3 and 1.3 mm bands. 

The main objective of this research is to compare the molecular line information of I19312 obtained from the present survey with that of known red nova remnants, 
the envelope of carbon-rich or oxygen-rich evolved stars, and YSOs. 
Through this comparative study, we will explore how we can reliably identify red nova remnants from their molecular chemical abundances, 
and at the same time, consider whether I19312 is a red nova remnant or not.

The structure of this paper is as follows: we present the observations and data reduction in Sect.~2, and the main results and a brief review of the previous studies on molecular chemical properties of I19312 are provided in Sect.~3. Then, we discuss the physical and chemical properties derived from the present observation in Sect.~4, and finally, summarize the results in Sect.~5.

\section{Observations and data reduction}

The observations towards the center of I19312 (RA 19 33 24.24, Dec 19 56 55.6; J2000) were performed with the Institutde Radioastronomie Millim{\'e}trique (IRAM) 30m-diameter telescope\footnote{IRAM is supported by INSU/CNRS (France), MPG (Germany), and IGN (Spain).} 
at Pico Veleta Observatory (Spain) in January 2017. 
We used the Eight Mixer Receiver (EMIR) with dual-polarization and the Fourier Transform Spectrometers (FTS) backend to cover frequency ranges from 83.9--91.8 GHz in the 3 mm window and 218.2--226.0 GHz in the 1.3 mm window. 
The instantaneous frequency coverage per sideband and polarization is 4 GHz with a frequency channel spacing of 195 kHz. 
The standard wobbler switching mode was carried out with a $\pm$ 110$\arcsec$ offset and the beam throwing rate is 0.5 Hz. 
The on-source times of each band are about one hour.
Pointing was checked frequently with strong millimeter emitting quasi-stellar objects.
The typical system temperatures are 124 K at the 3 mm band and 256 K at the 1.3 mm band. 

The antenna temperature $(T_{\rm A})$ is converted into the main beam temperature $(T_{\rm mb})$ through the relation $T_{\rm mb}=T_{\rm A}\times\frac{F_{\rm eff}}{B_{\rm eff}}$, 
where $F_{\rm eff}$ and $B_{\rm eff}$ are the forward efficiency and the main-beam efficiency of the IRMA 30 m telescope at the corresponding bands, 
as listed in Table~\ref{Table1}. 
The data reduction is carried out using the Continuum and Line Analysis Single-dish Software (CLASS) of the GILDAS package\footnote{http://www.iram.fr/IRAMFR/GILDAS}.
Individual spectra covering the same wavelength ranges are co-added with the weights proportional to the inverse square of the noise levels.
A 1st order polynomial baseline subtraction is performed.
We identify the molecular lines based on the information from 
the Jet Propulsion Laboratory database\footnote{https://spec.jpl.nasa.gov/ftp/pub/catalog/catform.html} \citep[JPL,][]{Pickett98},
the NIST database recommended rest frequencies for observed interstellar molecular microwave transitions\footnote{http://pml.nist.gov/cgi-bin/micro/table5/start.pl} \citep{Lovas04}, 
the Cologne Database for Molecular Spectroscopy catalog\footnote{https://cdms.astro.uni-koeln.de} \citep[CDMS,][]{Muller01,Muller05}, 
and the splatalogue database for astronomical spectroscopy\footnote{https://splatalogue.online//advanced.php}.
To validate the detection and identification of a given line, 
we checked the other transitions from the same species that are available in the current observations and those in the literature.
The root mean square (RMS) levels in $T_{\rm mb}$ scale are 19.8 and 89.8 mK for the frequency resolution of 0.1953 MHz at the rest frequency of 89.5 and 223.7 GHz, respectively. 
To measure weak lines, we smooth the corresponding spectra over 2--6 channels using the "boxcar" method to increase the signal-to-noise ratio, resulting in a velocity resolution of 1.3--3.9 and 0.5--1.6 km\,s$^{-1}$ at 90 and 220 GHz, respectively.

\section{Results}

The full spectra at the 3 and 1.3 mm bands with the strong lines marked are shown in Figs.~\ref{Figure1} and \ref{Figure2}, respectively.
Any identifiable feature above the 3$\sigma$ noise level is regarded as a real detection. 
If a fainter feature corresponds to a transition from a definitely identified molecule, we consider it a tentative detection.
The spectra show a line density of about 1.6 per GHz.
We detect a total of 28 emission lines belonging to 12 molecular species and 10 different isotopologues, 
which consist of 26 thermal lines and two SiO maser lines. 
The results are listed in Table~\ref{Table2}.
Seventeen thermal lines and a maser line are newly reported for this object. 
For the first time,  C$^{17}$O, $^{30}$SiO, HN$^{13}$C, HC$^{18}$O$^{+}$, H$_{2}$CO, and $c$-C$_{3}$H$_{2}$ have been discovered in this object; all are typical circumstellar molecules. 
The molecular compositions are dominated by oxygen-bearing species except for HCN, HNC, C$_{2}$H, HC$_{3}$N, and $c$-C$_{3}$H$_{2}$.
This confirms the oxygen-rich nature of this object, as suggested by \citet{Nakashima00} and \citet{Nakashima11} on the basis of the presence of SiO, H$_{2}$O, and OH maser lines.

We make use of the Gaussian profile to fit the emission lines utilizing the line analysis tool in CLASS. 
The fittings are shown in Figs.~\ref{Figure3}--\ref{Figure14}, 
which are ordered according to the molecular sizes and the rest frequencies of the transition lines. 
Table~\ref{Table2} presents the results along with parameters including the rest frequency, 
molecular species, upper and lower energy levels, velocity-integrated intensity, line center velocity, 
full-width at half-maximum ($\Delta V_{1/2}$) of the line, peak temperature, RMS of the baseline, velocity resolution, 
and half-power beam width (HPBW) of the telescope.

\subsection{Molecular lines}

Here we report the detection of molecular lines by species and compare them with the previous detections.

\subsubsection{Carbon Monoxide -- CO}

The $J=2 \rightarrow 1$ transitions of $^{13}$CO, C$^{17}$O, and C$^{18}$O are detected at the 1.3 mm band (see Fig.~\ref{Figure3}). 
The interferometric observations of CO, $^{13}$CO, and C$^{18}$O lines reveal a spherical broad-component region with an expanding velocity of $\sim$30 km\,s$^{-1}$ 
and a bipolar narrow-component region that is spatially larger than the broad-component region \citep{Nakashima05}.
The channel maps of the CO $J=1 \rightarrow 0$ and $2 \rightarrow 1$ lines indicate the presence of a strong self-shielding feature in the velocity range of $\sim$35--36 km\,s$^{-1}$.
Based on the wide-field CO maps, \cite{Nakashima16} showed that the extent of the central emission region is roughly 90$\arcsec$ in the north-south direction and 120$\arcsec$ in the east-west direction. 
As shown in Fig.~\ref{Figure3}, the $J=2 \rightarrow 1$ lines of $^{12}$CO, $^{13}$CO, and C$^{18}$O exhibit three narrow components and a broad component.
The peak-temperature ratios between the narrow and broad components are about 4.4, 16.0, and 15.4 for the $J=2 \rightarrow 1$ lines of $^{13}$CO, C$^{17}$O, and C$^{18}$O, respectively. 
The different peak-temperature ratios of these three lines may be attributed to their different optical depths.
The mean velocities of the red and blue narrow components are 
36.8 $\pm$ 0.1, 37.2 $\pm$ 0.3, and 36.9 $\pm$ 0.3 km\,s$^{-1}$ for $^{13}$CO, C$^{17}$O, and C$^{18}$O, respectively. 
These values are in good agreement with the peak velocities of the strongest narrow components
(36.3 $\pm$ 0.1, 
36.6 $\pm$ 0.3, 
and 35.7 $\pm$ 0.3 km\,s$^{-1}$, 
for $^{13}$CO, C$^{17}$O, and C$^{18}$O lines, respectively) 
and the broad components 
(35.7 $\pm$ 0.1, 35.7 $\pm$ 0.3, and 36.1 $\pm$ 0.3 km\,s$^{-1}$ for $^{13}$CO, C$^{17}$O, and C$^{18}$O lines, respectively),
implying that the regions emitting the red and blue lines are symmetrically distributed with respect to the central narrow-line emission region. 
The two weak components may arise from the bipolar outflows indicated by the H$_{2}$O $J_{K_a,K_c} = 6_{1,6} \rightarrow 5_{2,3}$ maser line \citep{Nakashima11}.

\subsubsection{Sulfur Monoxide -- SO}

Two SO transitions, $J_K=2_2 \rightarrow 1_1$ and $6_5 \rightarrow 5_4$, lie within the observational frequency range and have both been detected.
The line profiles consist of narrow and broad components, as shown in Fig.~\ref{Figure4}. 
Given a low signal-to-noise ratio, we are unable to adequately fit the SO $J_K=2_2 \rightarrow 1_1$ broad component. 
\citet{Nakashima00} have reported the detection of these lines, but with a lower signal-to-noise ratio, they only detected the narrow components peaking at 30--40 km\,s$^{-1}$. 
The presence of SO can be confirmed by previous observations of the $J_K=3_2 \rightarrow 2_1$ and $2_3 \rightarrow 1_2$ lines \citep{Deguchi04, Nakashima05, Nakashima15},  which lie outside our frequency range. 
The SO $J_K=3_2 \rightarrow 2_1$ map suggests that the broad-component region has an extent of <3$\arcsec$ 
while the narrow emission components peak at about 3$\arcsec$ north-west and 2$\arcsec$ south-east of the near-infrared emission peak position \citep{Nakashima05}.

\subsubsection{Silicon Monoxide -- SiO}
\label{Sect.3.1.3}

The thermal lines of SiO $J=2 \rightarrow 1$ and its $^{30}$Si isotopologue are detected, as shown in Fig.~\ref{Figure5}.
The $^{29}$SiO $J=2 \rightarrow 1$ line is only marginally seen.
The main lines clearly exhibit both narrow and broad components, 
which are ambiguous for the isotopic lines due to their low signal-to-noise ratios.
The peak temperature ratios between the narrow and broad components are 1.1 and 0.5 for the SiO and $^{30}$SiO lines, respectively. 
Using the NRO 45 m telescope, \cite{Deguchi04} obtained the main beam brightness temperature of the SiO $J=2 \rightarrow 1$ line to be about 0.20 K. 
Considering the different beam sizes, their measurements are consistent with ours.

Figure~\ref{Figure5} also presents the detection of two SiO maser lines at 86.24 and 85.64 GHz, respectively.
The SiO $J=2 \rightarrow 1, v=2$ line is a new detection in I19312.
The profiles of these two maser lines can be fit using two Gaussian components, but appear to be remarkably different.
A comparison with the previous observations reveals a temporal variation. 
The velocity range of the SiO $J=2 \rightarrow 1, v=1$ line varies 
from $\sim$10--30 km\,s$^{-1}$ in 2000 \citep{Nakashima00} 
to $\sim$20--40 km\,s$^{-1}$ in 2001 \citep{Deguchi04} 
to $\sim$40--60 km\,s$^{-1}$ in 2017 (this work), 
suggesting an acceleration.
The $v=2$ line shows a velocity component at 60--70 km\,s$^{-1}$. 
Such a high velocity has not been detected for other maser molecules (i.e. OH, H$_{2}$O, and CH$_{3}$OH, \citealt{Nakashima11,Deguchi04,Nakashima11}).

\subsubsection{Hydrogen Cyanide -- HCN}

The detections of the $J=1 \rightarrow 0$ transition lines of HCN and its $^{13}$C isotopologue are shown in Fig.~\ref{Figure6}. 
Both have three hyperfine structures.
A prominent broad component is visible for HCN. 
The peak temperature ratio between the narrow and broad components is about 2 for HCN, much smaller than that of H$^{13}$CN ($\sim$12). 
This may be attributed to different optical depths and/or different $^{12}$C/$^{13}$C isotopic ratios in the narrow- and broad-line regions.
The HC$^{15}$N $J=1 \rightarrow 0$ line is marginally detected, 
which seems to show a narrow feature peaked at about 35 km\,s$^{-1}$ and a broad feature in the velocity range of 20--50 km\,s$^{-1}$. 
Due to the low single-to-noise ratio ($\sim$2$\sigma$), 
the narrow and broad features cannot be distinguished and thus a single Gaussian fitting is employed. 
The HCN $J=1 \rightarrow 0$ map shows that 
the narrow and broad components coincide with the IRAS position of this object although the narrow component appears to be elongated along the direction perpendicular to the bipolar axis of the nebula within a size of 1$\rlap{.}\arcmin$9 × 1$\rlap{.}\arcmin$9 \citep{Nakashima04a}.

\subsubsection{Hydrogen Isocyanide -- HNC}

The $J=1 \rightarrow 0$ transition lines of HNC and its $^{13}$C isotopologue are detected, as shown in Fig.~\ref{Figure7}. 
To our best knowledge, this is the first time that HN$^{13}$C has been discovered in I19312.
The main line shows three narrow components. 
The central component is 15 and 47 times stronger than the blue- and red-side components, respectively.
The H$^{13}$NC line only shows a narrow feature. 
Presumably, its satellite features are too faint to be visible. 
\citet{Deguchi04} reported the observation of the HNC $J=1 \rightarrow 0$ line, in which only the central strong narrow component is detected.

\subsubsection{\texorpdfstring{Formylium Cation -- HCO$^{+}$}{}}

The $J=1 \rightarrow 0$ transition lines of HCO$^{+}$ and its $^{13}$C and $^{18}$O isotopologue are detected.
As shown in Fig.~\ref{Figure8}, the $J=1 \rightarrow 0$ lines of HCO$^{+}$ and H$^{13}$CO$^{+}$ show both narrow and broad components, 
while the HC$^{18}$O$^{+}$ $J=1 \rightarrow 0$ line exhibits a narrow feature only.
There are absorption features at about 31--35 and 43--47 km\,s$^{-1}$ superposed on the HCO$^{+}$ $J=1 \rightarrow 0$ line.
Interferometer observations of the HCO$^{+}$ $J=1 \rightarrow 0$ show a pronounced bipolar shape in the north-east to south-west direction 
and a broad component of this line arising from a relatively compact region of $\lesssim$3$\arcsec$ \citep{Nakashima04b}. 
The position-velocity diagram of \citet{Nakashima04b} suggests the presence of a bipolar outflow with an expansion velocity of about 10 km\,s$^{-1}$.
The map of the HCO$^{+}$ $J=3 \rightarrow 2$ suggests that it peaks at about 7$\arcsec$ east of the near-infrared emission peak \citep{Nakashima05}.
The peak temperature ratios between the narrow and broad components are 3.8 and 20.3 for the HCO$^{+}$ and H$^{13}$CO$^{+}$ lines, respectively, 
probably suggesting different optical depths and/or different $^{12}$C/$^{13}$C ratios between the narrow- and broad-line regions. 
Assuming that the HC$^{18}$O$^{+}$ line has the same peak temperature ratio as the H$^{13}$CO$^{+}$ $J=1 \rightarrow 0$ line, its broad component has a peak temperature of about 3.5 mK, well below the noise level.

\subsubsection{\texorpdfstring{Sulfur Dioxide -- SO$_{2}$}{}}

\cite{Deguchi04} for the first time reported a marginal detection of SO$_{2}$ through the $J_{K_a,K_c}=11_{1,11} \rightarrow 10_{0,10}$ line in this source. 
There are more than a dozen lines of SO$_{2}$ lying within our observed frequency range. 
However, we only detect the strongest one, i.e. the $J_{K_a,K_c}=11_{1,11} \rightarrow 10_{0,10}$ line.
Only a narrow feature is detected with an intensity of about 6$\sigma$ of the noise level, as shown in Fig.~\ref{Figure9}.

\subsubsection{\texorpdfstring{Ethynyl Radical -- C$_{2}$H}{}}

As far as we know, there is no previous report on the detection of C$_{2}$H in I19312.
We, therefore, present the first detection of this species through the C$_{2}$H $N=1 \rightarrow 0$ transition with intensity larger than 7$\sigma$ of the noise level.
As shown in Fig.~\ref{Figure10}, six isolated hyperfine lines appear to be narrow.

\subsubsection{\texorpdfstring{Formaldehyde -- H$_{2}$CO}{}}

There are three transition lines of H$_{2}$CO in the observed frequency range. 
They are the $J_{K_a,K_c}=3_{2,2} \rightarrow 2_{2,1}$ and $3_{2,1} \rightarrow 2_{2,0}$ lines of $p$-H$_{2}$CO and $o$-H$_{2}$CO $J_{K_a,K_c}=3_{1,2} \rightarrow 2_{1,1}$ line.
All are clearly detected with intensity larger than 5$\sigma$ of the noise level, as shown in Fig.~\ref{Figure11}. 
This is the first detection of H$_{2}$CO in this object. 
These lines show both narrow and broad components. 
H$_{2}$CO is a good tracer of the kinetic temperature ($T_{\rm k}$) of the environment because the flux ratio between the H$_{2}$CO lines on different $K_{\rm a}$ ladders are dominated by collisions. 
Once the $T_{\rm k}$ is restricted, the spatial density of the gas can be well determined from the flux ratio of H$_{2}$CO lines from the same $K_{\rm a}$ ladder \citep{Mangum93,Tang18}. 
Unfortunately, the only two $p$-H$_{2}$CO lines detected here have energy levels too close to estimate the physical conditions. 
Further observations of other H$_{2}$CO lines would be invaluable.

\subsubsection{\texorpdfstring{Cyanoacetylene -- HC$_{3}$N}{}}

There are two transition lines of HC$_{3}$N in the frequency coverage of our observations. 
The $J=10 \rightarrow 9$ line of HC$_{3}$N is clearly detected at the 3 mm band, 
while the HC$_{3}$N $J=24 \rightarrow 23$ line at the 1.3 mm band is only marginally detected with an intensity of about 2$\sigma$ above the noise level.
As shown in Fig.~\ref{Figure12}, both lines appear as a pronounced narrow feature. 
This is consistent with previous observations of other HC$_{3}$N lines \citep{Nakashima15,Deguchi04}.

\subsubsection{\texorpdfstring{Methanol -- CH$_{3}$OH}{}}

The thermal emission of CH$_{3}$OH in I19312 was first detected at 96.74 GHz by \cite{Deguchi04}.
Interestingly, the class I methanol maser emission, a shock tracer, was detected in this object too \citep{Nakashima15}.   
Only the two strongest thermal CH$_{3}$OH lines lying within our frequency coverage  
(the $J_{K_a,K_c}=5_{-1,0} \rightarrow 4_{0,0}\,E$ and $4_{2,0} \rightarrow 3_{1,0}\,E$ transitions) 
are detected. 
As shown in Fig.~\ref{Figure13}, both have only a narrow component.

\subsubsection{\texorpdfstring{Cyclopropenylidene -- $c$-C$_{3}$H$_{2}$}{}}

The simplest cyclic hydrocarbon species, $c$-C$_{3}$H$_{2}$, is the first detected in this object through the $J_{K_a,K_c}=2_{1,2} \rightarrow 1_{0,1}$ transition, 
as shown in Fig.~\ref{Figure14}.
Although dozens of transitions of $c$-C$_{3}$H$_{2}$ lay within our frequency range, 
the $J_{K_a,K_c}=2_{1,2} \rightarrow 1_{0,1}$ line has the lowest excitation temperature ($T_{\rm ex}$) at the upper energy level ($E_{\rm u}$ = 6.4 K) compared to those of the other lines (29.0--1024.1 K). 
The low temperature of this source probably hampers the detection of other $c$-C$_{3}$H$_{2}$ lines.

\subsection{Line profiles}
\label{Sect.3.2}

The molecular line profiles depend on the combined effects of kinematic structures, optical depths, the source size ($\theta_{\rm s}$), and the telescope-beam size ($\theta_{\rm b}$).
For a spherical envelope with a constant expansion velocity, the line profiles could be grouped into four classes: 
flat-topped, parabolic, double-peaked, and smoothed parabolic. 
However, the small widths of the narrow components and the blending nature of the broad components do not allow us to make a definite profile classification. 
Based on the line profile only, we cannot be sure whether these lines arise from an expanding envelope surrounding an evolved star or accretion inflows of YSO.

The $Herschel$ PACS O\,{\sc i} ${\rm ^{3}{P}_{0} \rightarrow {^3P}_{1}}$, 170 $\mu$m continuum, CO $J=14 \rightarrow 13$, 
and H$_{2}$O $J_{K_a,Kc}=1_{1,2} \rightarrow 1_{0,1}$ images of I19312 indicate that 
the dusty circumstellar envelope is symmetric with respect to the center on a scale of about 54$\arcsec$ $\times$ 54$\arcsec$ \citep{Cordiner16}.
However, high-resolution interferometer observations of CO and its $^{13}$C and $^{18}$C isotopic lines show that 
the envelope has an oval shape with a long axis in the north-west to south-east direction. 
For instance, the $^{13}$CO emission has an angular size of about 50$\arcsec$ $\times$ 30$\arcsec$ with a position angle of about 130$^{\circ}$ \citep{Nakashima05}. 
The H$_{2}$O maser observation indicates a bipolar flow structure \citep{Nakashima11}.
It follows that I19312 has complex dynamic structures. 
The broad component of the emission lines shows a symmetric profile, 
suggesting a symmetric dynamical structure, 
whereas it is hard to tell the symmetry property of the narrow component given its small width.

In Fig.~\ref{Figure15}, we compare the line-center velocities of the narrow and broad components. 
All narrow features have a consistent velocity of about 36.5 km\,s$^{-1}$ with a standard deviation of 0.7 km\,s$^{-1}$. 
The line-center velocities of the broad components are more scattered with a mean value of 33.2 km\,s$^{-1}$ and a standard deviation of 5.3 km\,s$^{-1}$.
Given the close velocities between the narrow and broad components, we can rule out the possibility that 
the two components originate from completely unrelated objects coinciding with the line of sight. 
We infer that the broad and narrow lines mainly trace the active inner regions and the quiet outer envelope, respectively.
The structure of the broad-line regions can be explained in terms of a merger event (red nova), as originally proposed by \cite{Nakashima11}.
In this scenario, the outer envelope is formed through the early mass-loss of an evolved star. 
The slightly different velocities between the components can be attributed to the self-shielding of the broad lines (see Sect.~\ref{ShapeX} for the details).

In Fig.~\ref{Figure16}, we compare the widths of these broad lines. 
It is interesting to note that 
O-bearing molecular transitions HCO$^{+}$ $J=1\rightarrow0$, SiO $J=2\rightarrow1$, and SO $J=6\rightarrow5$ 
generally exhibit a larger width of >25 km\,s$^{-1}$, 
while C-bearing ones 
HCN $J=1\rightarrow0$, H$_{2}$CO $J_{K_a,K_c}=3_{2,2}\rightarrow2_{2,1}$, and the $J=2\rightarrow1$ transitions of $^{13}$CO, C$^{17}$O, and C$^{18}$O have generally smaller width (<15 km\,s$^{-1}$). 
The shock can be well traced by SiO species and HCO$^{+}$ is sensitive to UV field intensity.
One plausible explanation is that the merger event generates intense shocks and UV radiations which destruct dust and/or dissociate CO, 
and subsequently enhance the O-bearing molecules in the inner regions.  
However, this cannot be taken as evidence against the YSO hypothesis since SiO and HCO$^{+}$ are also good tracers of YSO outflow \citep[e.g.][]{Csengeri16,Gregersen97,Yang21}.
 
It is instructive to compare the observed line widths with those of different types of objects in order to ascertain the nature of I19312. 
Through an observation of H$_{2}$CO lines towards 100 high-mass star-formation regions at an extremely early stage, 
\citet{Tang18} obtain an average line width of 3.9 $\pm$ 0.4 km\,s$^{-1}$. 
This value is close to the mean width of the narrow lines in I19312 ($\Delta V_{1/2}$ = 3.3 $\pm$ 0.5 km\,s$^{-1}$).
Nevertheless, such a narrow line profile is also exhibited among low luminosity AGB stars. 
For instance, the profile of CO lines in EP\,Aqr show a narrow feature ($\sim$1.4 km\,s$^{-1}$) superimposed on a broader one ($\sim$10.8 km\,s$^{-1}$) \citep{Winters03}.
The mean width of the broad lines in I19312 is $\Delta V_{1/2}$ = 21.6 $\pm$ 3.2 km\,s$^{-1}$, which is larger than the typical velocity of an AGB wind (5--15 km\,s$^{-1}$, \citealt{Hofner18}), 
but obviously lower than that of a red nova (>100 km\,s$^{-1}$, \citealt{Kaminski18}).
YSO outflows have velocities of a few to $\sim$100 km\,s$^{-1}$ \citep{Wu04}, 
encompassing the widths of both narrow and broad features of I19312.

All molecules exhibiting a broad feature are detected with the narrow feature, but not vice versa. 
CO, SO, SiO, HCN, HCO$^{+}$, and H$_{2}$CO show both broad and narrow features, 
while HNC, SO$_{2}$, C$_{2}$H, HC$_{3}$N, CH$_{3}$OH, and $c$-C$_{3}$H$_{2}$ are detected only in the narrow one. 
Generally, the former are dense gas tracers, and the latter prefer to form in cold environments.
To better understand the origins of the two features, we perform a morpho-kinematical simulation (See Appendix~\ref{ShapeX}), 
which shows that the observations of representative emission lines could be reproduced by 
a nebula composed of an inner spherical expanding component, an outer expanding elliptical component, 
and a bipolar outflow.

\subsection{Optical depths}
\label{Sect.3.3}

The optical depths of the HCN $J=1\rightarrow0$ hyperfine lines can be estimated from their intensity ratios. 
$T_{\rm mb}$ is proportional to $1-e^{-\tau}$, 
where the optical depth $\tau$ of a given hyperfine line is proportional to its line strength.
The line strengths given in the CDMS catalogue \citep{Muller01,Muller05} show 
that the relative optical depth between the $J=1\rightarrow0$ transition of HCN  $F=1\rightarrow1$, $2\rightarrow1$, and $0\rightarrow1$ hyperfine lines is $10.2 : 17.0 : 3.4$.
Thus we obtain $\tau(F=2\rightarrow1) = 1.3$ and 3.9 from the observed $T_{\rm mb}(F=1\rightarrow1)/T_{\rm mb}(F=2\rightarrow1)$ and $T_{\rm mb}(F=2\rightarrow1)/T_{\rm mb}(F=0\rightarrow1)$ ratios, respectively.
These values are roughly consistent with that derived by \cite{Deguchi04}.
For the H$^{13}$CN $J=1\rightarrow0$ transition, 
the relative optical depth between the $F=1\rightarrow1$, $2\rightarrow1$, and $0\rightarrow1$ hyperfine lines is $9.4 : 15.7 : 3.1$. We obtain a lower optical depth of $\tau(F=2\rightarrow1)$ = 0.6--0.8. 
For the calculations, we have assumed that both lines have the same excitation temperature, 
which however may not strictly be true. 
An accurate determination of $\tau$ requires a detailed model with an accurate knowledge of the physical and thermal structure of the source, 
which is beyond the scope of this paper. 
Therefore, the $\tau$ values obtained here are merely tentative ones.

The same technique is employed to calculate the optical depths of C$_{2}$H $N =1 \rightarrow0$.
The relative line strengths of the six hyperfine lines 
$N_{J,F}=1_{3/2,1}\rightarrow0_{1/2,1}$, $1_{3/2,2}\rightarrow0_{1/2,1}$, $1_{3/2,1}\rightarrow0_{1/2,0}$, $1_{1/2,1}\rightarrow0_{1/2,1}$, $1_{1/2,0}\rightarrow0_{1/2,1}$, and $1_{1/2,1}\rightarrow0_{1/2,0}$ are $6.6 : 64.8 : 32.3 : 32.3 : 13.0 : 6.6$. 
Consequently, we obtain the optical depth of the strongest line $\tau(1_{3/2,2}\rightarrow0_{1/2,1})$ = 1.0--2.3.

Under the assumption of local thermal equilibrium (LTE), 
the optical depths of the $J=2\rightarrow1$ transition lines of $^{13}$CO and C$^{18}$O can be estimated based on the formulae given by \citet{Nishimura15} (see their Eqs. 5 and 6). 
Using $T_{\rm ex}$ = 32 and >111 K for the narrow and broad components (see Sect.~\ref{HCN-to-HNC}) as an input,
we obtain the line-center optical depths of $^{13}$CO and C$^{18}$O $J=2\rightarrow1$ to be 0.62 and <0.11 for the narrow component and 0.11 and <0.01 for the broad component, respectively.

\subsection{Rotation diagram analysis and molecular abundances}
\label{Sect.3.4}

The column densities ($N$) and $T_{\rm ex}$ of the narrow components of SO, C$^{17}$O, HC$_{3}$N, and CH$_{3}$OH and the broad component of SO are estimated through the rotation diagram analysis.
Under the assumption of LTE and optical thinness, the level populations can be expressed with 
\begin{equation}
{\rm ln} \frac{N_{\rm u}}{g_{\rm u}} = {\rm ln} \frac{8\pi k \, \nu^{2} \, \int T_{\rm S} \, {\rm d}{v}}{hc^{3} \, A_{\rm ul} \, g_{\rm u}}\,  =  {\rm ln} \frac{N}{Q(T_{\rm ex})} - \frac{E_{\rm u}}{kT_{\rm ex}}, 
\label{equation2}
\end{equation}  
where $k$ and $h$ are the Boltzmann constant and Planck constant,
$\nu$ is the rest frequency in Hz, 
$E_{\rm u}/k$ is the upper-level energy in K,
$T_{\rm S}$ is the source brightness temperature of the line, 
and $Q(T_{\rm ex})$, $A_{\rm ul}$, and $g_{\rm u}$ are the partition function, spontaneous emission coefficient, and upper state degeneracy taken from the CDMS catalog \citep{Muller01,Muller05}, respectively.
Under the assumption of a Gaussian distribution of the surface brightness,
the conversion between $T_{\rm S}$ and $T_{\rm mb}$ follows the expression 
$T_{\rm S}=T_{\rm mb}(\theta_{\rm b}^{2} + \theta_{\rm s}^{2})/\theta_{\rm s}^{2}$,
where $\theta_{\rm b}$ is the antenna HPBW (see Table~\ref{Table2}) and $\theta_{\rm s}$ is the source size.
$\int T_{\rm S}\,{\rm d}v$ is the velocity-integrated intensity. 
Our knowledge of the $\theta_{\rm s}$ value of I19312 is poor, 
which may  differ from species to species. 
Based on the mappings of the $J=1 \rightarrow 0$ and $2 \rightarrow 1$ transitions of CO, $^{13}$CO, and C$^{18}$O, CS $J=2 \rightarrow 1$, HCO$^{+}$ $J=3 \rightarrow 2$, and SO $J_{K}=3_2 \rightarrow 2_1$ \citep{Nakashima05}, 
we adopt $\theta_{\rm s}$ = 15$\arcsec$ and 5$\arcsec$ for the narrow and broad components, 
resulting in the 
$(\theta_{\rm b}^{2} + \theta_{\rm s}^{2})/\theta_{\rm s}^{2}$
values of 1.5--4.7 and 6.0--33.7 for the two components, respectively. 
As the beam-dilution factor increases with decreasing frequency, 
the correction will steepen the 
$\ln({N_{\rm u}}/{g_{\rm u})}$ versus $E_{\rm u}$ relation, 
and thus lower the $T_{\rm ex}$. 
The effect is less significant for the narrow features given the larger $\theta_{\rm s}$.

If a molecule has more than one transition detected, its $T_{\rm ex}$ and $N$ can be determined via a linear fitting to Eq.~\ref{equation2}. 
To make a more reliable rotational diagram analysis, we retrieve the archive data obtained by the NRO 45 m telescope \citep{Deguchi04}. 
For the complemented data, the antenna temperature ($T_{\rm a}^{*}$) is converted into $T_{\rm mb}$ scale through $T_{\rm mb}$ = $T_{\rm a}^{*}/{B_{\rm eff}}$ and $T_{\rm mb}$ = $T_{\rm a}^{*}/{A_{\rm eff}}$ for the narrow and broad features of the emission lines respectively arising from the extended and compact regions, 
where ${B_{\rm eff}}$ and ${A_{\rm eff}}$ are the main beam efficiency and the aperture efficiency, respectively.
The resultant rotation diagrams are presented in Fig.~\ref{Figure17} and the obtained results are reported in Table~\ref{Table3}.
Generally, the fit is good. 
The narrow component of HC$_{3}$N can be better fit using two straight lines corresponding to a cold and warm component. 
The mean $T_{\rm ex}$ of the narrow component is 15.0 $\pm$ 1.0 K, 
which is lower than the $T_{\rm k}$ of CH$_{3}$OH (36 K) estimated by \cite{Nakashima15} based on an LVG approximation, 
probably indicating a deviation from LTE or a stratified temperature structure.

For the molecules with only one transition detected or with detected transitions having too close $E_{\rm u}$ to perform the rotational diagram analysis, 
we estimate their column densities using 
\begin{equation}
N = \frac{8\pi k \, \nu^{2} \, Q(T_{\rm ex})}{hc^{3} \, A_{\rm ul} \, g_{\rm u}} \, e^{E_{\rm u}/kT_{\rm ex}} \, C_{\tau} \int T_{\rm S} \, {\rm d}v ,
\label{N_t}
\end{equation}  
where $C_{\tau}$ = $\tau / (1-e^{-\tau})$ is an optical-depth correction factor. 
For the calculations, we use the average $T_{\rm ex}$ estimated from the rotation diagram analysis and drop the optically thin assumption.
The results along with the adopted $\tau$ are listed in Table~\ref{Table3}. 
In the optically thick case, the error of $N$ would be large because of
the large uncertainty of $\tau$ (see Sect.~\ref{Sect.3.3}).
For instance, if $\tau$ $\sim$ 1, a 50\% uncertainty of $\tau$ could cause a $\sim$30\%
uncertainty of $N$.

The fractional abundances relative to H$_{2}$ listed in Table~\ref{Table3} are calculated using 
\begin{equation}
f({\rm X}) = N({\rm X}) \frac{f({\rm HCO^{+}})}{N({\rm HCO^{+}})},
\end{equation}  
where $f({\rm HCO^{+}})$, according to an LVG calculation \citep{Deguchi04},
are equal to 1.2 $\times$ 10$^{-9}$ and 5.7 $\times$ 10$^{-8}$ for the narrow and broad components, respectively.

We can estimate the  masses of the broad and narrow components using
\begin{equation} 
M = \frac{N({\rm HCO^{+}})  A}{f({\rm HCO^{+}})}  \mu m_{\rm H_2} ,
\label{equation4}
\end{equation}  
where $\mu=1.33$ as the mean molecular weight of 
the interstellar gas relative to the hydrogen molecule \citep{Hildebrand83},
$m_{\rm H_2}$ is the mass of the hydrogen molecule, and the source
area $A$ = $0.25 \pi (D\theta_{\rm s})^{2} $ with $D$ of 2.4 kpc from Gaia DR3. 
We get  $M$ = 4.9 and 0.45 $M_{\odot}$ for the narrow and broad components, respectively. 
The obtained masses thus are compatible with the hypothesis of a stellar origin.

However, the mass estimation and interpretation should be treated with caution. 
It is common for the HCO$^+$ line to be optically thick towards YSOs \citep[e.g.][]{Ungerechts97, Yang21}, 
and thus the mass could have been severely underestimated if I19312 was actually a YSO. 
Adopting a distance of 6.6 kpc for I19312, 
\cite{Dunham11} obtained an isothermal mass of 1275 $\pm$ 1100 $M_{\odot}$ based on 1.1 mm continuum observations. 
Subsequently, adopting a distance of 3.8 kpc, 
\cite{Urquhart18} obtained an isothermal mass of 186 $M_{\odot}$ based on dust emission observations. 
They did not specify the uncertainty in the mass estimation of I19312, 
but claimed a typical error of 20 percent for their sample. 
If correcting for the distance effect and keeping aside the large uncertainties in the mass estimations 
and the different beam sizes of those observations, 
these masses in the literature are one order of magnitude higher than the obtained value for the narrow component. 
It follows that the mass derived from Eq.~\ref{equation4} should be taken as a lower limit, 
and this source might be surrounded by a cold molecular cloud.

\subsection{Isotopic abundance ratios}

The isotopic abundance ratios in CSEs of evolved stars are largely affected by the nucleosynthesis in the stellar interior and subsequent dredge-up processes \citep{Herwig05}. 
The present observations allow us to derive the isotopic ratios of $^{12}$C/$^{13}$C, $^{14}$N/$^{15}$N, $^{16}$O/$^{17}$O/$^{18}$O, and $^{28}$Si/$^{29}$Si/$^{30}$Si in I19312 using the velocity-integrated intensities of the molecular lines arising from the same upper energy levels. 
The estimated values, along with those of the red nova CK\,Vul, the carbon-rich AGB star IRC+10216, 
the massive star formation region Orion\,KL, and the Sun \citep{Sutton85, Kahane00,Lodders03, Frayer15, Kaminski17}, 
are listed in Table~\ref{Table4}. 
It should be pointed out that if the main line is optically thick, 
the resultant isotopic ratios would be only tentative, 
in which the low limits derived from the line ratios are presented. 

We find that all the rare isotopes of the CNO elements we detected, $^{13}$C, $^{15}$N, $^{17}$O, and $^{18}$O, 
are enhanced in both the narrow and broad components of I19312 compared to IRC+10216, Orion\,KL, and the Sun. 
Such a behavior has also been found in the remnant of Nova Vul\,1670 with $^{12}$C/$^{13}$C = 2--6, $^{14}$N/$^{15}$N $\approx$ 26, $^{16}$O/$^{18}$O $\approx$ 23, and $^{16}$C/$^{17}$O > 225 \citep{Kaminski15}. 
\citet{Kaminski17} suggested that hot CNO cycles and partial He burning instead of the standard stellar-nucleosynthesis scenario should be invoked to explain the extreme enhancement of $^{15}$N and $^{18}$O.

Another CSE with extreme $^{13}$C, $^{15}$N, and $^{17}$O isotopic enrichment is the carbon-rich young PN K4-47 \citep{Schmidt18a}, 
which has isotopic ratios of $^{12}$C/$^{13}$C = 2.2 $\pm$ 0.8, $^{16}$O/$^{17}$O = 21.4 $\pm$ 10.3 and $^{14}$N/$^{15}$N = 13.6 $\pm$ 6.5. 
The bipolar outflow of K4-47 is highly collimated and the dynamic age is about 400--900 yr. 
\cite{Schmidt18a} propose two possible explanations for the rare CNO isotope enrichment, namely, 
its progenitor is a $J$-type AGB star that underwent a helium-shell flash, or alternatively, 
it resulted from a merger of a binary system as in the case of CK\,Vul \citep{Kaminski15,Kaminski17}. 
We note that the narrow component of I19312 has CNO isotopic ratios encompassed by those of $J$-type stars ($^{12}$C/$^{13}$C = 3--19, $^{14}$N/$^{15}$N = 153--282, and $^{17}$O/$^{18}$O = 270--850; see \citealt{Schmidt18a} and references therein).

For the more heavy element Si, 
we do not find prominent differences in its isotopic ratios between I19312, CK\,Vul, IRC+10216, Orion\,KL, and the Sun. 
This suggests that the progenitor of I19312 is unlikely to be a massive evolved star.

Although the pattern of isotopic ratios is consistent with the red nova hypothesis, 
some points ought to be cautioned. 
For the first place, optical depths significantly affect the determination of isotopic ratios in the nova CK\,Vul \citep{Kaminski17}, 
causing some uncertainty in the comparison. 
Moreover, as the main lines of HCO$^{+}$, HCN, HNC, and SiO in YSOs are likely to be optically thick \citep[e.g.][]{Schneider10}, 
if I19312 is a YSO, the isotopic ratios derived from these lines would be highly uncertain. 
Therefore, we cannot firmly conclude that the isotopic ratios are against the YSO hypothesis.

\section{Discussion}

The intensity ratios of molecular radio lines provide very useful information for probing the physical state and chemical composition of circumstellar envelopes \citep[e.g.][]{Bujarrabal94, Olofsson98}. 
In the following, we discuss the properties of I19312 based on the intensity ratios of three representative cases 
and statistical comparison of the spectra of different types of sources.

\subsection{Molecular abundance ratios} \label{HCN-to-HNC}

The velocity-integrated intensity line ratio of HCN/SiO is a good probe for distinguishing oxygen-rich stars from carbon-rich stars \citep{Bieging00, Olofsson98}. 
In our measurements, the $I$(HCN 1--0)/$I$(SiO 2--1) are 1.9 $\pm$ 0.1 and 10.6 $\pm$ 1.3 for the broad and narrow components, respectively. 
The 3-D model given in Sect.~\ref{ShapeX} suggests that the broad and narrow components correspond to the inner spherical and outer spheroidal components (see, Fig.~\ref{Figure_A_1}). 
This result indicates that the inner component exhibits oxygen-rich chemistry and the outer component exhibits carbon-rich chemistry. 
This clear difference in molecular chemical composition between the inner and outer components will provide an important clue when considering the origins of I19312. 
There are several possible physical processes that would affect the HCN/SiO: e.g. shock-induced chemistry and different agglomeration degrees between HCN and SiO onto dust grains \citep{Olofsson98,Bieging00,Gonzalez03,Schoier06}. 

HCN and its geometrical isomer HNC can be formed through the dissociative recombination reaction of HCNH$^{+}$ + $e$ $\rightarrow$ HCN/HNC + H in almost equal amounts. 
HNC is readily consumed by the endothermic reaction of HNC + H $\rightarrow$ HCN + H, 
which leads to a decreasing [HNC]/[HCN] with increasing temperature because the backward reaction has a much higher energy barrier (\citealt{Graninger14} and the references therein). 
The observations of dark clouds show an average [HCN]/[HNC] of 2.1 $\pm$ 1.2 \citep{Hirota98}, 
while hot cores exhibit a value of $\approx$80 \citep{Schilke92}. 
Therefore, the [HNC]/[HCN] can serve as a thermometer (e.g. \citealt{Jin15,Hacar20,Long21}).
A large-scale observation of the HCN and HNC $J=1 \rightarrow 0$ lines towards the Integral Shape Filament in Orion show that 
[HCN]/[HNC] systematically increase with increasing temperature 
from a value of nearly unity at $T_{\rm k}$ $\sim$ 10 K to larger than $13$ at $T_{\rm k}$ > 70 K.
The narrow-line region of I19312 has a [HCN]/[HNC] of 3.2 $\pm$ 0.9. 
Based on the empirical relation between [HCN]/[HNC] and $T_{\rm k}$ provided by \citet{Hacar20}, 
we obtain $T_{\rm k}$ = 32 $\pm$ 9 K, 
which is quite close to the value derived with an LVG calculation for CH$_{3}$OH \citep[36 K;][]{Nakashima15}.  
The broad component of the HNC line is too faint to be detected, 
and thus we can only estimate the lower limit of [HCN]/[HNC] to be 27.8,
resulting in $T_{\rm k}$ > 111 K for the broad-line region.
This clearly indicates different physical conditions between the broad- and narrow-line regions.

Because of associated formation/destruction routes, there is a tight correlation between the column densities of C$_{2}$H and $c$-C$_{3}$H$_{2}$.
The average [C$_{2}$H]/[$c$-C$_{3}$H$_{2}$] of diffuse and translucent clouds is 27.7 $\pm$ 8.0 \citep{Lucas00}. 
\citet{Gerin11} subsequently obtained a similar result (28.0 $\pm$ 1.4) based on larger sample observations. 
\cite{Schmidt18b} found that such a ratio even holds for the rather old PN NGC\,7293 with [C$_{2}$H]/[$c$-C$_{3}$H$_{2}$] = 29 $\pm$ 6, 
where C$_{2}$H and $c$-C$_{3}$H$_{2}$ have similar spatial distributions and their abundances are 10--100 times higher than those in diffuse nebulae. 
Therefore, they hypothesized that PNe are the main contributor to the two molecules in interstellar space. 
The narrow-line region of I19312 has a [C$_{2}$H]/[$c$-C$_{3}$H$_{2}$] of 23.2 $\pm$ 1.9, 
in excellent agreement with those of diffuse and planetary nebulae. 
Their abundances are $f$(C$_{2}$H) = 6.72 $\times$ 10$^{-8}$ and $f$($c$-C$_{3}$H$_{2}$) = 2.89 $\times$ 10$^{-9}$, 
which lie between those diffuse and planetary nebulae.
At this moment, the origin of C$_{2}$H and $c$-C$_{3}$H$_{2}$ molecules in I19312 is unclear. 
However, the facts that the [C$_{2}$H]/[$c$-C$_{3}$H$_{2}$] of the narrow component is similar to those of PNe and diffuse clouds, 
and that the C$_{2}$H and $c$-C$_{3}$H$_{2}$ abundance rates are intermediate between those of PNe and diffuse clouds, 
may provide some clues as to the origin of I19312. 
For example, if the emission sources of the narrow component include material originating in PNe, I19312 may have formed in a stellar merger including a PN.

\subsection{Comparison with other sources}
\begin{figure}
\centering
        \includegraphics[width=16cm]{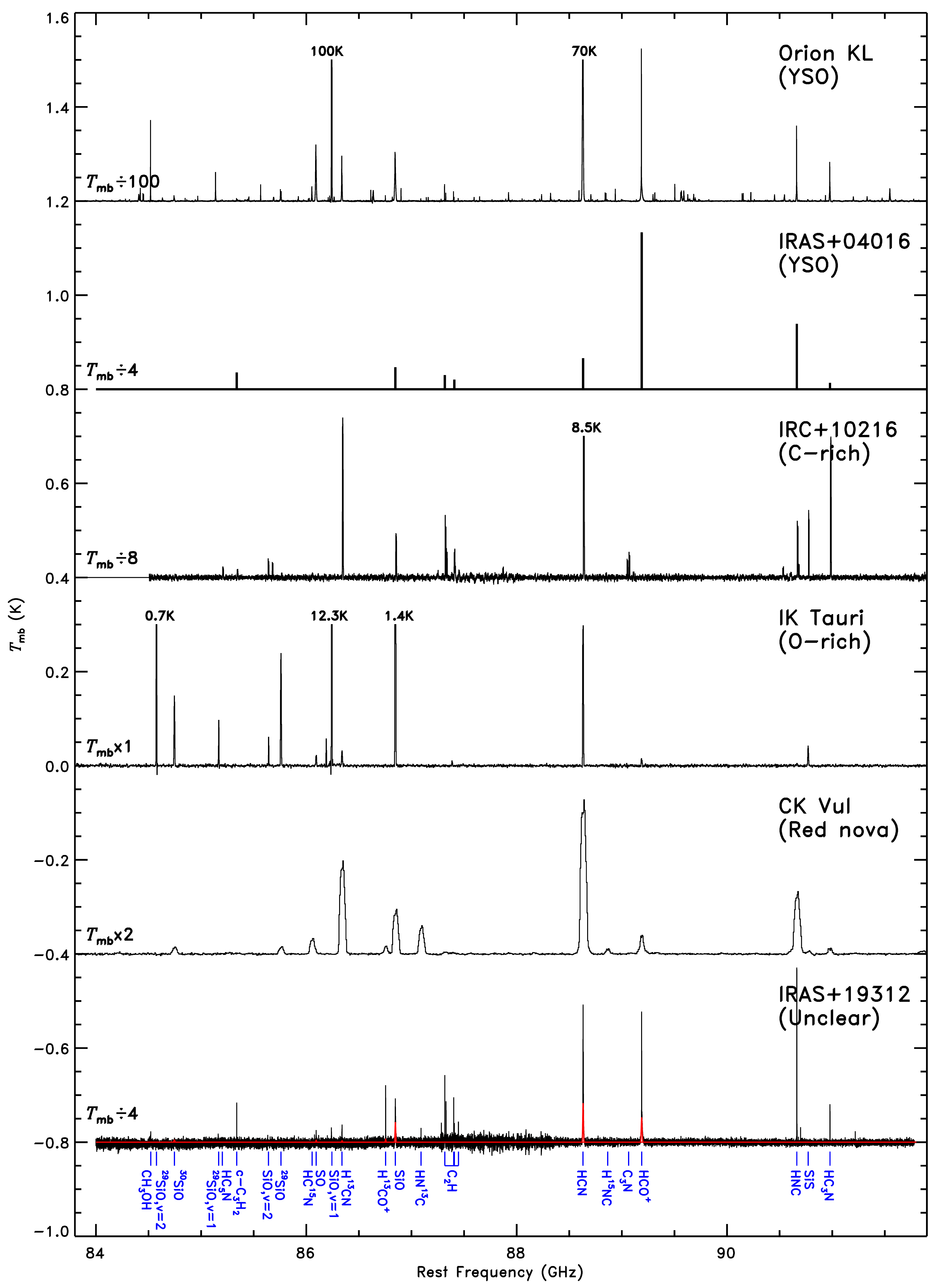}
 \caption{The 84--92 GHz spectrum of I19312 (the bottom panel),
 where the decomposed broad component is represented
 in red. The spectra of Orion\,KL, IRAS+04016, IRC+10216, IK\,Tauri, and CK\,Vul obtained by \citet{Frayer15}, \citet{Le20}, Tuo et al. (2022, in preparation), \citet{Velilla17}, and \citet{Kaminski17} are overplotted for comparison.
}
\label{Figure1}
\end{figure}

\begin{figure}
\centering
        \includegraphics[width=16cm]{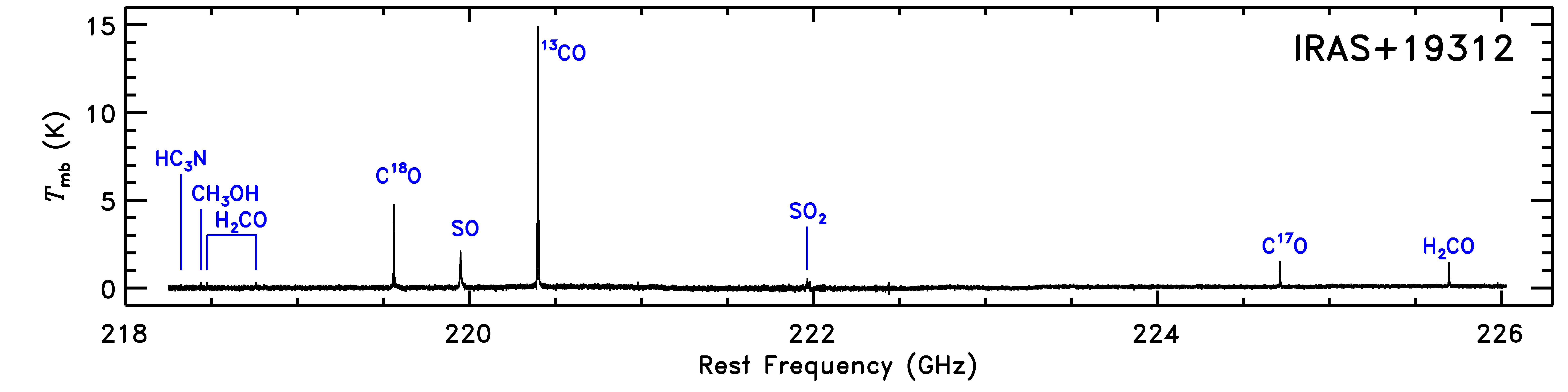}
 \caption{The 218--226 GHz spectrum of I19312.
}
\label{Figure2}
\end{figure}

\begin{figure}
\centering
        \includegraphics[width=8cm]{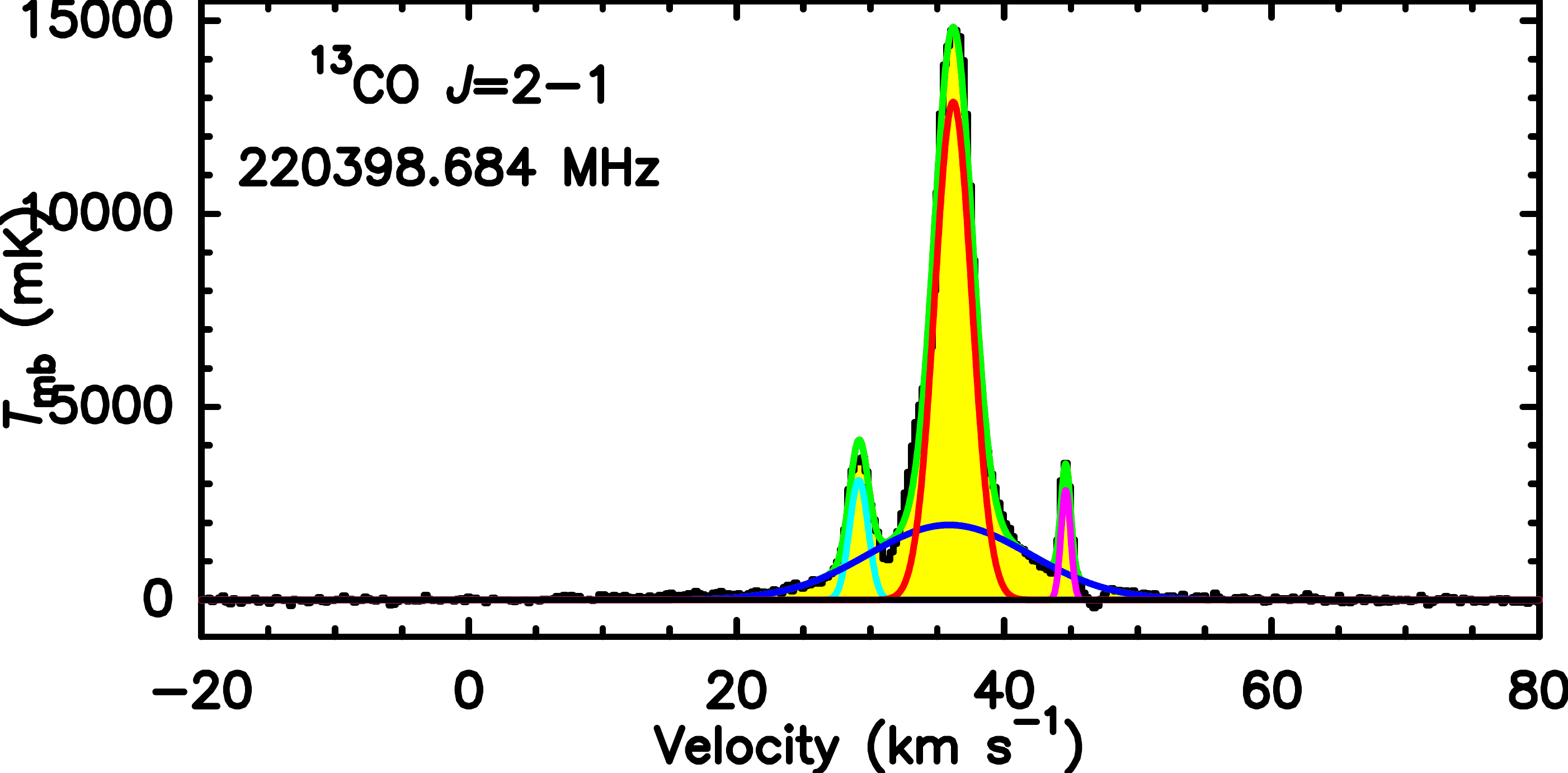}
        \includegraphics[width=8cm]{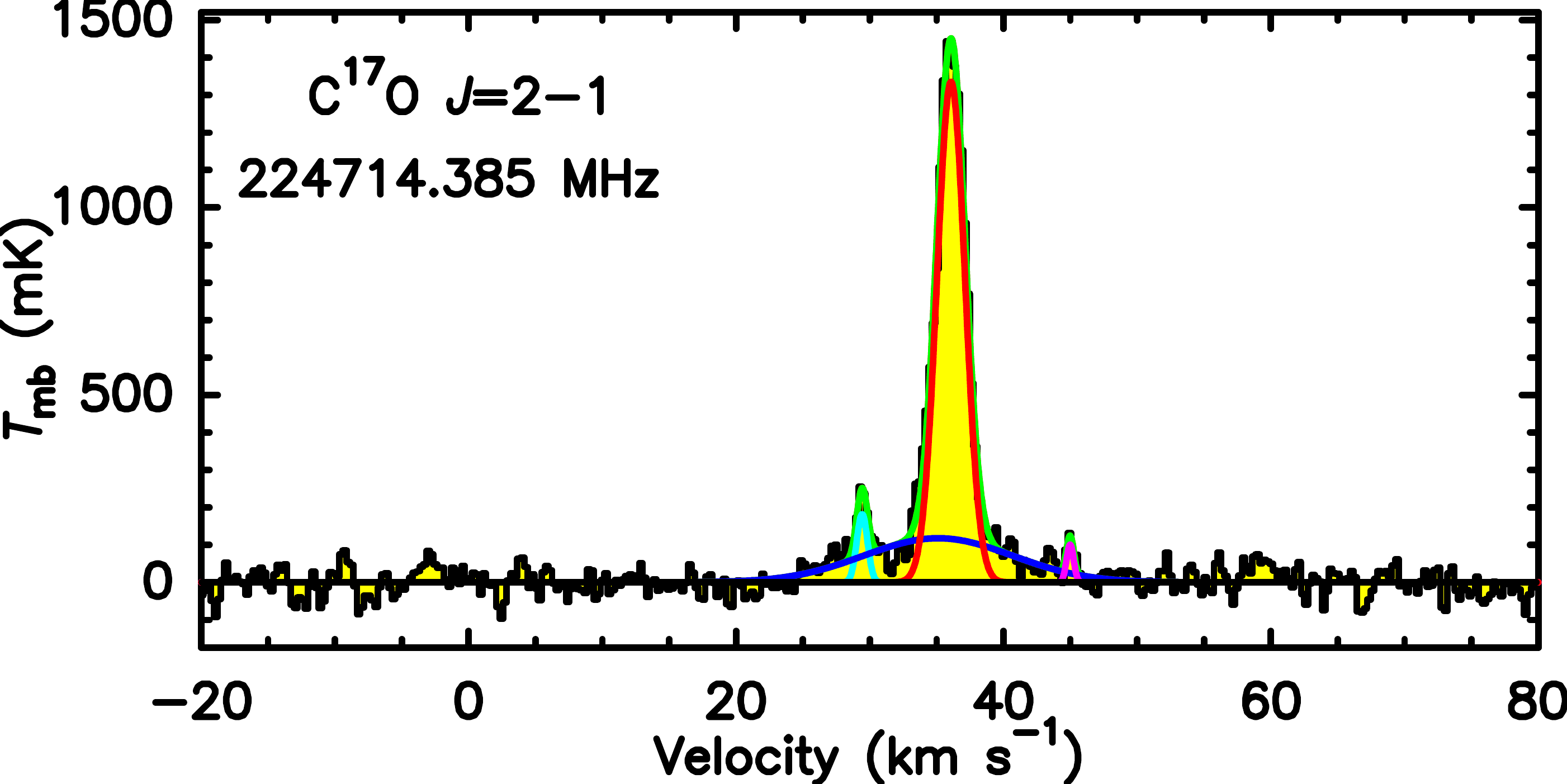}
        \includegraphics[width=8cm]{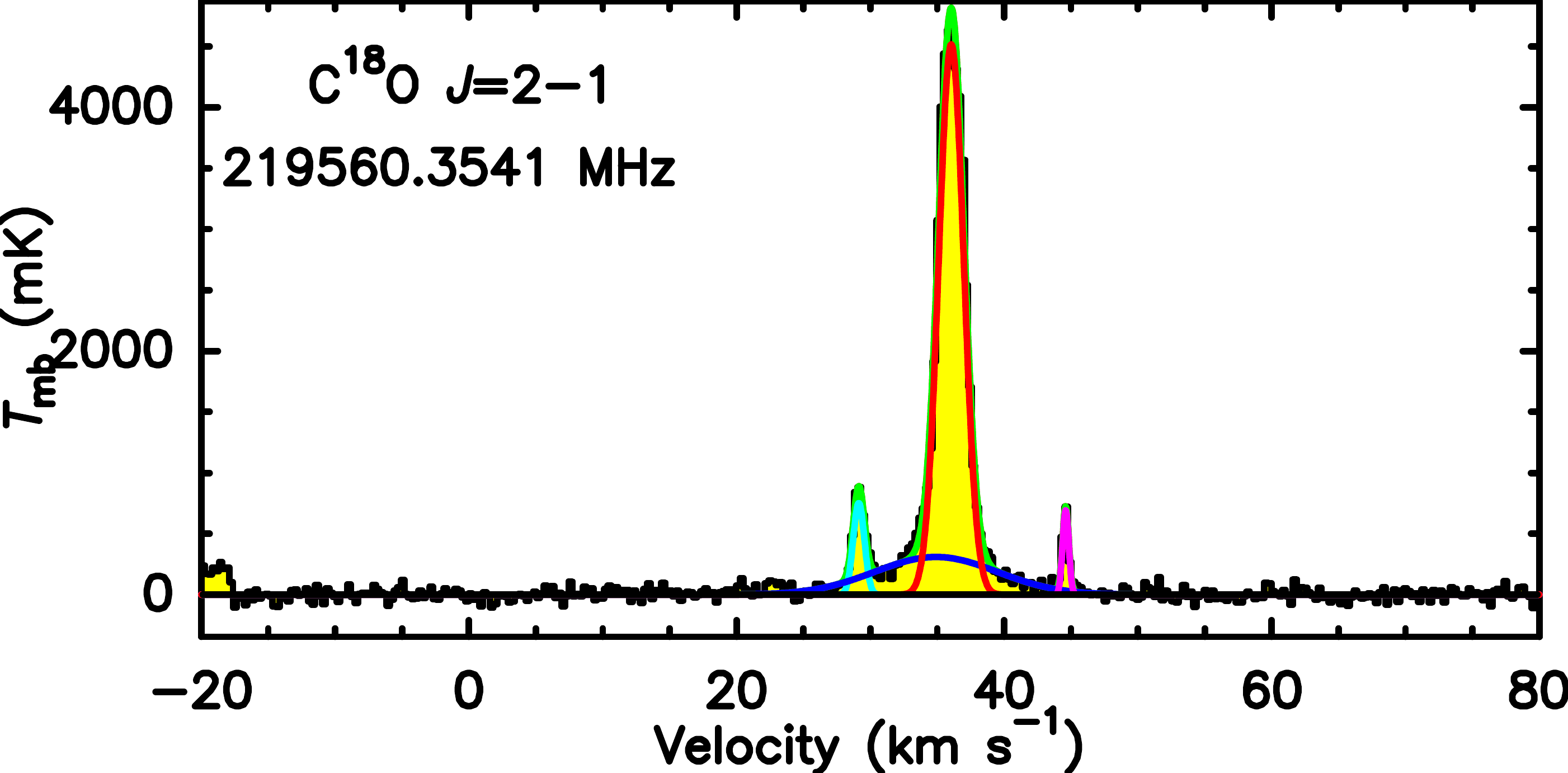}
 \caption{The spectrum (black and filled yellow) of I19312 showing the emission lines of $^{13}$CO and C$^{17}$O. 
 The cyan, red, and pink curves represent the Gaussian fitting profiles of the narrow composition. 
 The blue curve represents the Gaussian fitting profile of the broad composition. 
 The green curves are the total fitting profiles.
}
\label{Figure3}
\end{figure}

\begin{figure}
\centering
        \includegraphics[width=8cm]{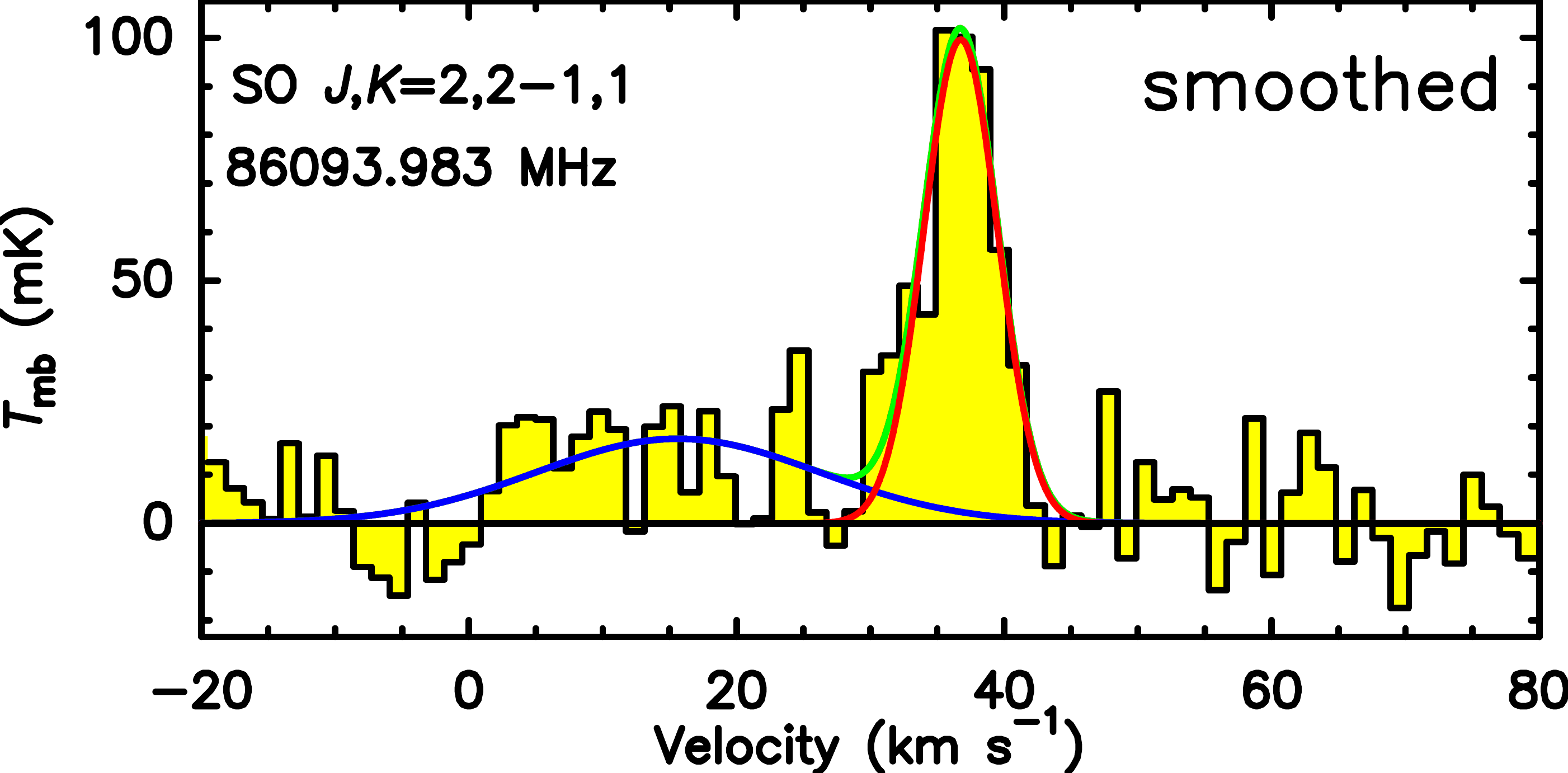}
        \includegraphics[width=8cm]{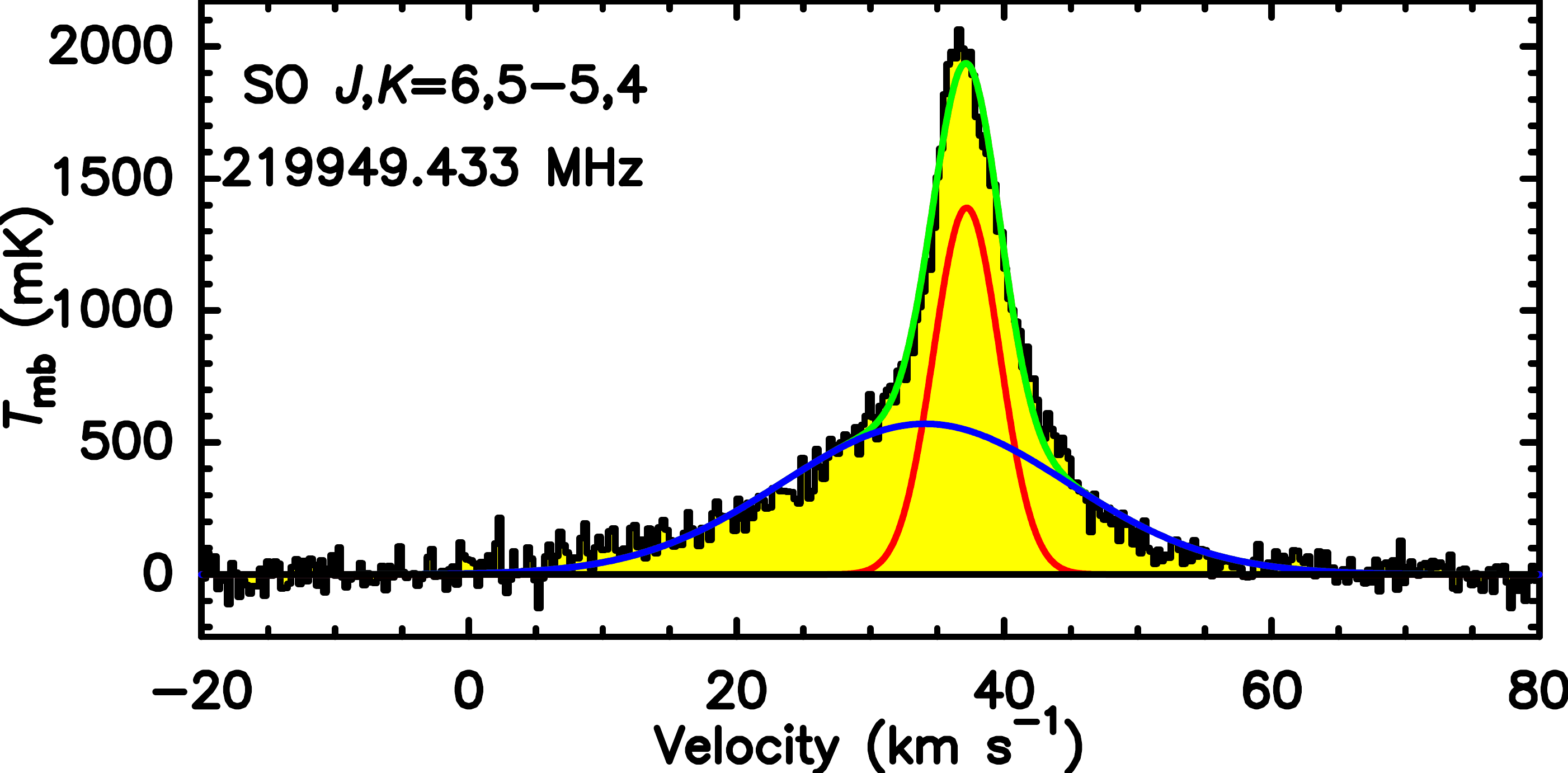}
 \caption{Same as Fig.\ref{Figure3} but for SO.
}
\label{Figure4}
\end{figure}

\begin{figure}
\centering
        \includegraphics[width=8cm]{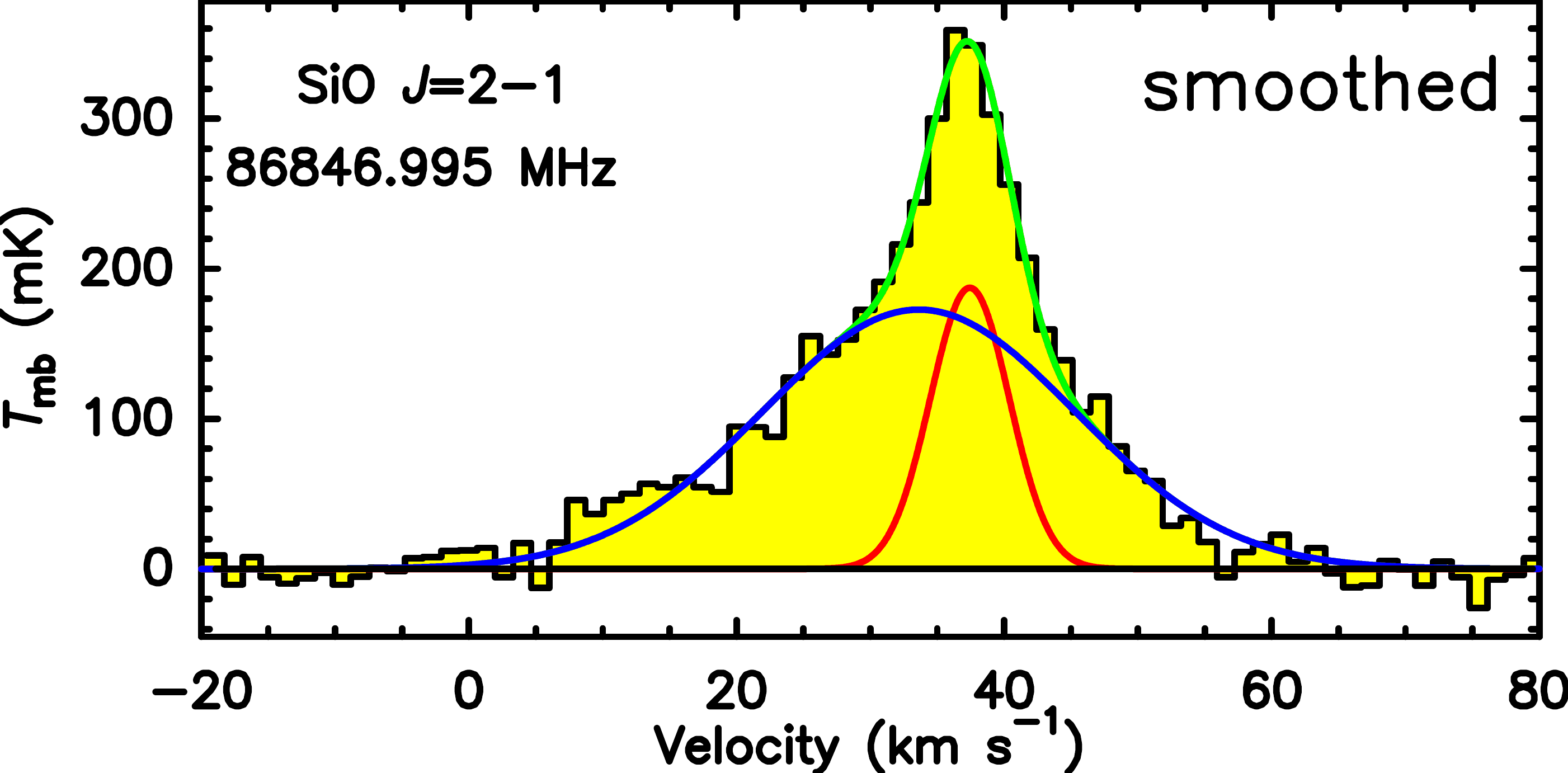}
        \includegraphics[width=8cm]{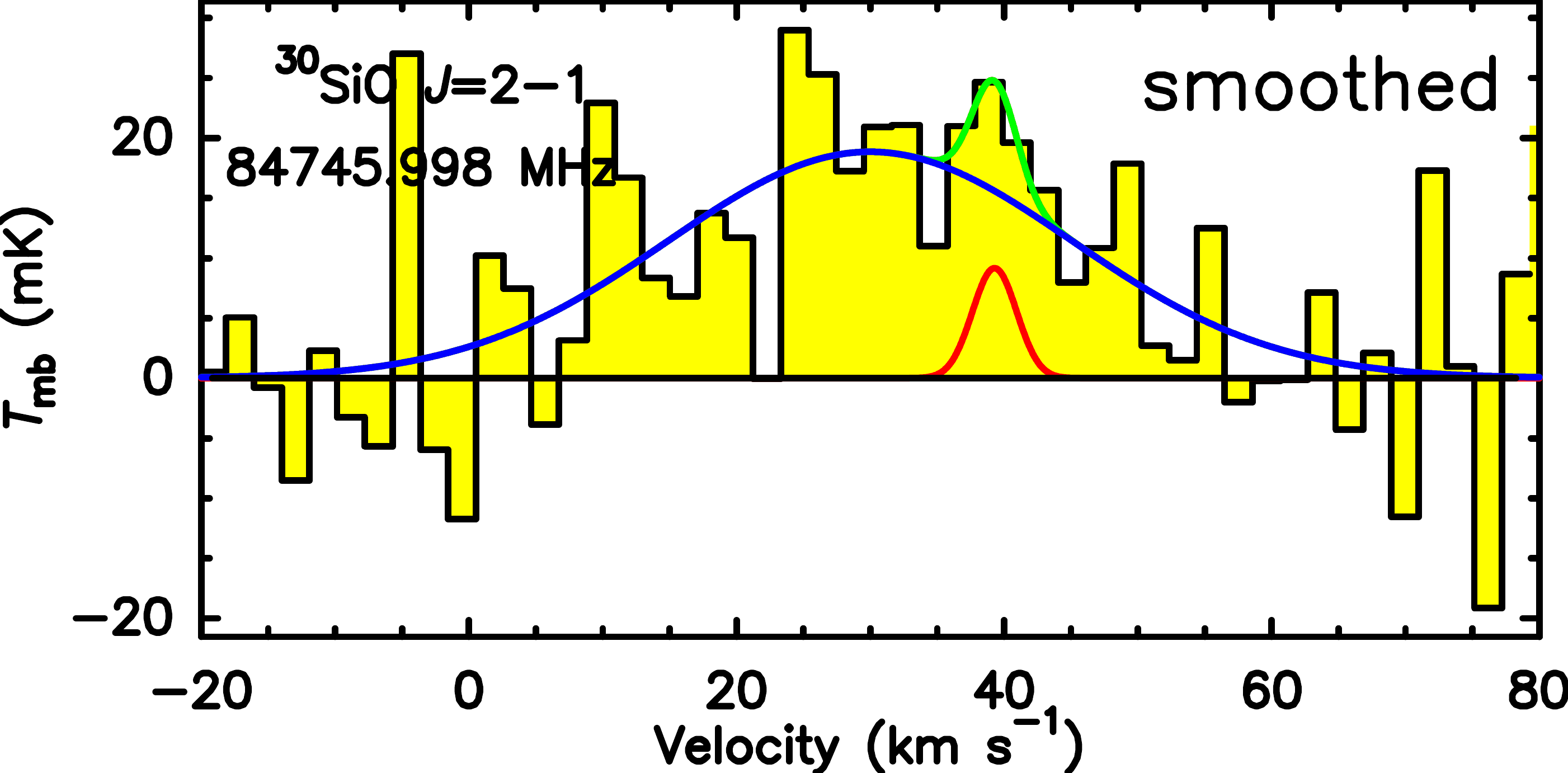}
        \includegraphics[width=8cm]{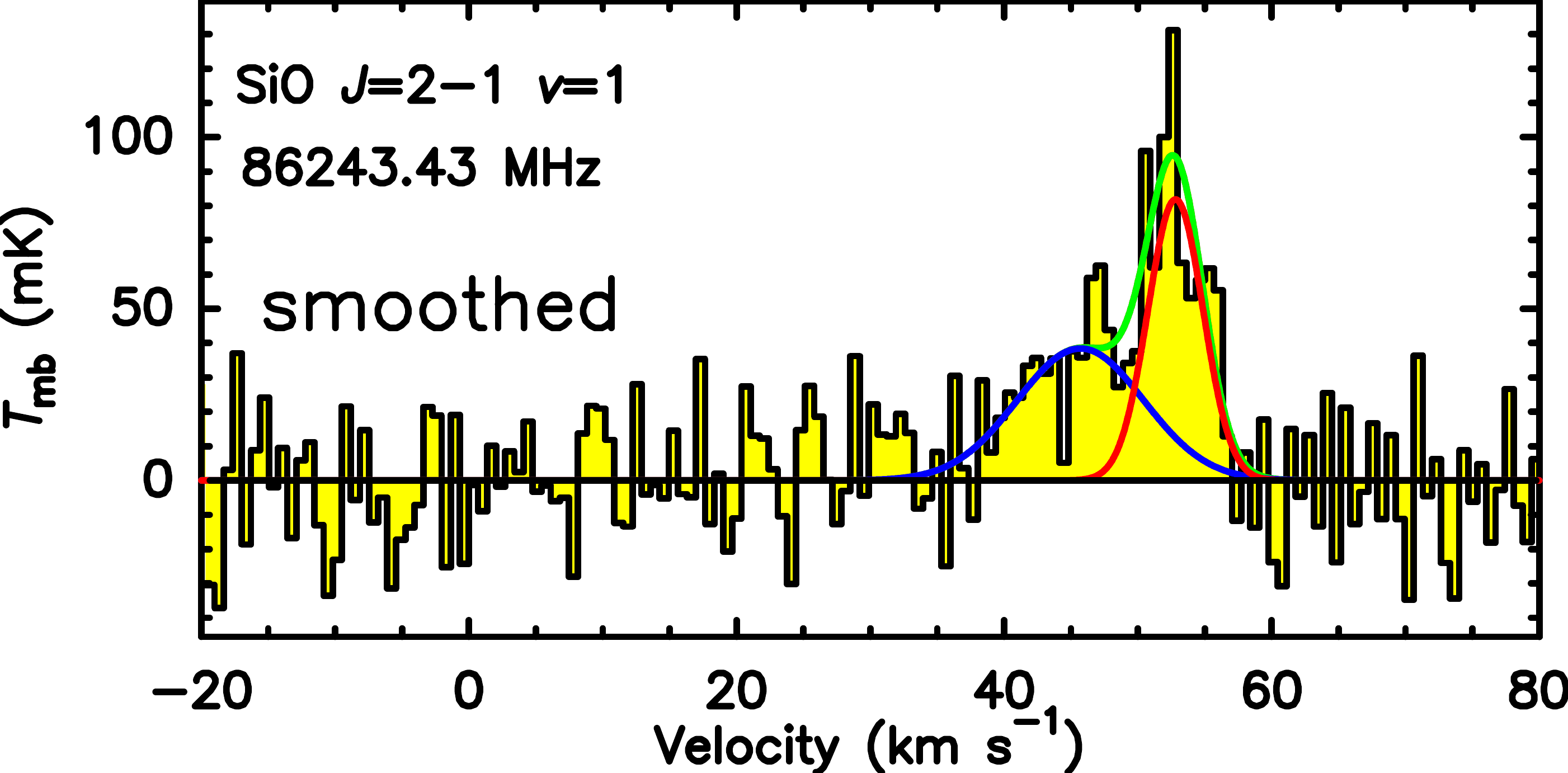}
        \includegraphics[width=8cm]{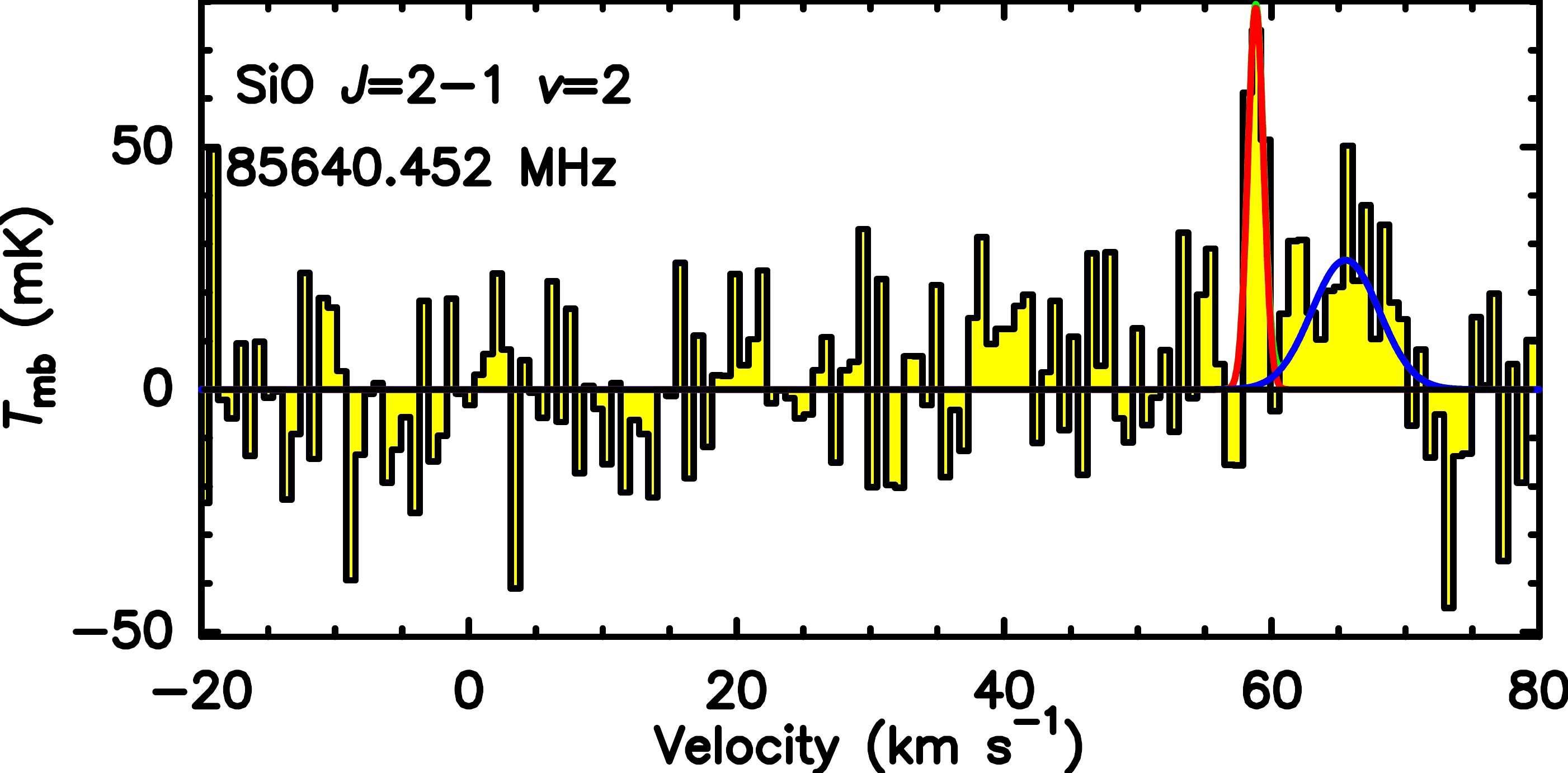}
        \includegraphics[width=8cm]{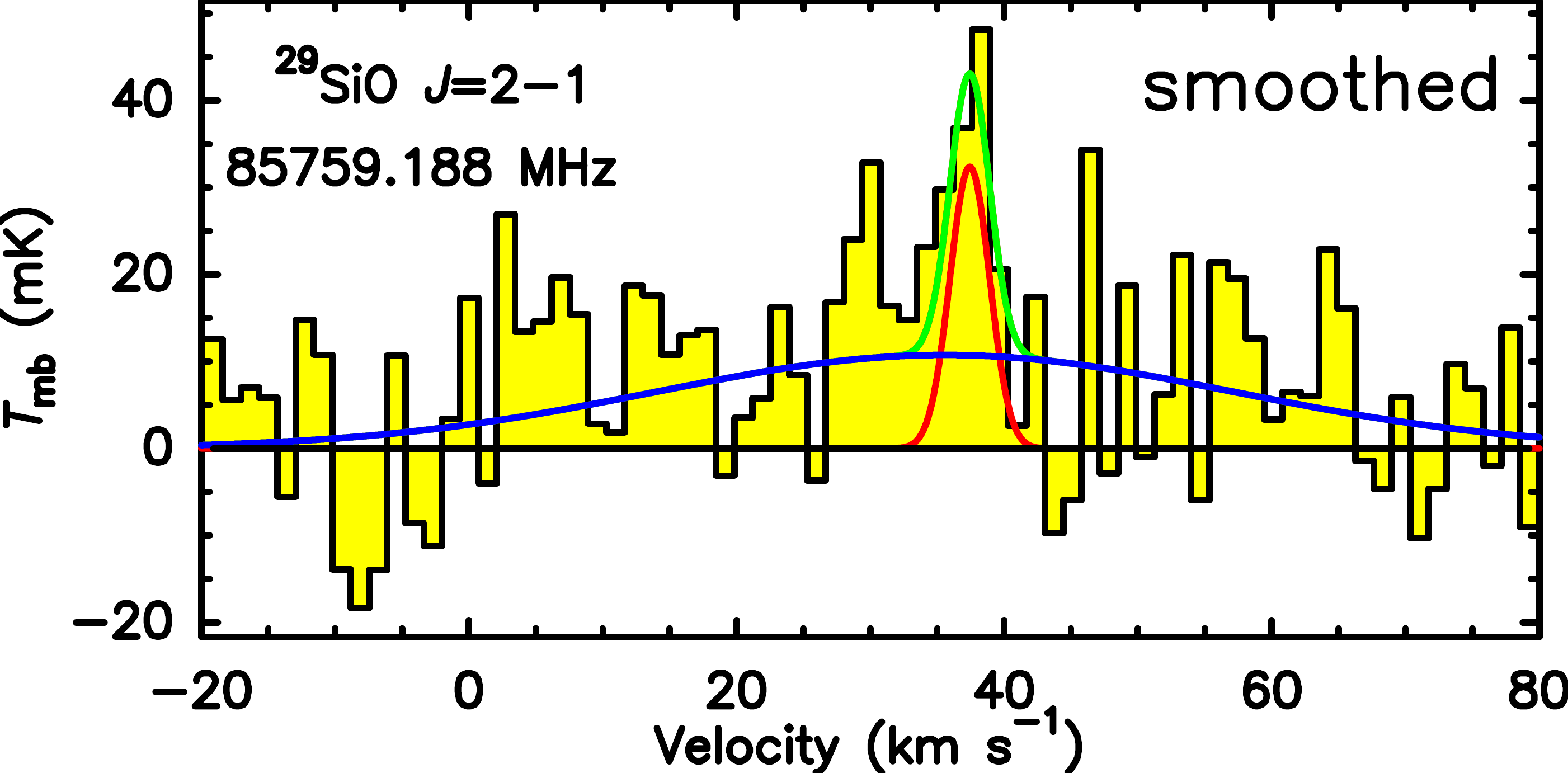}
 \caption{Same as Fig.~\ref{Figure3} but for SiO and $^{30}$SiO.
}
\label{Figure5}
\end{figure}

\begin{figure}
\centering
        \includegraphics[width=8cm]{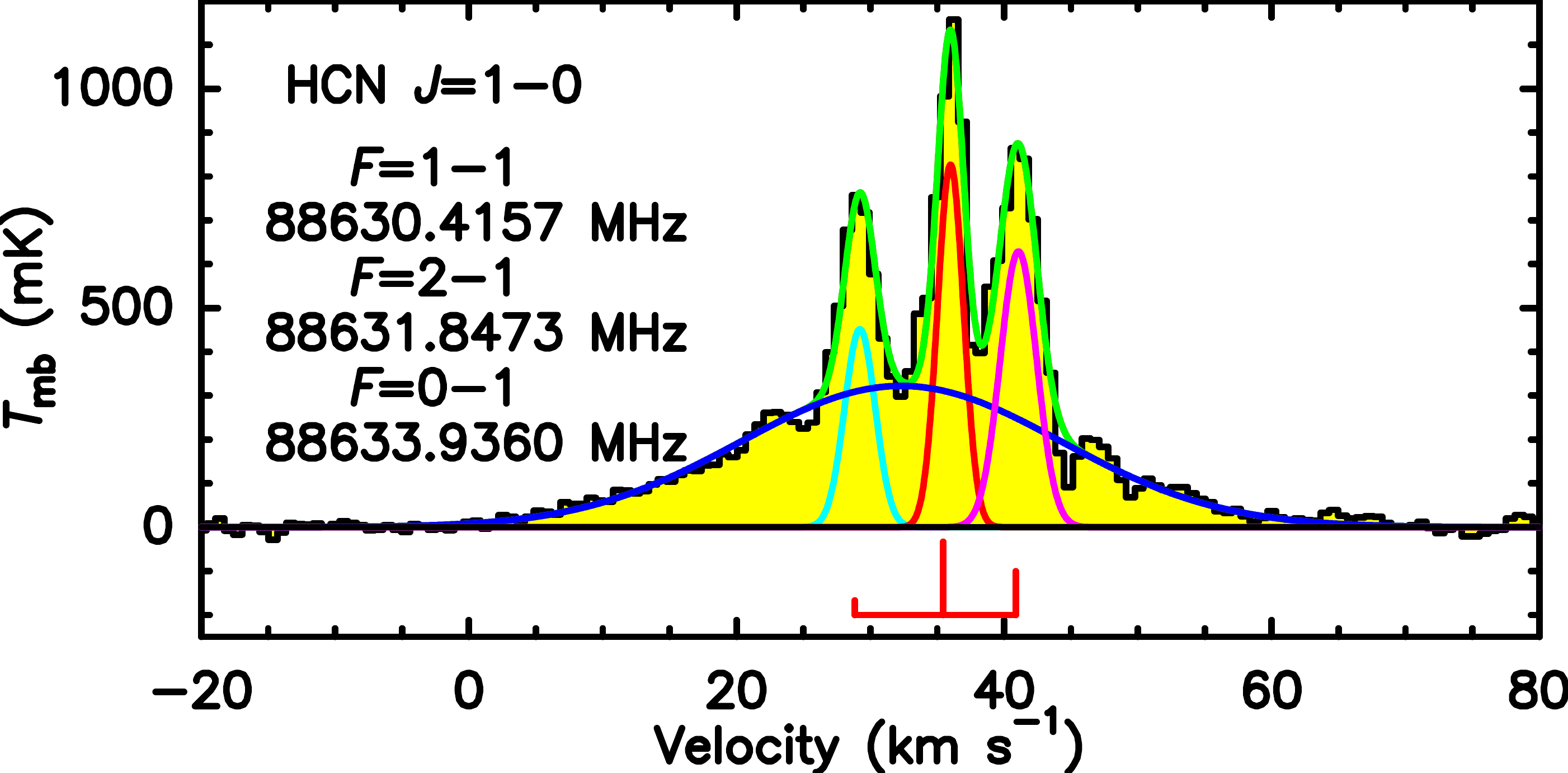}
        \includegraphics[width=8cm]{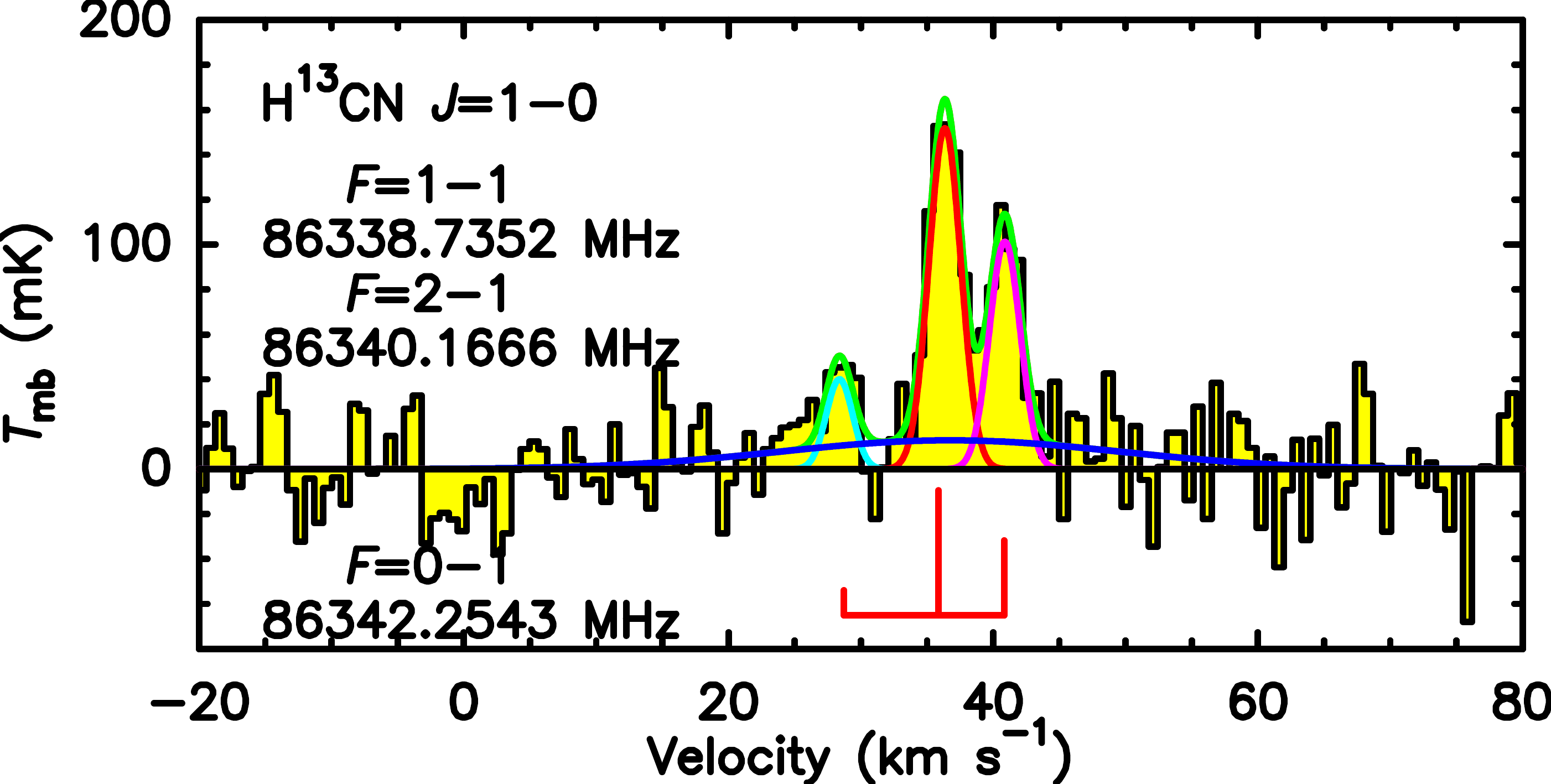}
        \includegraphics[width=8cm]{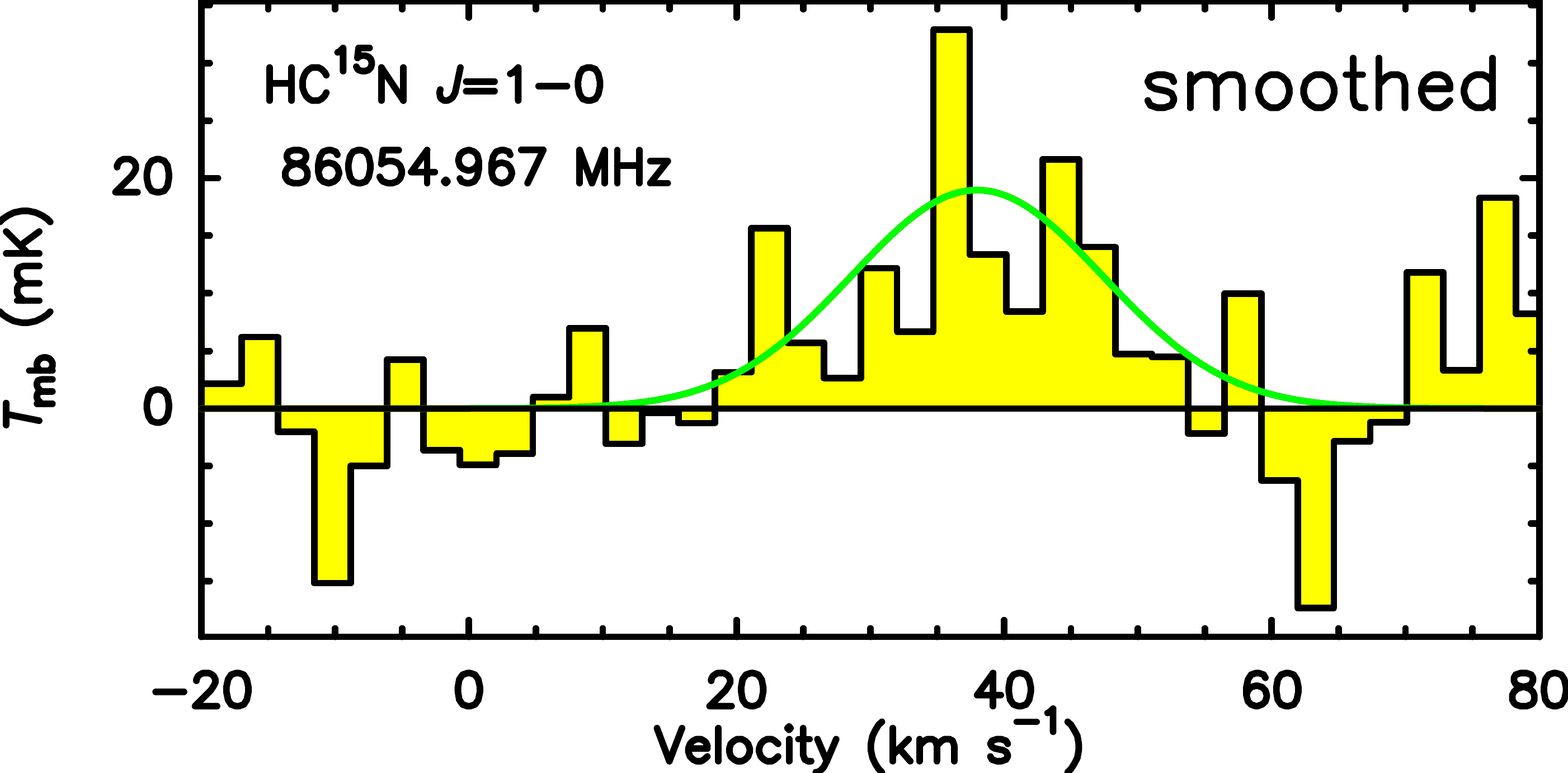}
 \caption{Same as Fig.~\ref{Figure3} but for HCN, H$^{13}$CN, and HC$^{15}$N.
 The vertical lines mark the positions and relative intensities of hyperfine components. 
}
\label{Figure6}
\end{figure}

\begin{figure}
\centering
        \includegraphics[width=8cm]{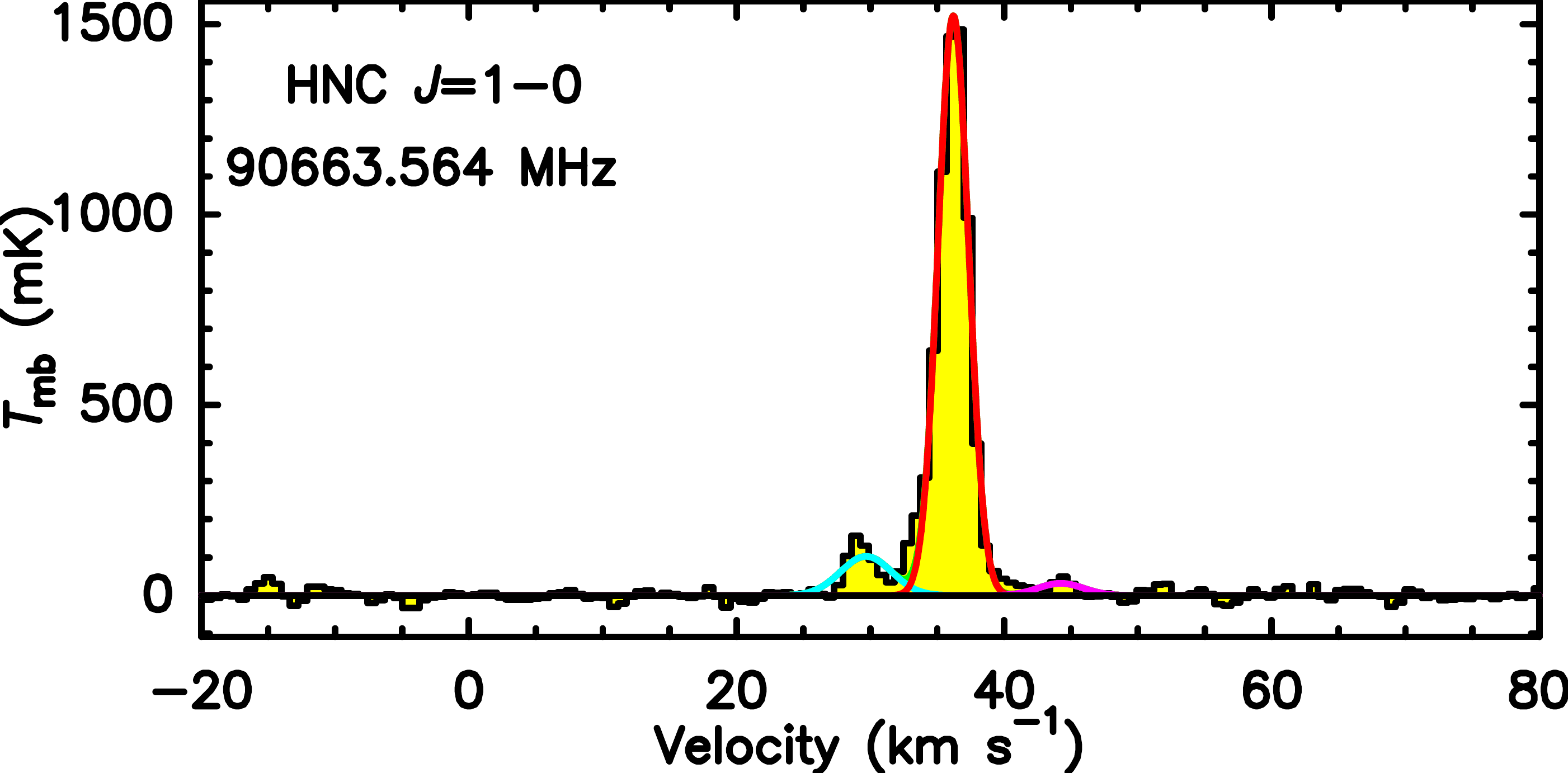}
        \includegraphics[width=8cm]{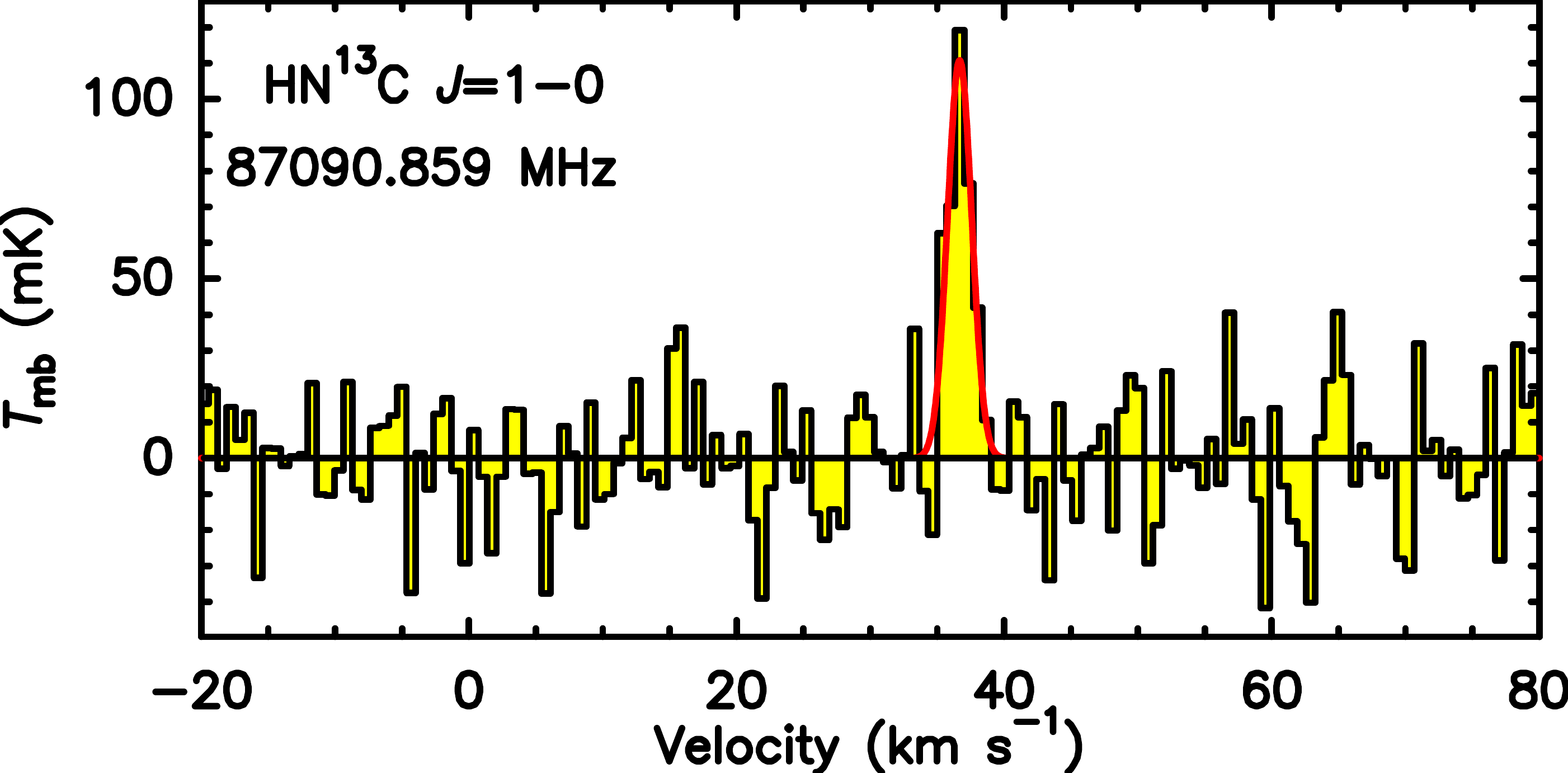}
 \caption{Same as Fig.~\ref{Figure3} but for HNC and HN$^{13}$C.
}
\label{Figure7}
\end{figure}

\begin{figure}
\centering
        \includegraphics[width=8cm]{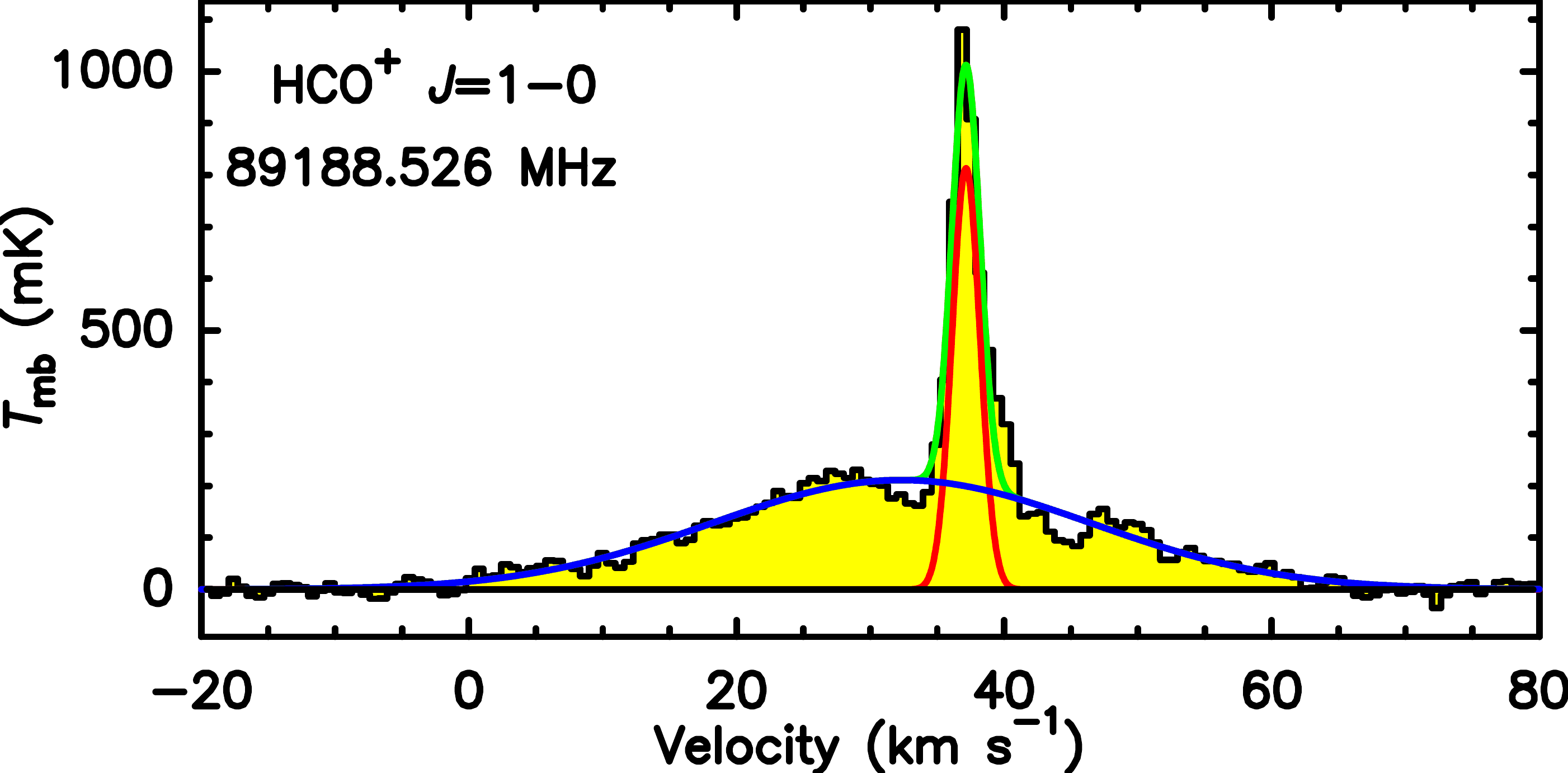}
        \includegraphics[width=8cm]{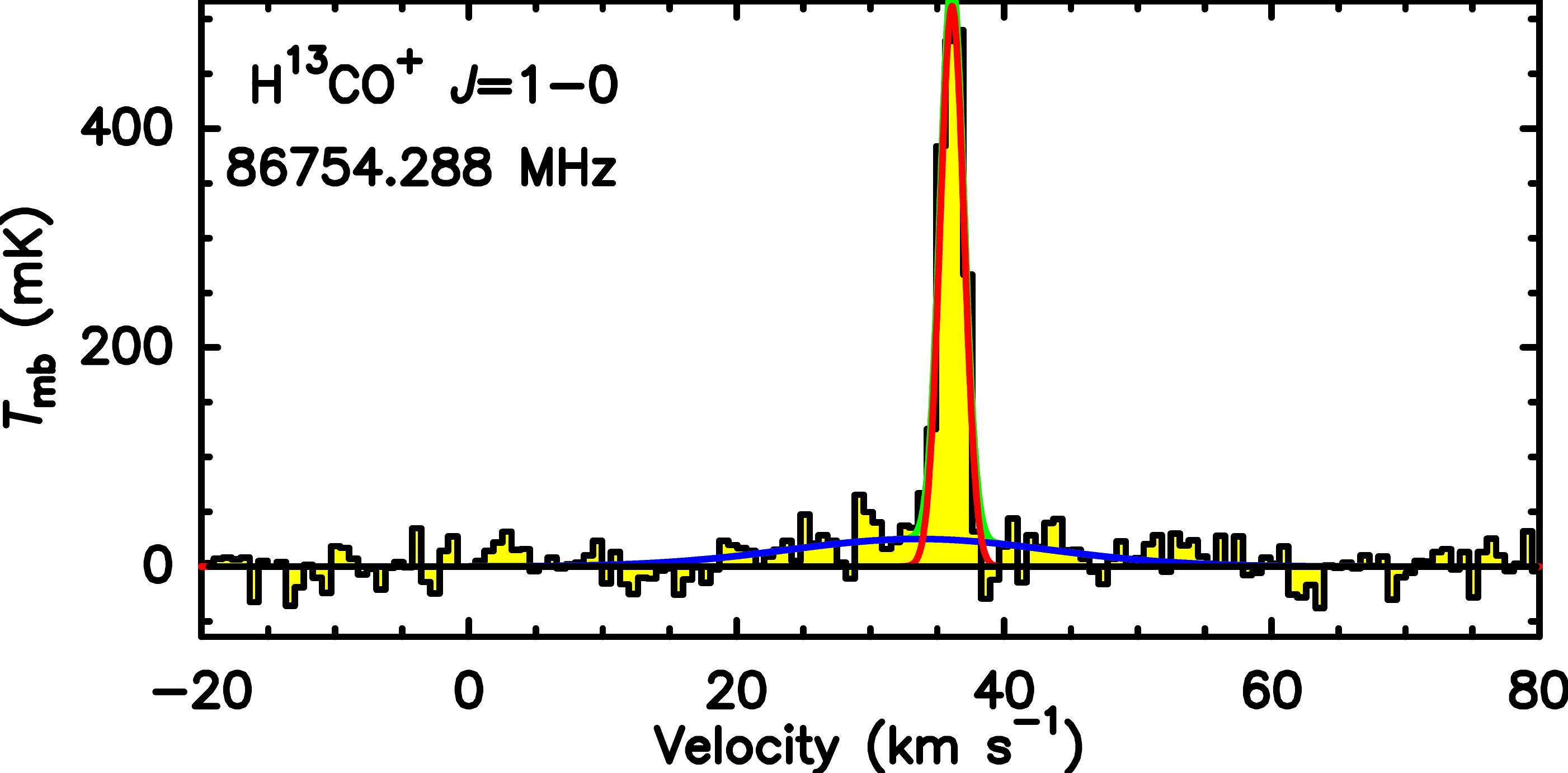}
        \includegraphics[width=8cm]{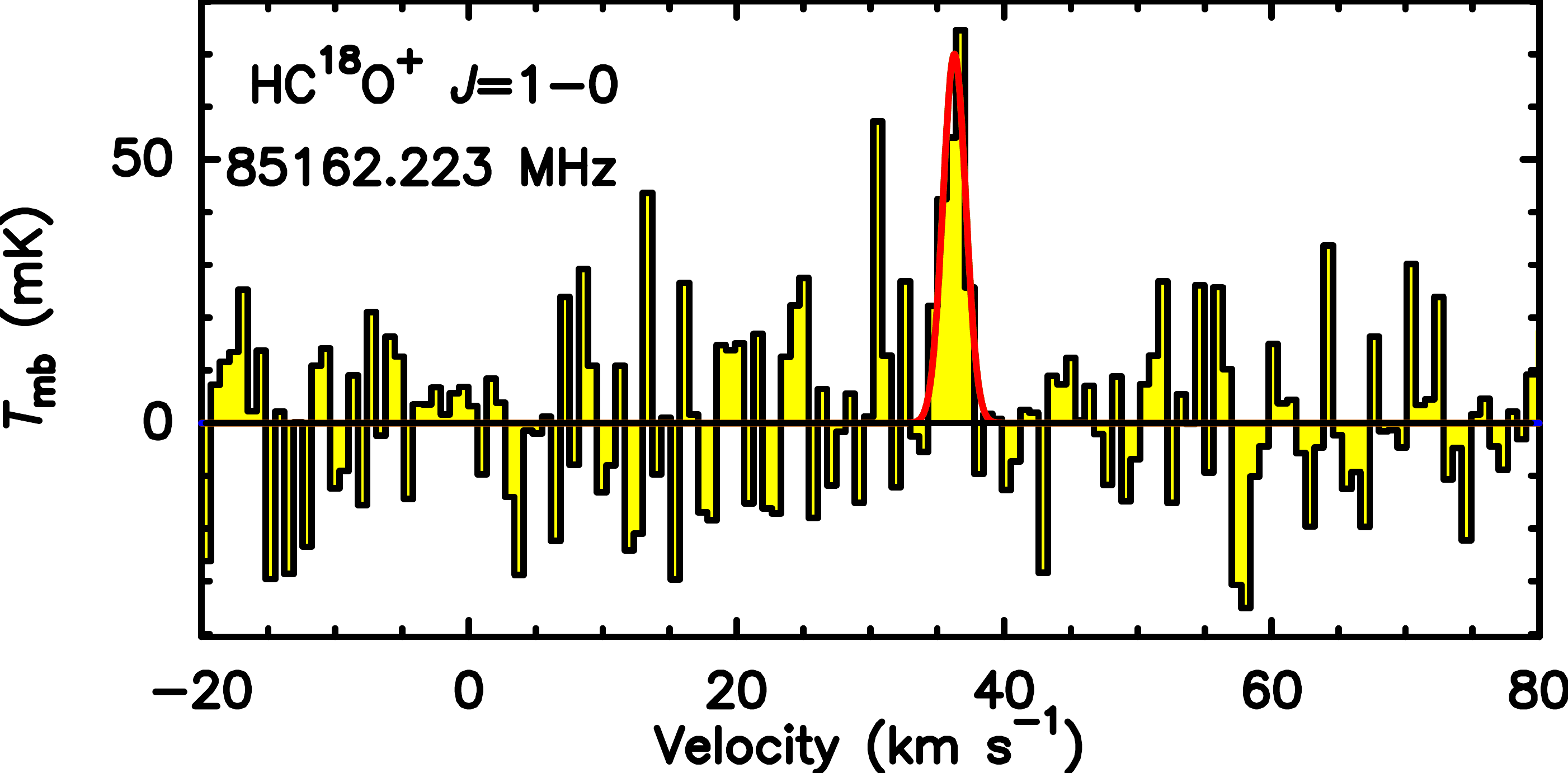}
 \caption{Same as Fig.~\ref{Figure3} but for HCO$^{+}$, H$^{13}$CO$^{+}$, and HC$^{18}$O$^{+}$.
}
\label{Figure8}
\end{figure}

\begin{figure}
\centering
        \includegraphics[width=8cm]{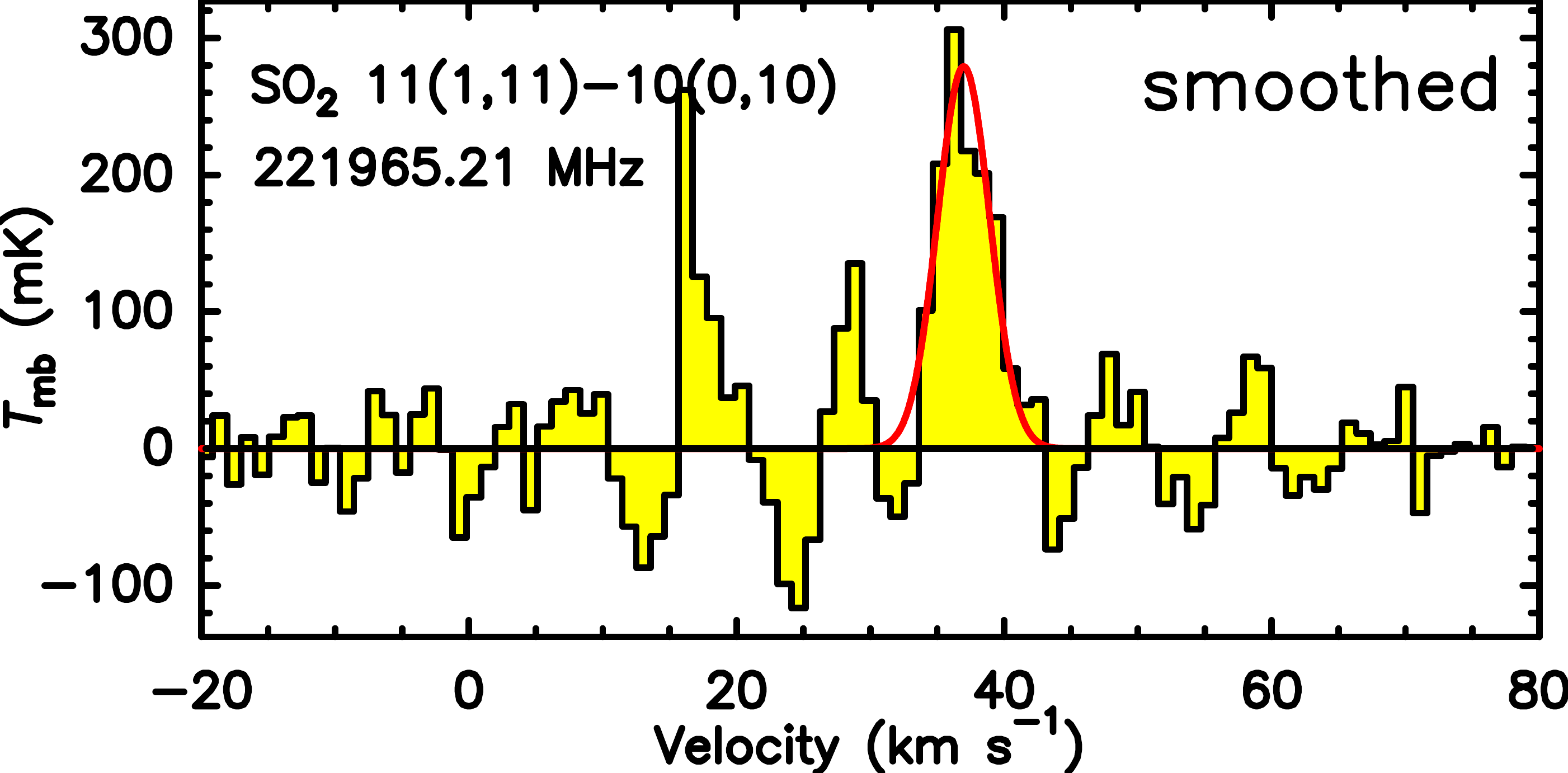}
 \caption{Same as Fig.~\ref{Figure3} but for SO$_{2}$.
}
\label{Figure9}
\end{figure}

\begin{figure}
\centering
        \includegraphics[width=8cm]{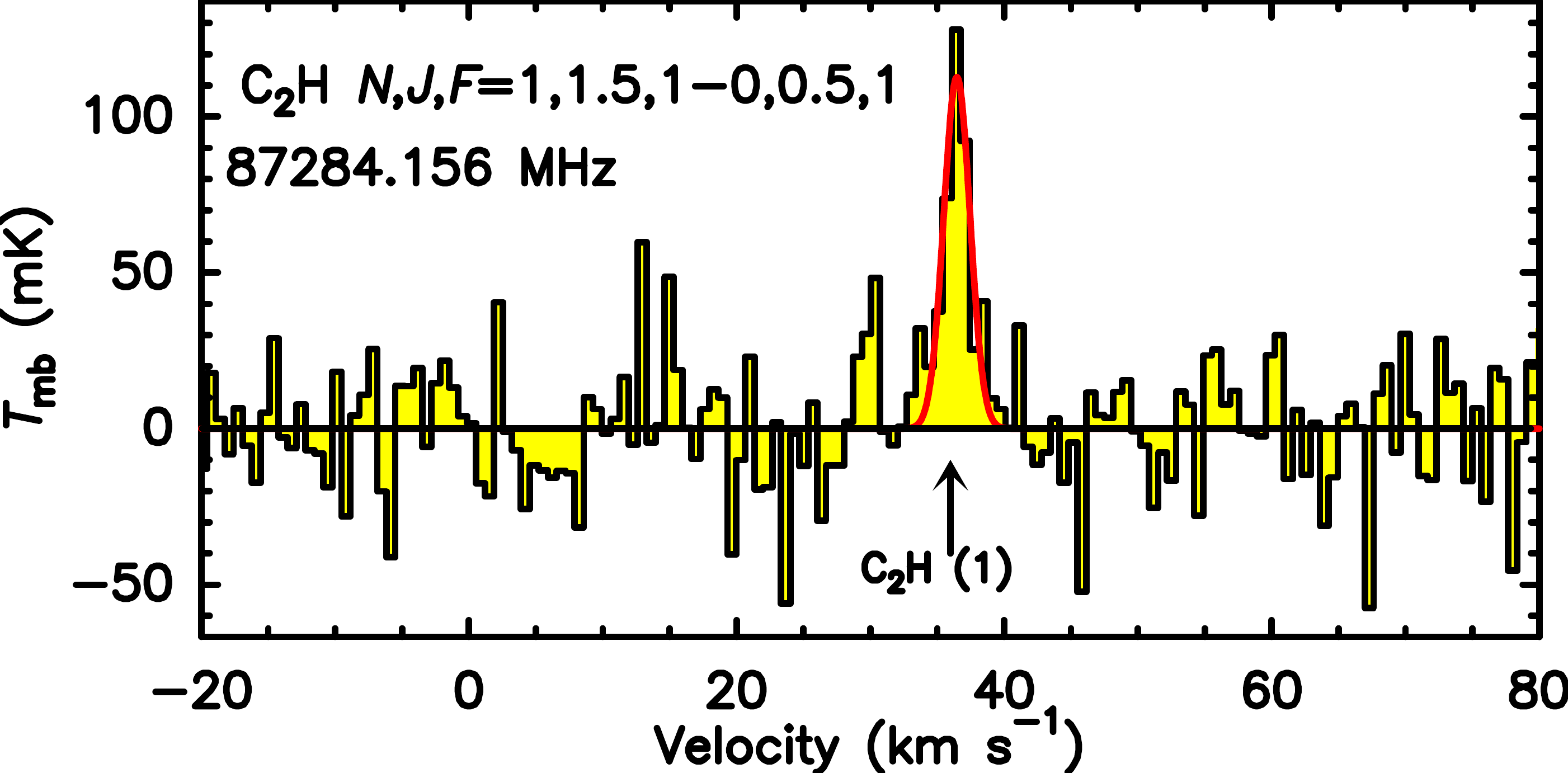}
        \includegraphics[width=8cm]{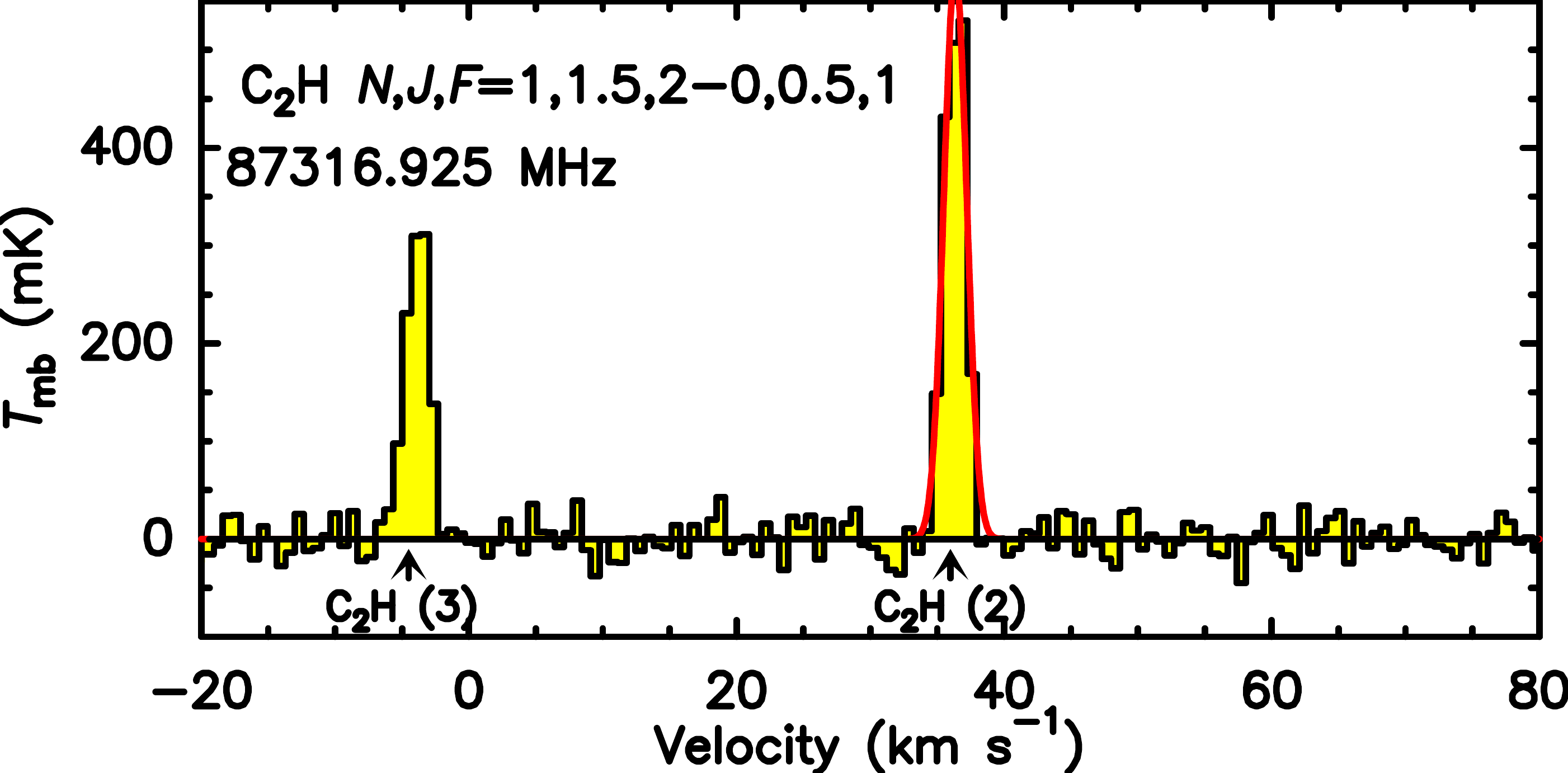}
        \includegraphics[width=8cm]{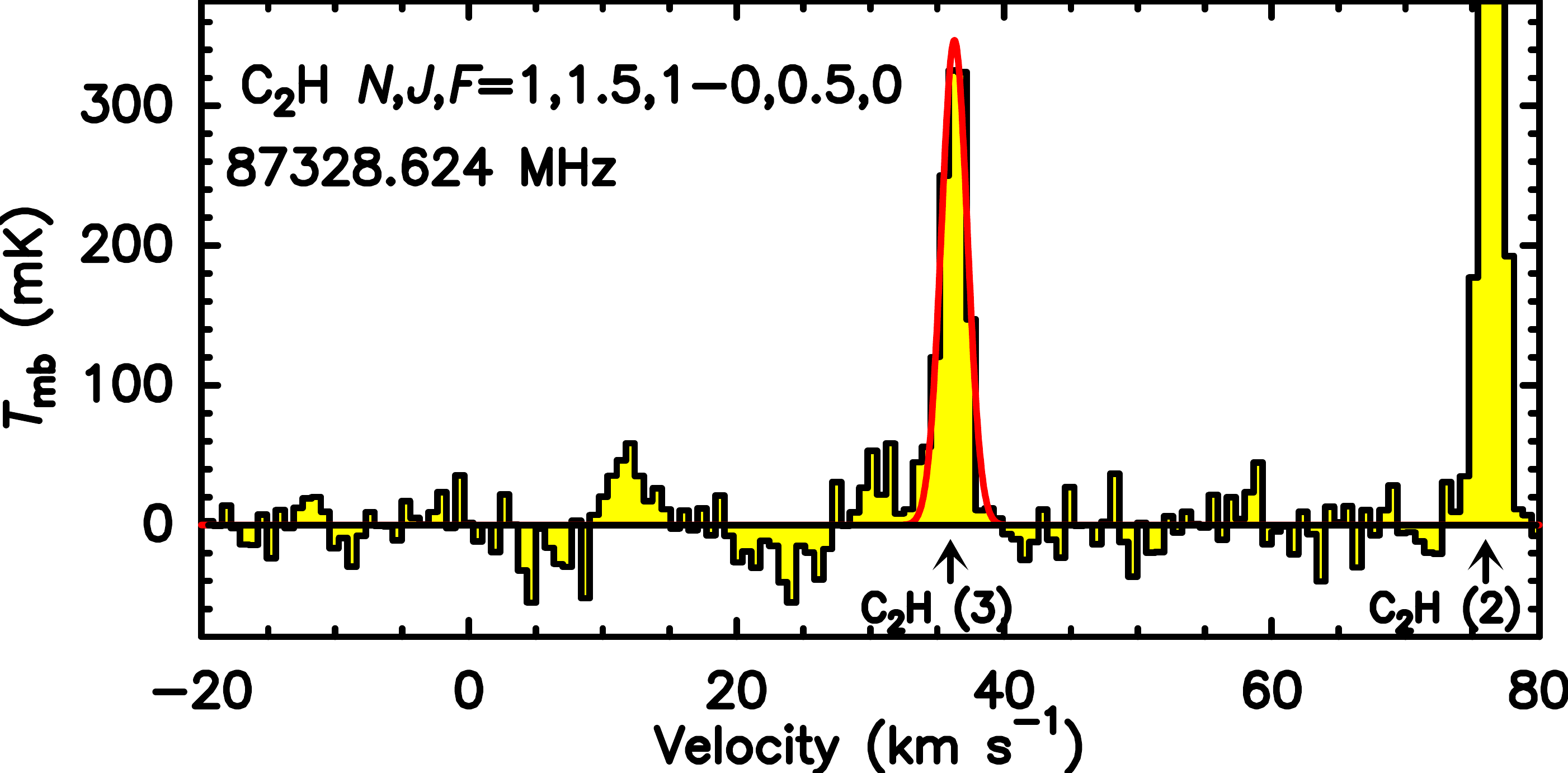}
        \includegraphics[width=8cm]{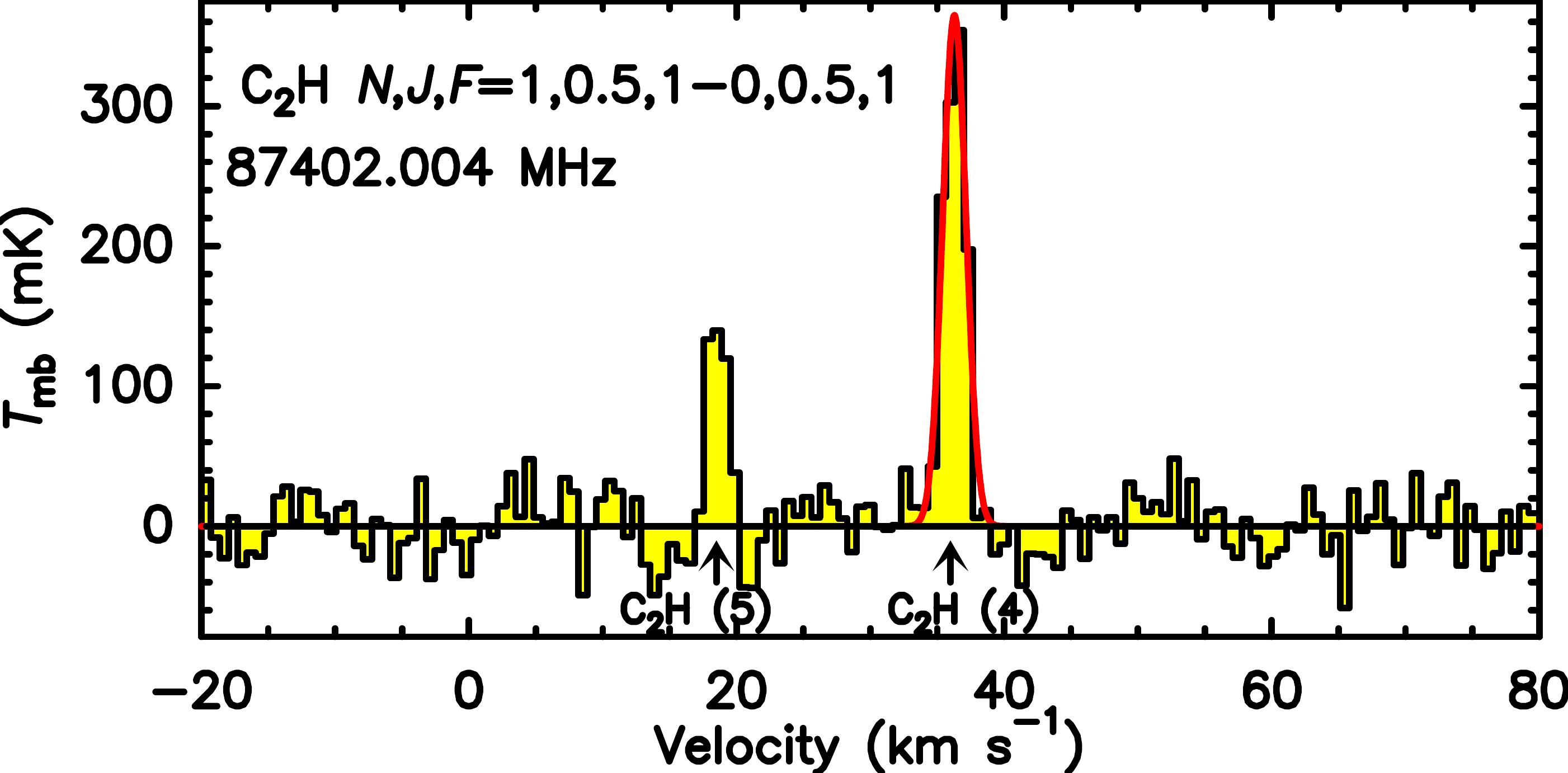}
        \includegraphics[width=8cm]{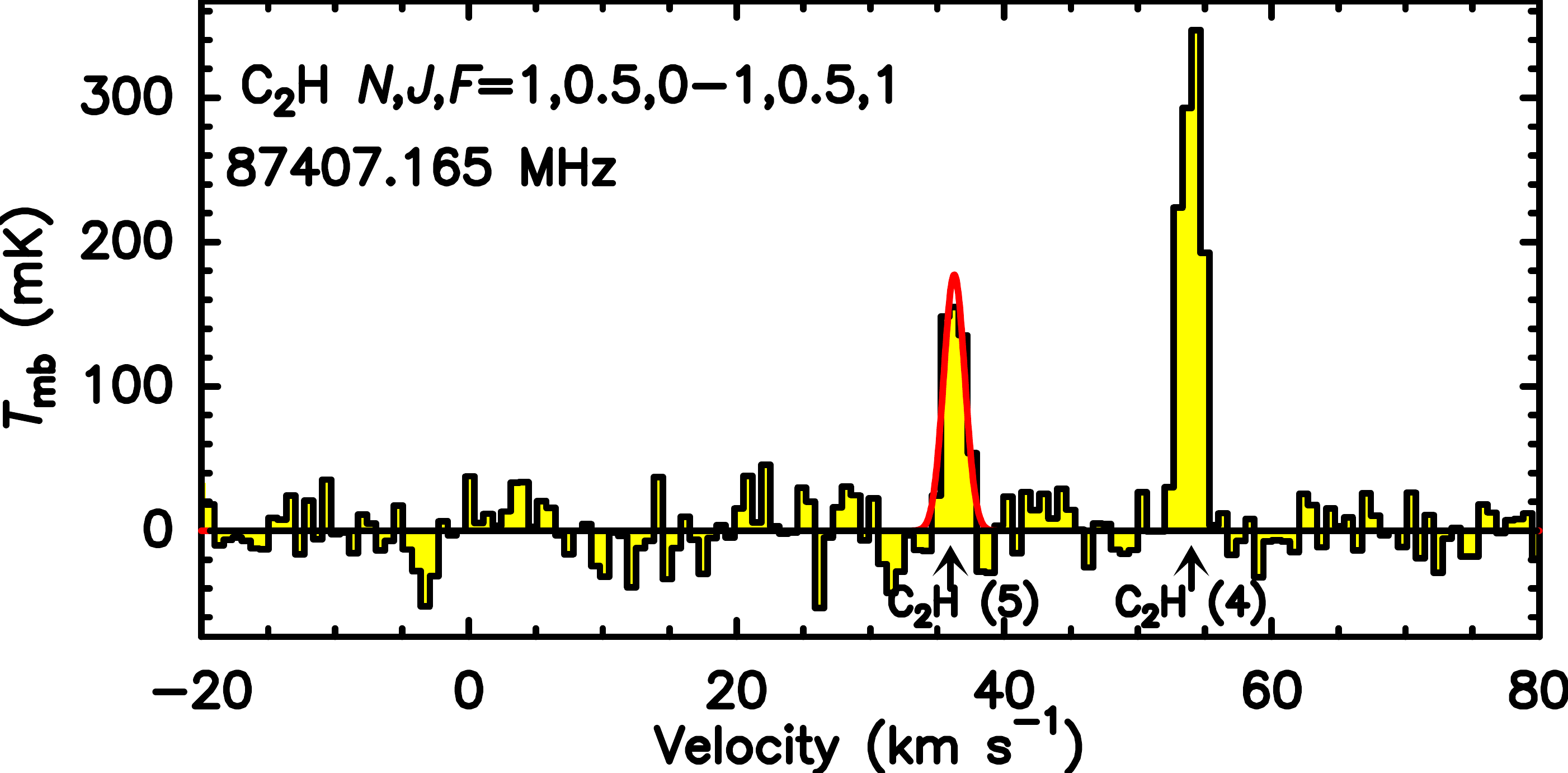}
        \includegraphics[width=8cm]{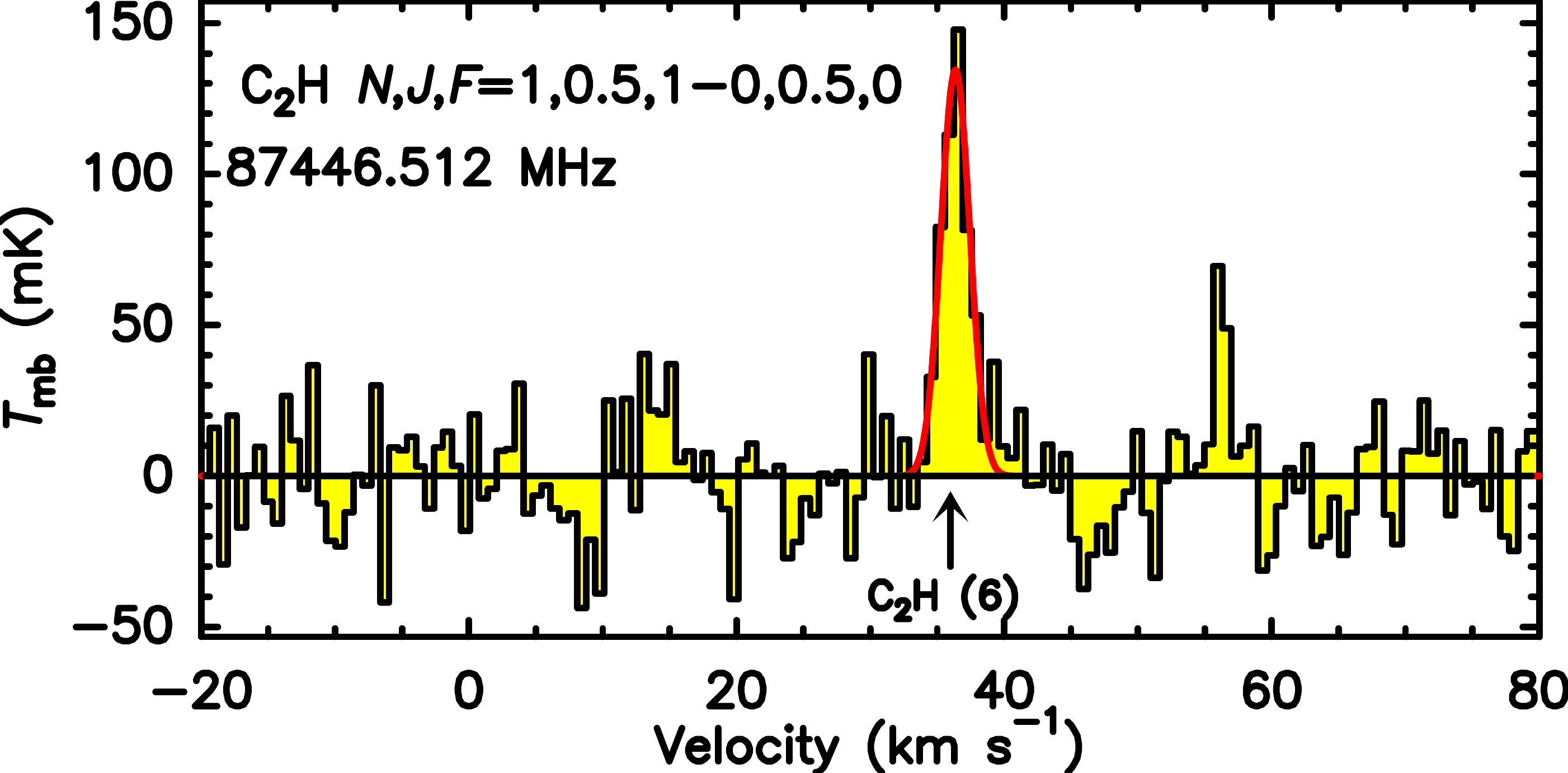}
 \caption{Same as Fig.~\ref{Figure3} but for C$_{2}$H.
}
\label{Figure10}
\end{figure}

\begin{figure}
\centering
        \includegraphics[width=8cm]{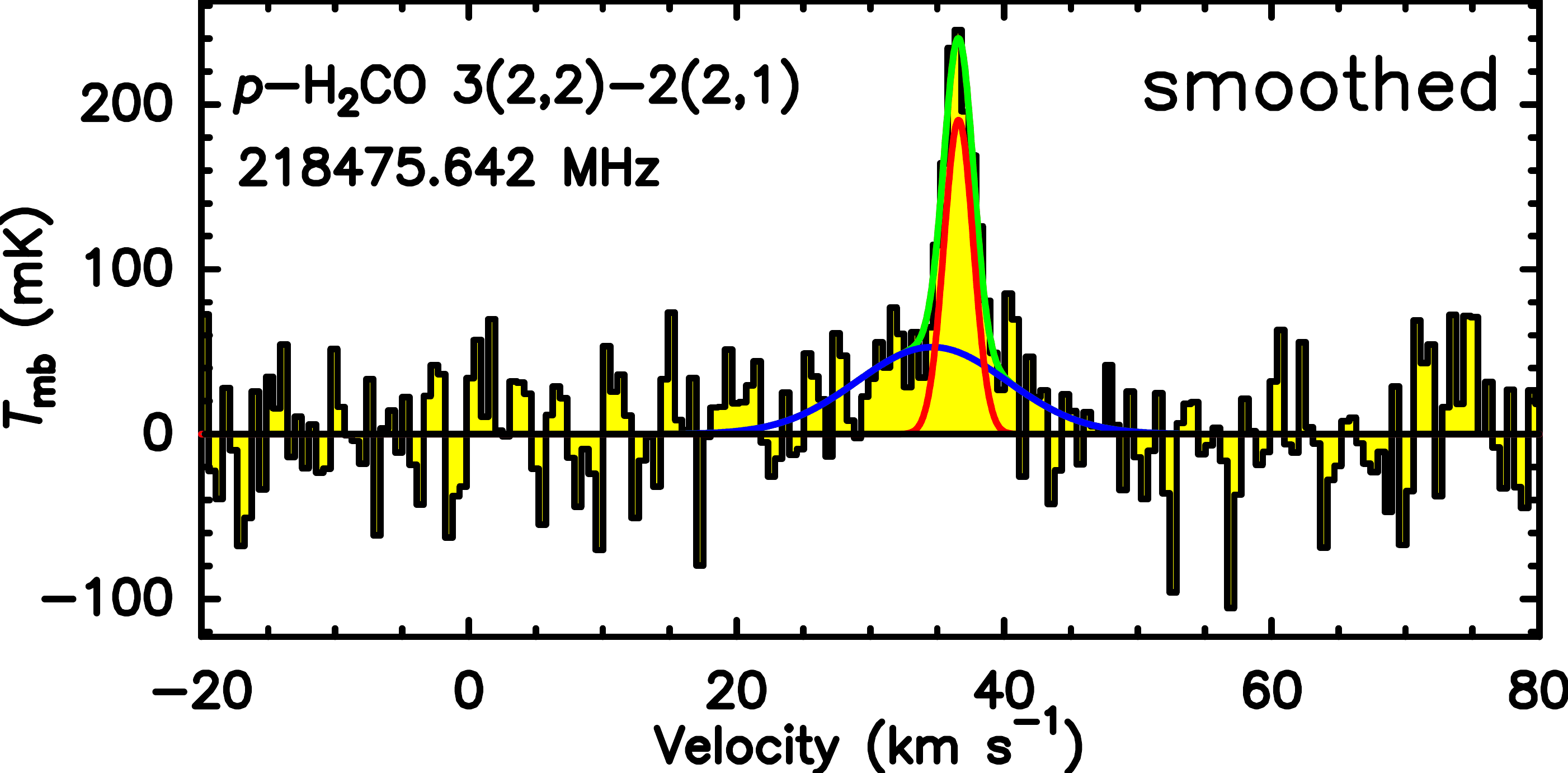}
        \includegraphics[width=8cm]{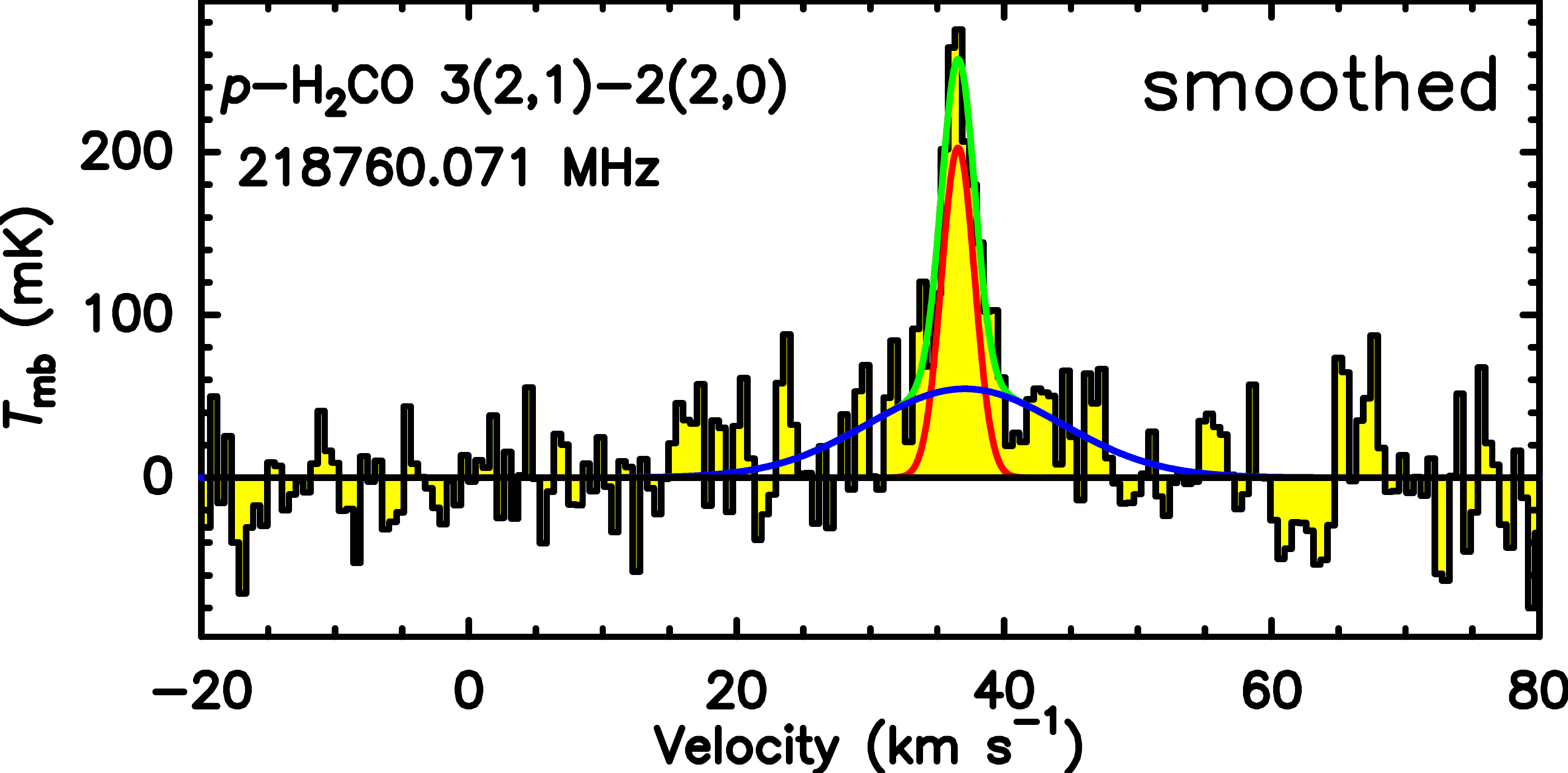}
        \includegraphics[width=8cm]{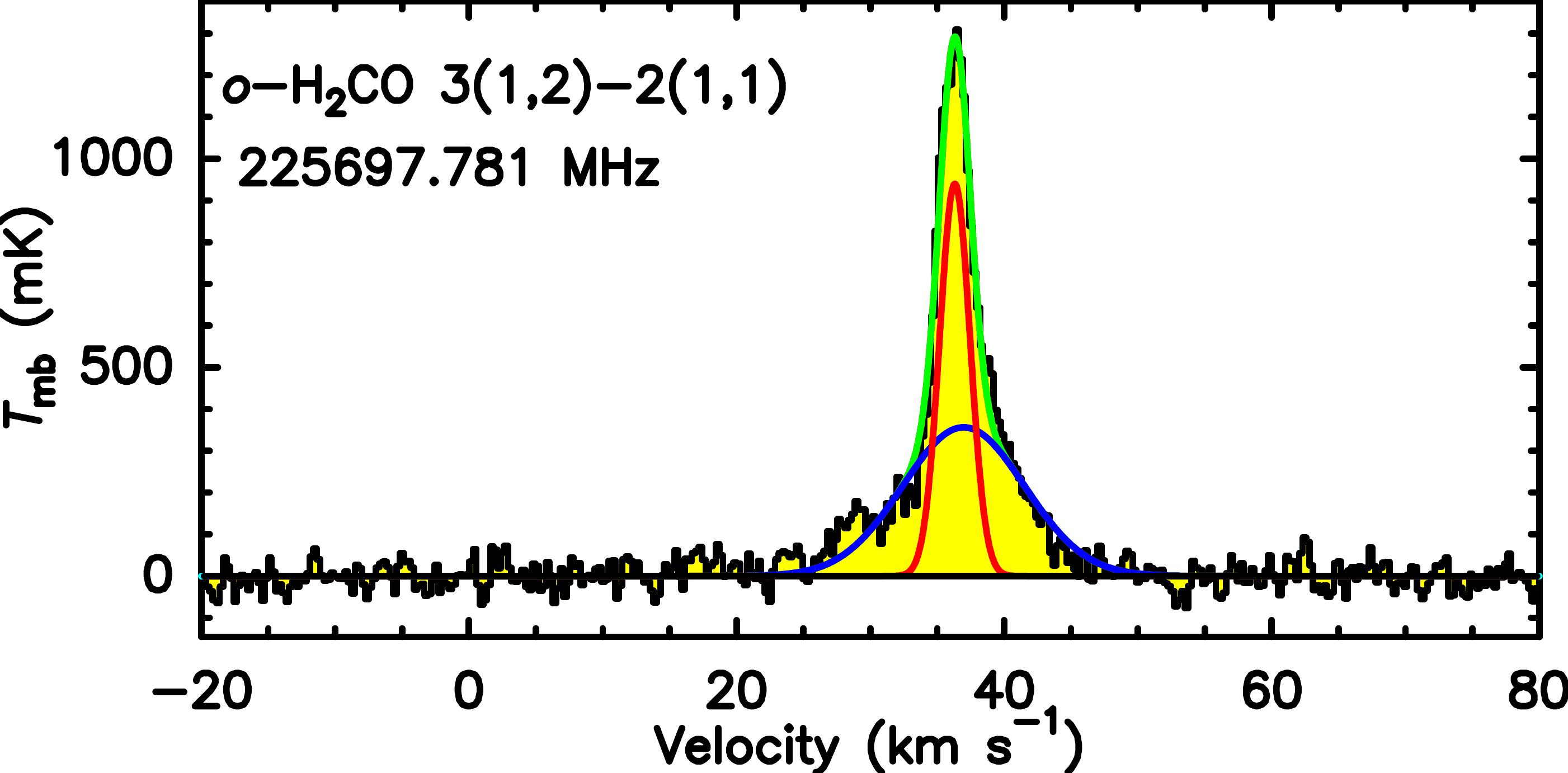}
 \caption{Same as Fig.~\ref{Figure3} but for $p$-H$_{2}$CO and $o$-H$_{2}$CO.
}
\label{Figure11}
\end{figure}

\begin{figure}
\centering
        \includegraphics[width=8cm]{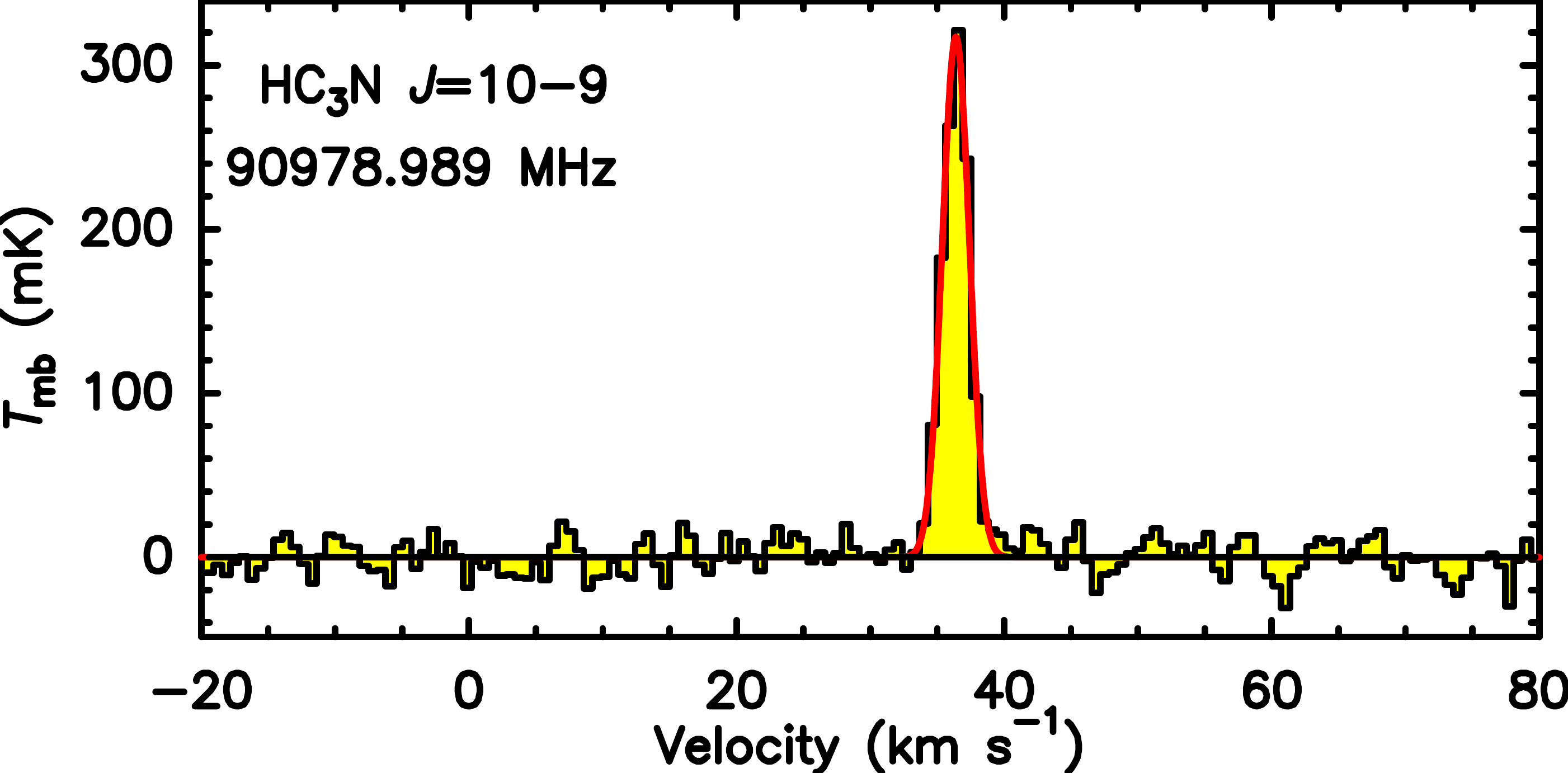}
        \includegraphics[width=8cm]{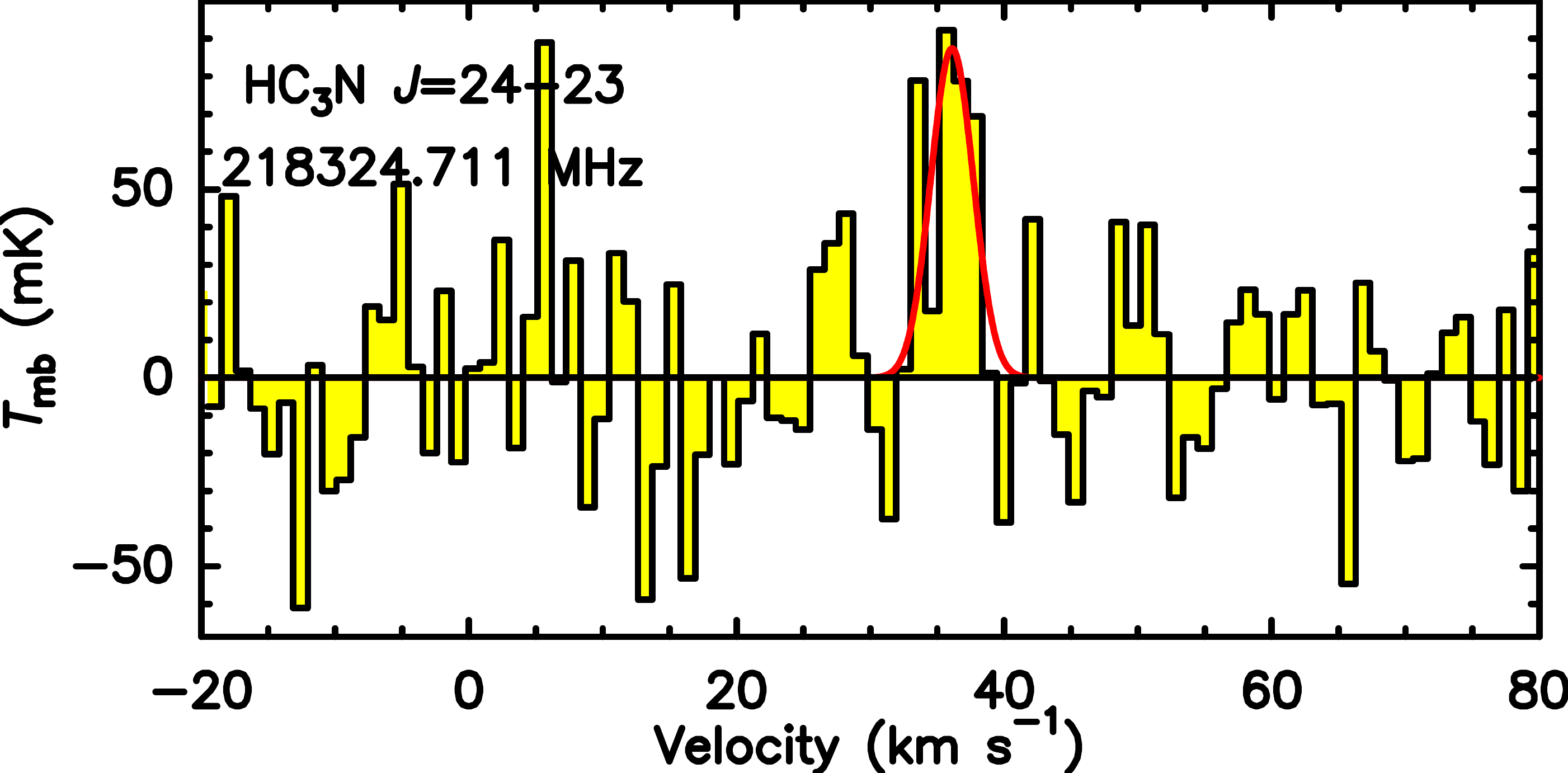}
 \caption{Same as Fig.~\ref{Figure3} but for HC$_{3}$N.
}
\label{Figure12}
\end{figure}

\begin{figure}
\centering
        \includegraphics[width=8cm]{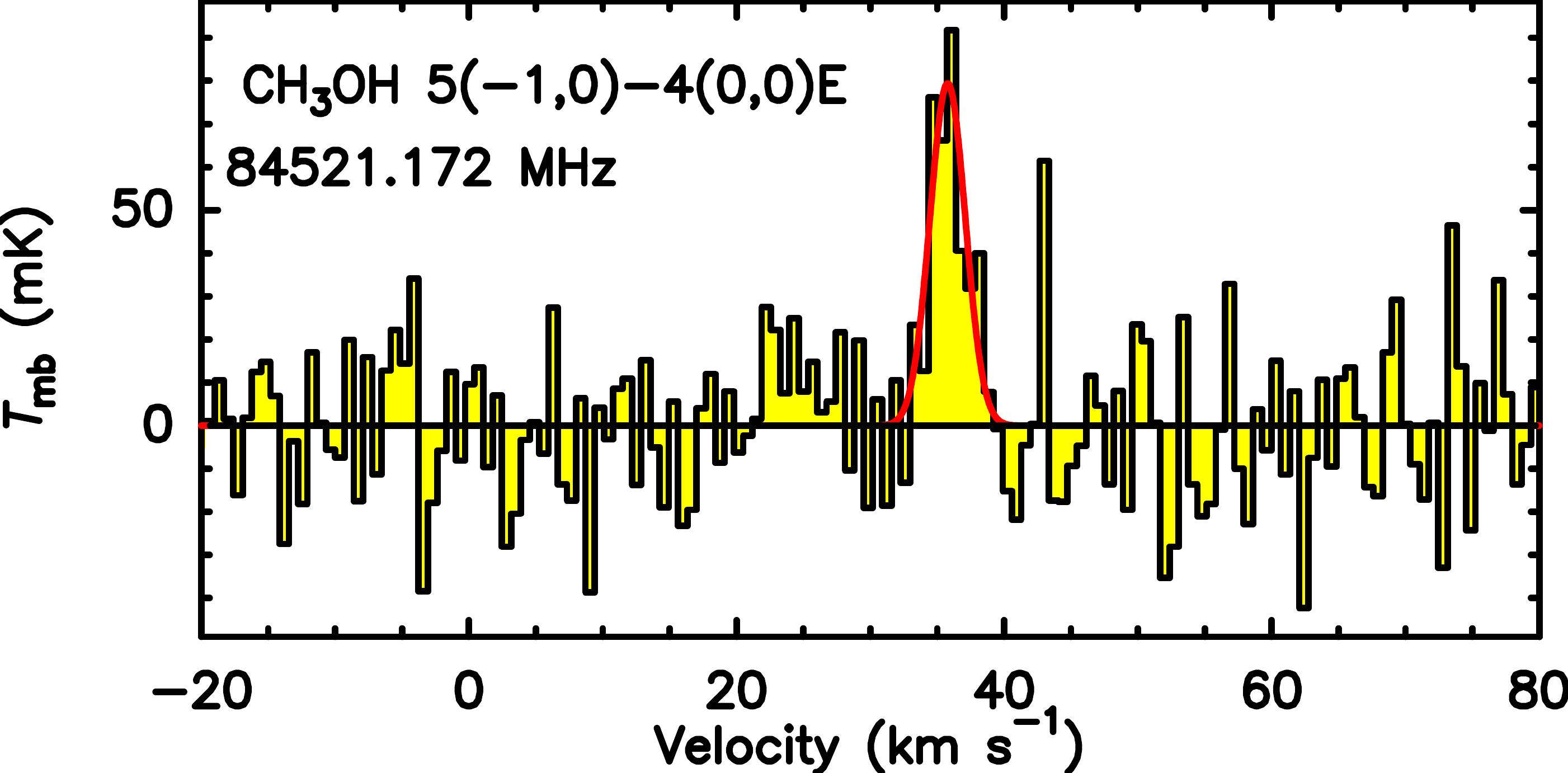}
        \includegraphics[width=8cm]{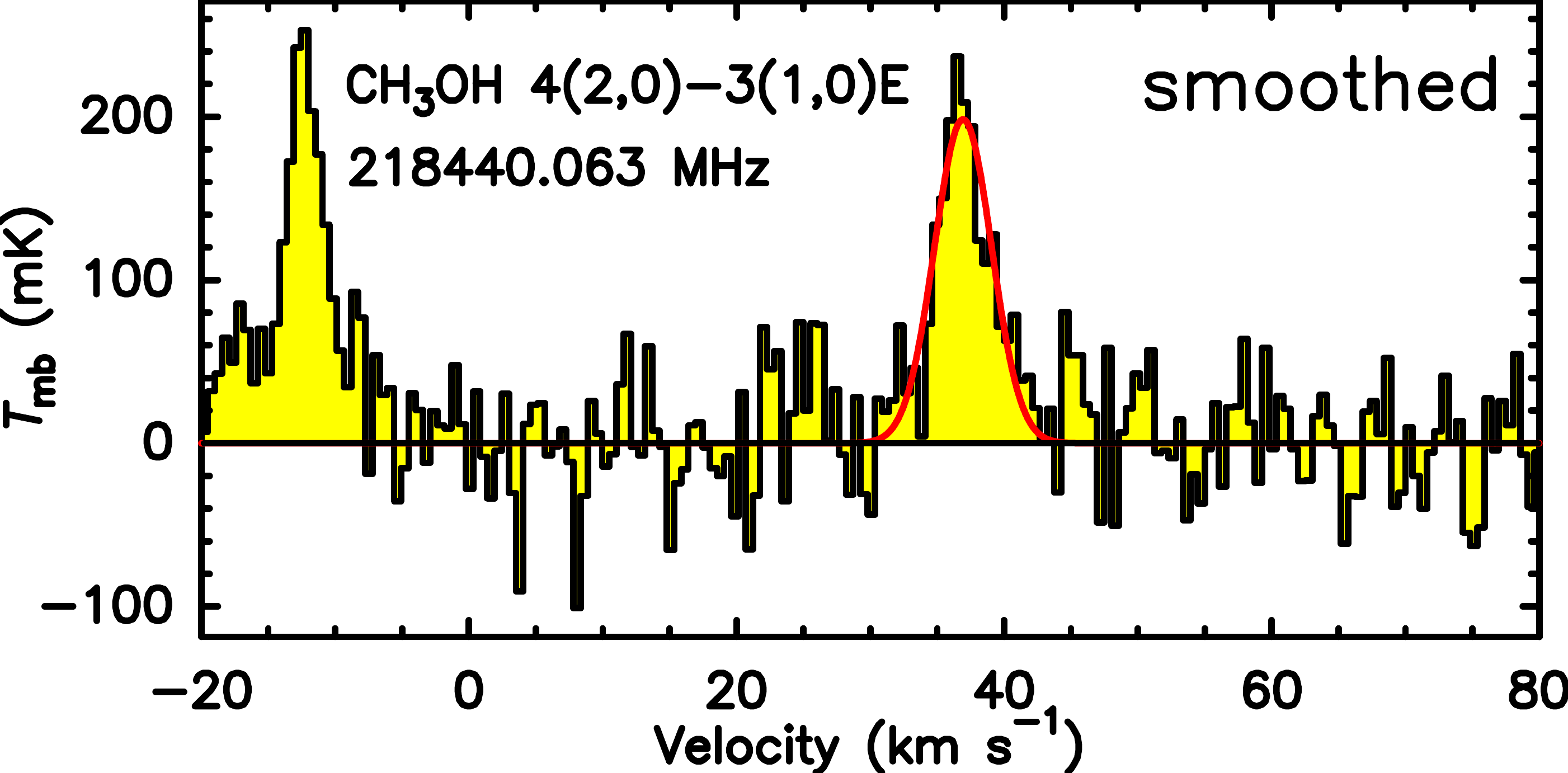}
 \caption{Same as Fig.~\ref{Figure3} but for CH$_{3}$OH.
}
\label{Figure13}
\end{figure}

\begin{figure}
\centering
        \includegraphics[width=8cm]{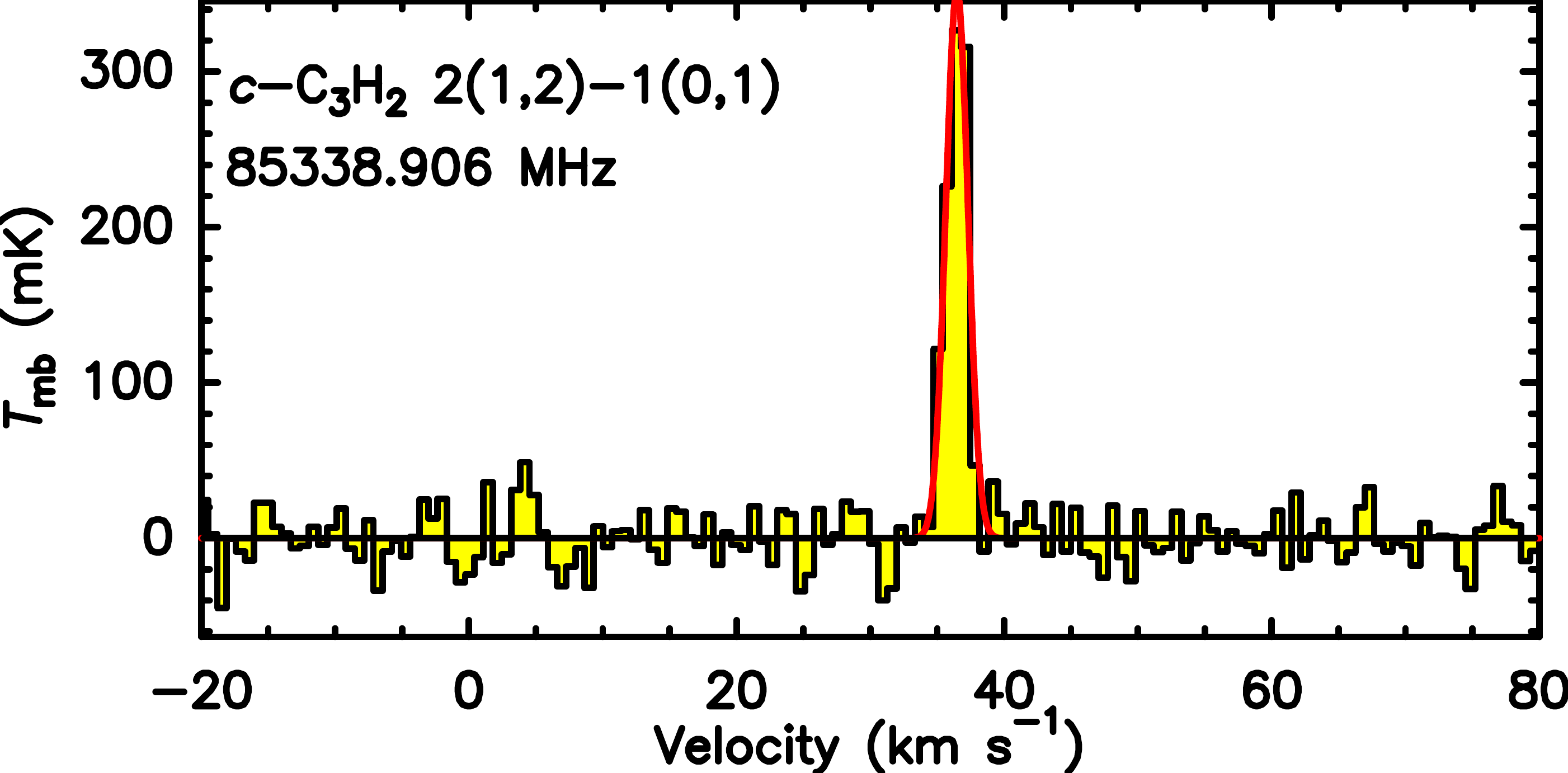}
 \caption{Same as Fig.~\ref{Figure3} but for $c$-C$_{3}$H$_{2}$.
}
\label{Figure14}
\end{figure}

\begin{figure}
\centering
        \includegraphics[width=16cm]{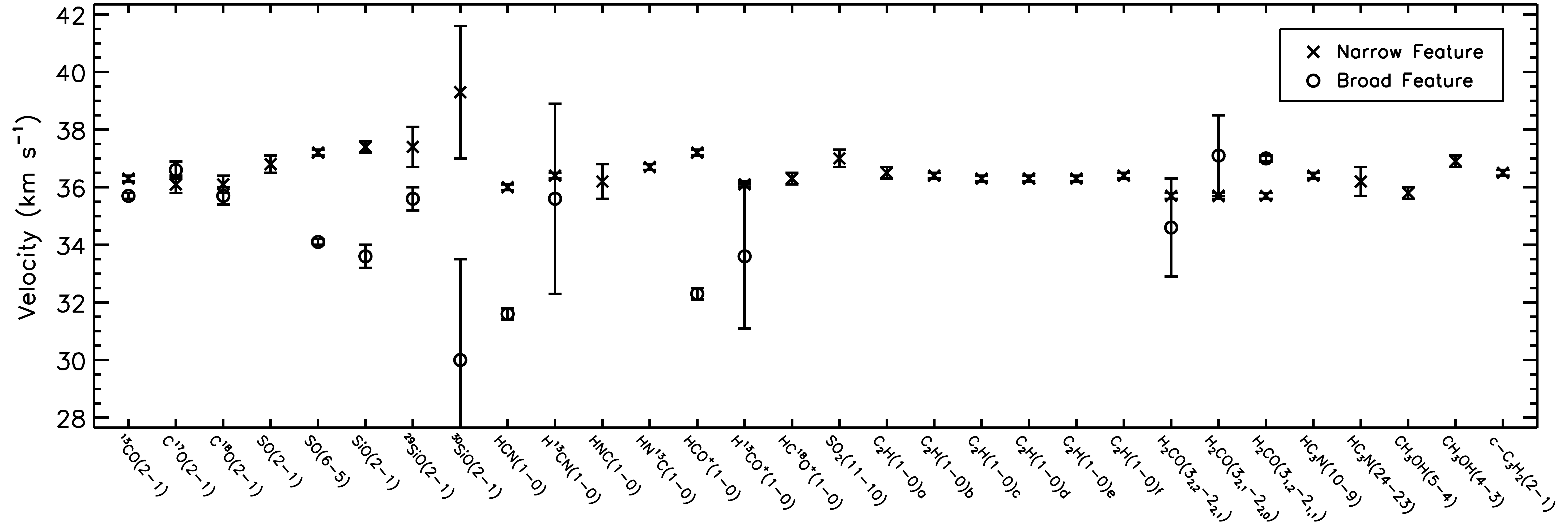}
 \caption{Line-center velocities of the thermal lines. 
}
\label{Figure15}
\end{figure}

\begin{figure}
\centering
        \includegraphics[width=16cm]{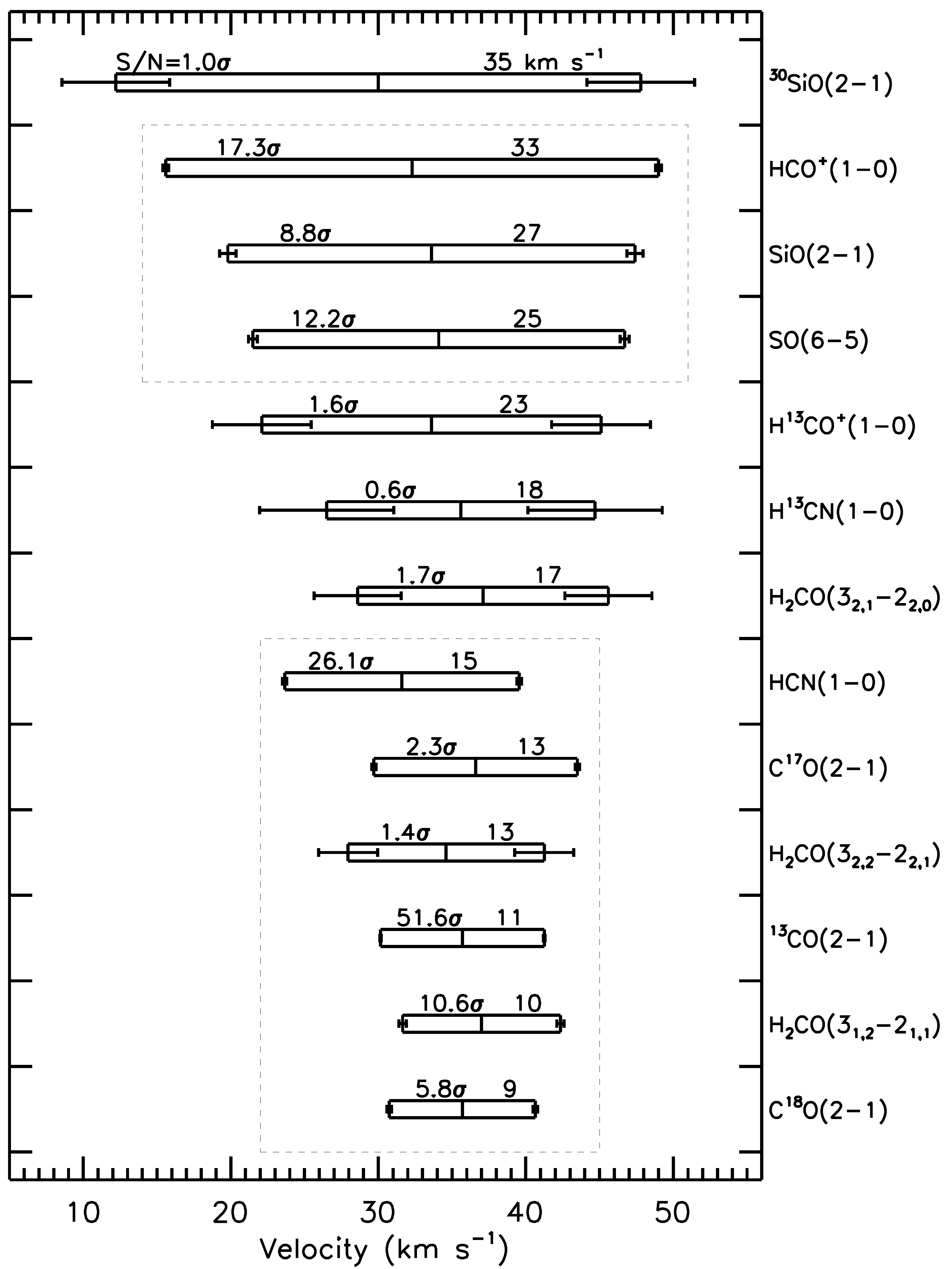}
 \caption{The widths ($\Delta V_{1/2}$; indicated by the horizontal span of the rectangles)
and the peak velocities relative to the source (represented by 
the vertical short bars in the middle of the rectangles) of the broad features. 
The error bars exhibited on both left and right sides of each rectangle show the uncertainties of $\Delta V_{1/2}$. 
The S/N number on the upper left of each rectangle is the ratio of the peak temperature of each line to the RMS. 
The number on the upper right of each rectangle is $\Delta V_{1/2}$ in km\,s$^{-1}$.
The dashed boxes indicate where the features have an S/N larger than 5$\sigma$.
}
\label{Figure16}
\end{figure}

\begin{figure}
\centering
        \includegraphics[width=8cm]{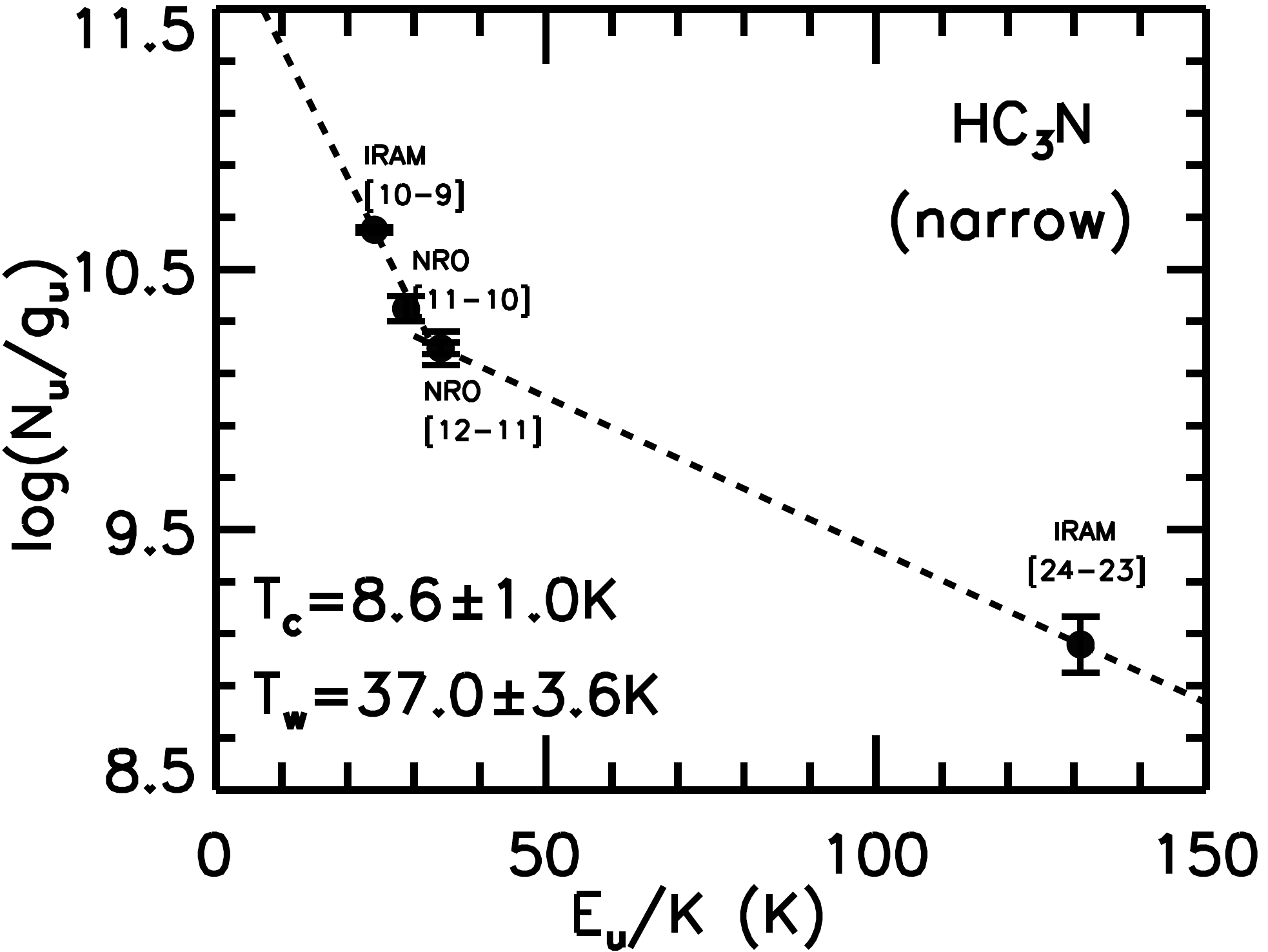}
        \includegraphics[width=8cm]{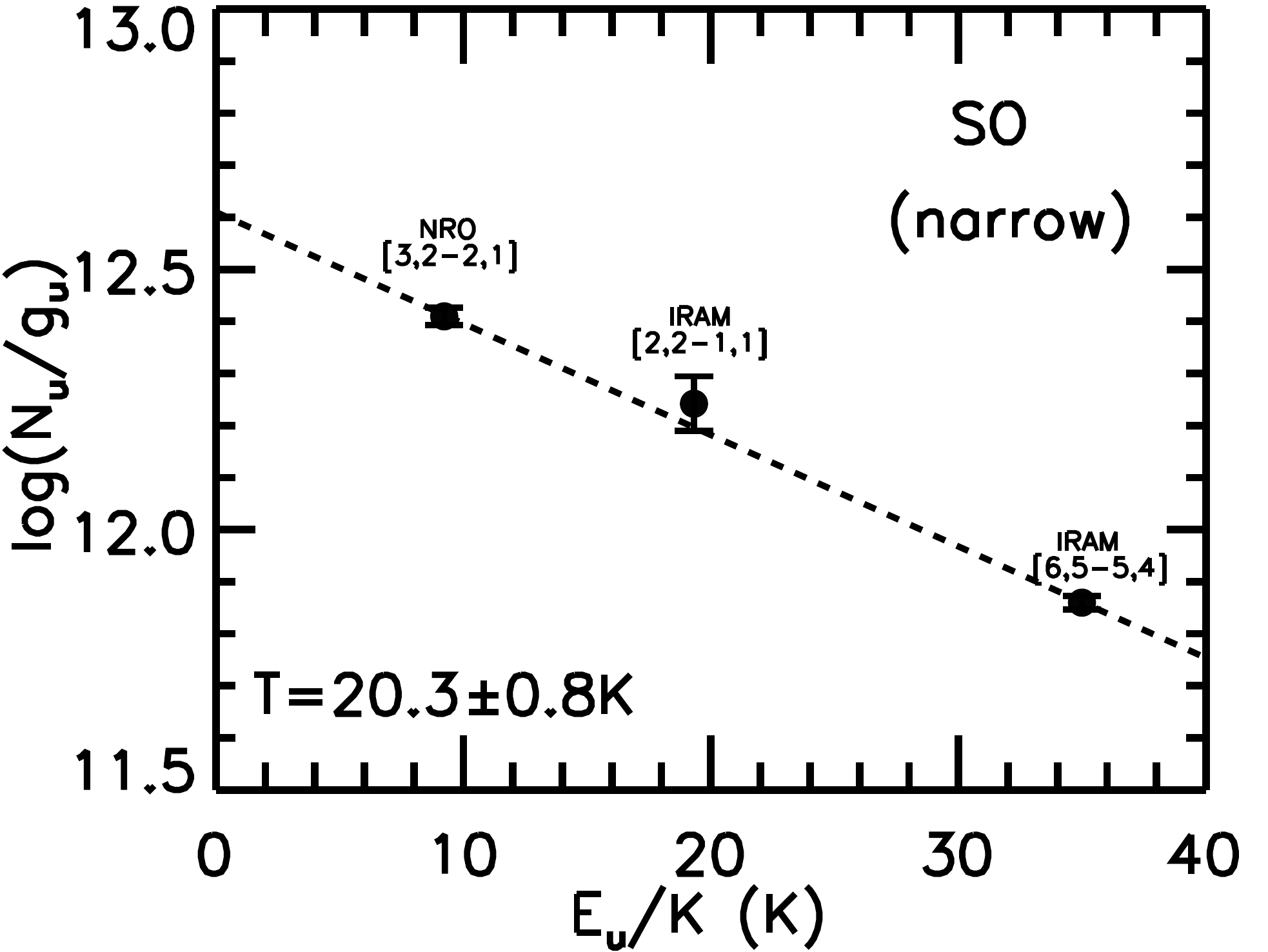}
        \includegraphics[width=8cm]{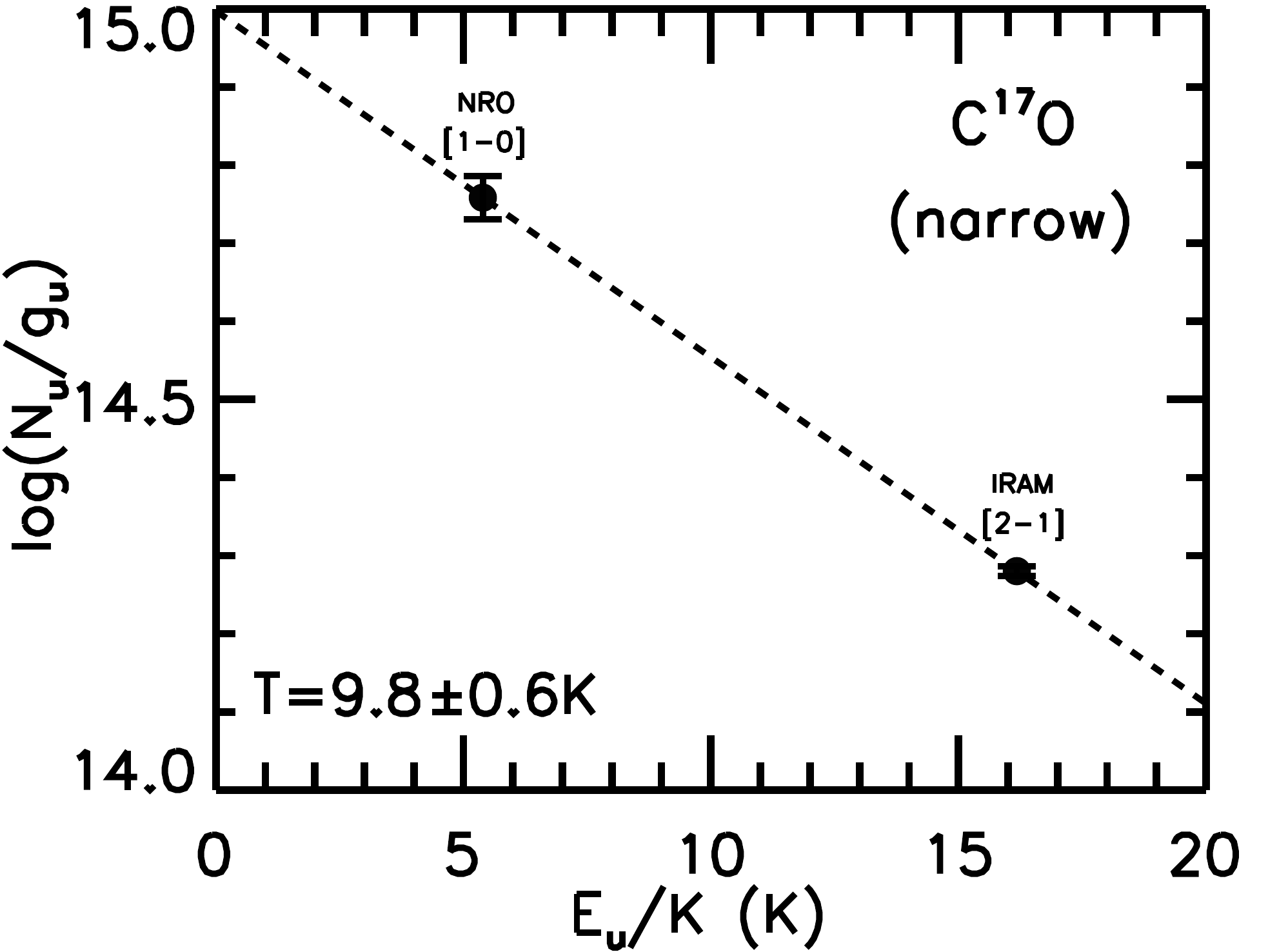}
        \includegraphics[width=8cm]{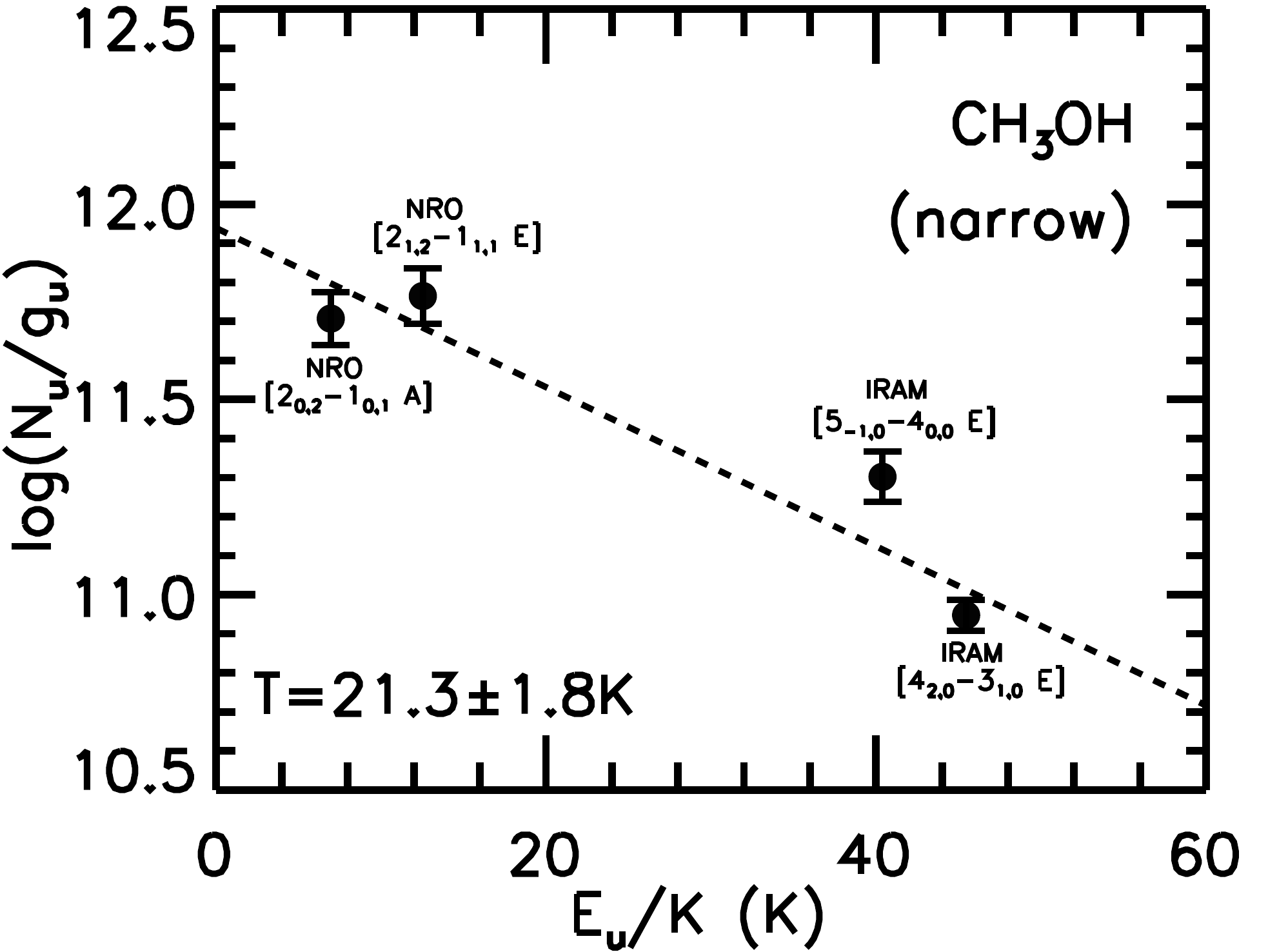}
        \includegraphics[width=8cm]{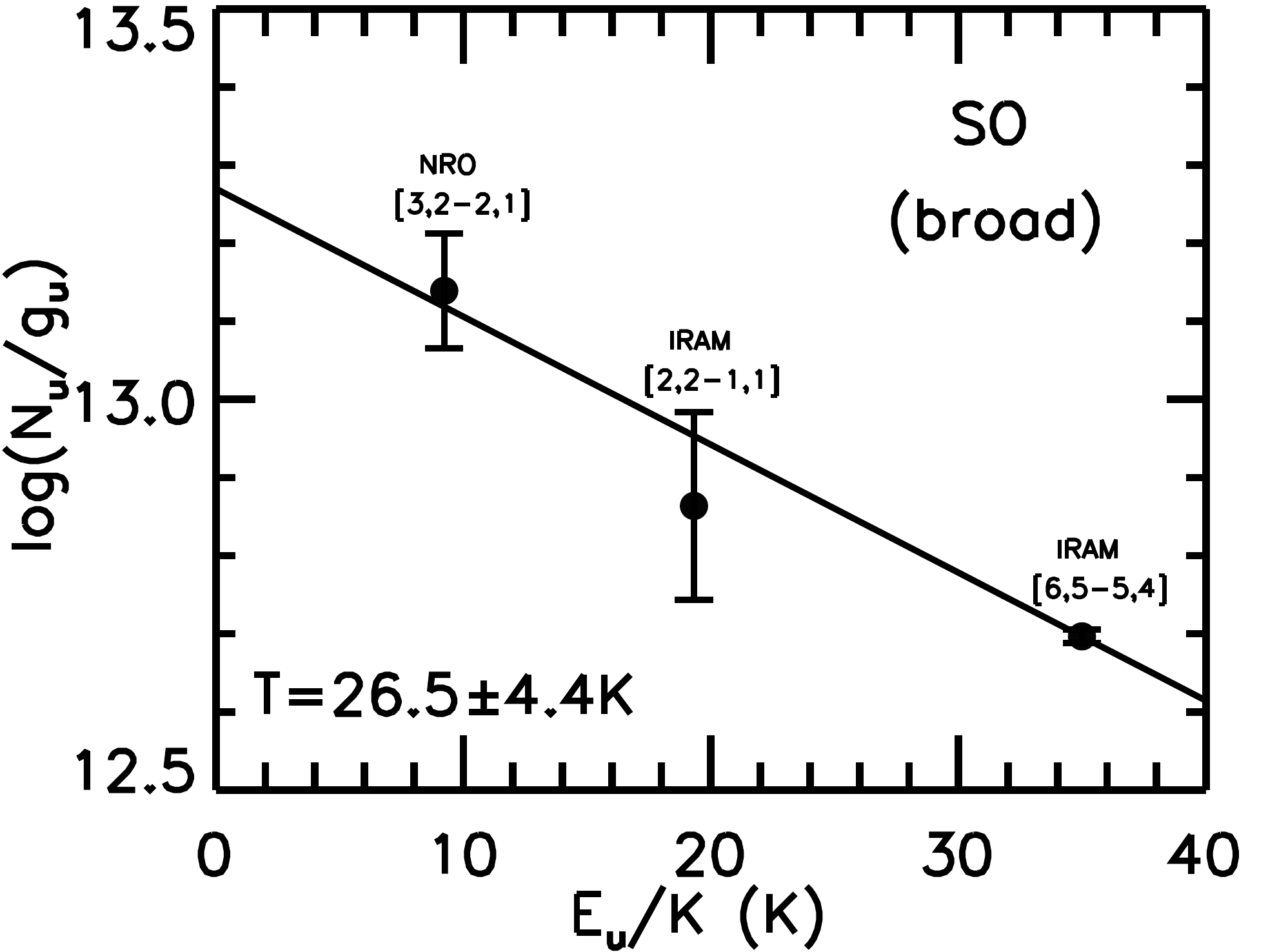}
 \caption{Rotation diagrams. 
The dashed and filled lines are linear fittings for the narrow and broad features, respectively. 
For HC$_{3}$N we use two components with excitation temperature $T_{\rm c}$ and $T_{\rm w}$.
}
\label{Figure17}
\end{figure}

We statistically analyze the similarity between the molecular spectra of I19312 and other types of stars. 
For that, we collect the 84--92 GHz spectral data of five sources belonging to four types that are available in the literature, 
as shown in Fig.~\ref{Figure1}.
The archetypal Kleinunmann-Low Nebula in the Orion (Orion\,KL, \citealt{Kleinmann67}), 
which contains numerous embedded young stellar objects, 
is one of the brightest and better-known massive star formation regions with a distance of about 400 pc \citep{Kounkel17}. 
IRAS\,04016+2610 (also named L1489\,IRS) is a low-mass protostar with a bolometric temperature of 226 K and a luminosity of 7 $L_{\odot}$ \citep{Green13,Sheehan17}. 
This object has a Keplerian rotational disk and a bipolar outflow \citep{Brinch07,Yen14}. 
IRC+10216 is a prototype carbon-rich star with a mass-loss rate of 2 $\times$ 10$^{-5}$ $M_{\odot}$\,yr$^{-1}$ \citep{Agundez12}. 
IK\,Tauri (also named NML\,Tau) is a Mira-type variable star with a mass-loss rate of 3.8--30 $\times$ 10$^{-6}$ $M_{\odot}$\,yr$^{-1}$ \citep{Neri98, Kim10}, surrounded by an oxygen-rich molecular envelope. 
CK\,Vulpeculae is the remnant of a red nova outburst \citep{Kaminski15} that is thought to be caused by a stellar merger event in 1670 \citep{Hevelius71}.
 
A two-sample Kolmogorov-Smirnov (K-S) test, 
carried out using the kstest2\footnote{see https://www.mathworks.com/help/stats/kstest2.html} function in the MATLAB software, 
is employed to assess the spectral similarity between the narrow/broad component and the comparison objects in a quantitative manner. 
Table~\ref{Table5} lists the velocity-integrated intensities of the emission lines detected within the frequency range of 84--92 GHz in I19312 and the five comparison objects (shown in Fig.~\ref{Figure1}), 
which are used as the inputs for the K-S test. 
The non-detected lines are regarded as having zero intensity. 
To correct for the distance effect, we first normalize the velocity-integrated intensities 
to those of the HCN $J=1 \rightarrow 0$ line as well as the strongest line of the individual objects. 
The observations of IRAS+04016, IK\,Tauri, CK\,Vul, and I19312 were carried out with the IRAM 30 m telescope with similar sensitivity, 
while Orion\,KL and IRC+10216 were observed by using different instruments. 
For a meaningful comparison, 
we need to correct the effect of different sensitivities of individual spectra. 
For that, we calculate the intensity ratios between the faintest lines and the HCN $J=1 \rightarrow 0$ line or the strongest line for each spectrum, 
as listed in the last two rows of Table~\ref{Table5}. 
When comparing two spectra, 
we set the lowest intensity ratio as a threshold; 
all the lines with relative intensity ratios lower than the threshold are assigned to have a zero intensity. 
Furthermore, to examine the reliability of the statistics, 
we prepare the input data sets in two ways: 
1) all the lines which are detected in at least one object listed in Table~\ref{Table5} are included; 
2) only those with non-zero intensity in one or two of the compared objects are included.

Table~\ref{Table6} lists the results of the two-sample K-S test. 
Presumably, the most similar objects are those with the largest p-value and lowest $D$ values. 
Although the four sets of data do not give consistent results, 
the spectral patterns of the broad feature seem to resemble that of the red nova, 
while the p-values appear to favor an AGB nature for the narrow-line region. 
Nevertheless, we cannot rule out the possibility of I19312 being a low-mass YSO, 
which also exhibits a high similarity with the broad-line region of I19321.

All molecules, except HCN, in the broad-line region are O-bearing. 
A similar situation has been found by \cite{Deguchi85} in IK\,Tau and OH231.8+4.2, 
which are oxygen-rich evolved stars but have strong HCN lines. 
They speculated that the enhancement of HCN molecules in an oxygen-rich environment may be caused by the ejection from a carbon-rich companion. 
We could presume the same for the inner region of I19312.
The narrow feature emission may be contaminated by a surrounding molecular cloud. 
However, we do not detect any complex or deuterated molecules that are commonly seen in the spectra of dark clouds. 
If the cold gas is associated with the remaining natal cloud of I19312, 
this might indicate that fragile molecules have been largely destructed by stellar radiation.

\section{Summary}
\label{sec5}

We present a spectral line survey towards I19312 in the frequency ranges of 84--92 and 218--226 GHz using the IRAM 30 m telescope.
A total of 28 spectral lines belonging to 22 different molecular species and isotopologues are detected. 
The spectra exhibit both carbon- and oxygen-bearing molecules, 
which can be decomposed into a narrow and a broad components. 
Six molecules, C$^{17}$O, $^{30}$SiO, HN$^{13}$C, HC$^{18}$O$^{+}$, H$_{2}$CO, and $c$-C$_{3}$H$_{2}$, are detected for the first time in this object. 
We construct a 3-D morpho-kinematical model to account for the observed line profiles, 
which reveals the existence of collimated bipolar outflows and complex kinematic structures in the inner regions. 
By comparing the chemical compositions of I19312 with those of 
the high-mass YSO Orion\,KL, the low-mass YSO IRAS\,04016+2610, carbon-rich AGB star IRC+10216, oxygen-rich AGB star IK\,Tauri, and red nova CK\,Vul, 
we come to the conclusion that the chemical compositions of the narrow and broad components of I19312 show different chemistry.
A statistics of the similarity between the spectra of different sources is presented, 
which shows the possibility that the broad-line region has a nature of red nova or low-mass YSO, 
while the spectral behavior of the narrow-line region is similar to that of AGB envelopes. 
Considering the chemical properties and enhancement of isotope species such as $^{15}$N and $^{13}$C, the red nova hypothesis may explain many of I19312's peculiarities. 
Nevertheless, it remains unclear why the progenitor stars have not completely dispersed the surrounding cold gas.

Although we cannot yet firmly conclude the true nature of I19312, 
this work demonstrates that the molecular line survey provides a meaningful contribution towards that goal. 
In the future, further line surveys with much wider frequency coverage, spatial resolutions, and deeper sensitivities would be valuable.

\begin{acknowledgements}
We thank the anonymous referee for the useful comments that improved the manuscript.
This work is supported by the National Science Foundation of China (NSFC, grant Nos. 11973099, 12003080, 12041302, and 12073088), 
the Guangdong Basic and Applied Basic Research Foundation (No. 2019A1515110588), 
and the Fundamental Research Funds for the Central Universities (Sun Yat-sen University, No. 22qntd3101). 
We also acknowledge the science research grants from the China Manned Space Project with Nos. CMS-CSST-2021-A09 and CMS-CSST-2021-A10.
X.D.T acknowledges the support of the NSFC under grant No. 11903070, the Natural Science Foundation of Xinjiang Uygur Autonomous Region under grant No. 2022D01E06, and the Heaven Lake Hundred Talent Program of Xinjiang Uygur Autonomous Region of China.
We wish to express our gratitude to the staff at the IRAM 30 m telescope for their kind help and support during our observations. 
This work presents results from the European Space Agency (ESA) space mission Gaia. Gaia data are being processed by the Gaia Data Processing and Analysis Consortium (DPAC). Funding for the DPAC is provided by national institutions, in particular the institutions participating in the Gaia MultiLateral Agreement (MLA). The Gaia mission website is https://www.cosmos.esa.int/gaia. The Gaia archive website is https://archives.esac.esa.int/gaia.
\end{acknowledgements}

%

\begin{thebibliography}{}
\bibitem[Ag{\'u}ndez et al.(2012)]{Agundez12} Ag{\'u}ndez, M., Fonfr{\'\i}a, J.~P., Cernicharo, J., et al.\ 2012, \aap, 543, A48. doi:10.1051/0004-6361/201218963
\bibitem[Aladro et al.(2015)]{Aladro15} Aladro, R., Mart{\'\i}n, S., Riquelme, D., et al.\ 2015, \aap, 579, A101. doi:10.1051/0004-6361/201424918
\bibitem[Beck et al.(2010)]{Beck10} Beck, T.~L., Bary, J.~S., \& McGregor, P.~J.\ 2010, \apj, 722, 1360. doi:10.1088/0004-637X/722/2/1360
\bibitem[Bieging et al.(2000)]{Bieging00} Bieging, J.~H., Shaked, S., \& Gensheimer, P.~D.\ 2000, \apj, 543, 897. doi:10.1086/317129
\bibitem[Borkowski \& Harrington(2001)]{Borkowski01} Borkowski, K.~J. \& Harrington, J.~P.\ 2001, \apj, 550, 778. doi:10.1086/319812
\bibitem[Brinch et al.(2007)]{Brinch07} Brinch, C., Crapsi, A., Hogerheijde, M.~R., et al.\ 2007, \aap, 461, 1037. doi:10.1051/0004-6361:20065473
\bibitem[Bujarrabal et al.(1994)]{Bujarrabal94} Bujarrabal, V., Fuente, A., \& Omont, A.\ 1994, \aap, 285, 247
\bibitem[Cernicharo et al.(2011)]{Cernicharo11} Cernicharo, J., Ag{\'u}ndez, M., \& Gu{\'e}lin, M.\ 2011, The Molecular Universe, 280, 237. doi:10.1017/S1743921311025014
\bibitem[Clark \& Stephenson(1982)]{Clark82} Clark, D.~H. \& Stephenson, F.~R.\ 1982, Supernovae: A Survey of Current Research, 90, 355
\bibitem[Cooper et al.(2013)]{Cooper13} Cooper, H.~D.~B., Lumsden, S.~L., Oudmaijer, R.~D., et al.\ 2013, \mnras, 430, 1125. doi:10.1093/mnras/sts681
\bibitem[Cordiner et al.(2016)]{Cordiner16} Cordiner, M.~A., Boogert, A.~C.~A., Charnley, S.~B., et al.\ 2016, \apj, 828, 51. doi:10.3847/0004-637X/828/1/51
\bibitem[Csengeri et al.(2016)]{Csengeri16} Csengeri, T., Leurini, S., Wyrowski, F., et al.\ 2016, \aap, 586, A149. doi:10.1051/0004-6361/201425404
\bibitem[Deguchi \& Goldsmith(1985)]{Deguchi85} Deguchi, S. \& Goldsmith, P.~F.\ 1985, \nat, 317, 336. doi:10.1038/317336a0
\bibitem[Deguchi et al.(2004)]{Deguchi04} Deguchi, S., Nakashima, J.-I., \& Takano, S.\ 2004, \pasj, 56, 1083. doi:10.1093/pasj/56.6.1083
\bibitem[Deguchi et al.(2005)]{Deguchi05} Deguchi, S., Matsunaga, N., \& Fukushi, H.\ 2005, \pasj, 57, L25. doi:10.1093/pasj/57.5.L25
\bibitem[Dunham et al.(2011)]{Dunham11} Dunham, M.~K., Rosolowsky, E., Evans, N.~J., et al.\ 2011, \apj, 741, 110. doi:10.1088/0004-637X/741/2/110
\bibitem[Frayer et al.(2015)]{Frayer15} Frayer, D.~T., Maddalena, R.~J., Meijer, M., et al.\ 2015, \aj, 149, 162. doi:10.1088/0004-6256/149/5/162
\bibitem[Gaia Collaboration et al.(2021)]{Gaia21} Gaia Collaboration, Brown, A.~G.~A., Vallenari, A., et al.\ 2021, \aap, 649, A1. doi:10.1051/0004-6361/202039657
\bibitem[Gaia Collaboration et al.(2018)]{Gaia18} Gaia Collaboration, Brown, A.~G.~A., Vallenari, A., et al.\ 2018, \aap, 616, A1. doi:10.1051/0004-6361/201833051
\bibitem[Garc{\'\i}a-D{\'\i}az et al.(2018)]{Garcia18} Garc{\'\i}a-D{\'\i}az, M.~T., Steffen, W., Henney, W.~J., et al.\ 2018, \mnras, 479, 3909. doi:10.1093/mnras/sty1590
\bibitem[Gerin et al.(2011)]{Gerin11} Gerin, M., Ka{\'z}mierczak, M., Jastrzebska, M., et al.\ 2011, \aap, 525, A116. doi:10.1051/0004-6361/201015050
\bibitem[Gonz{\'a}lez Delgado et al.(2003)]{Gonzalez03} Gonz{\'a}lez Delgado, D., Olofsson, H., Kerschbaum, F., et al.\ 2003, \aap, 411, 123. doi:10.1051/0004-6361:20031068
\bibitem[Graninger et al.(2014)]{Graninger14} Graninger, D.~M., Herbst, E., {\"O}berg, K.~I., et al.\ 2014, \apj, 787, 74. doi:10.1088/0004-637X/787/1/74
\bibitem[Green et al.(2013)]{Green13} Green, J.~D., Evans, N.~J., J{\o}rgensen, J.~K., et al.\ 2013, \apj, 770, 123. doi:10.1088/0004-637X/770/2/123
\bibitem[Gregersen et al.(1997)]{Gregersen97} Gregersen, E.~M., Evans, N.~J., Zhou, S., et al.\ 1997, \apj, 484, 256. doi:10.1086/304297
\bibitem[Habing(1996)]{Habing96} Habing, H.~J.\ 1996, \aapr, 7, 97. doi:10.1007/PL00013287
\bibitem[Hacar et al.(2020)]{Hacar20} Hacar, A., Bosman, A.~D., \& van Dishoeck, E.~F.\ 2020, \aap, 635, A4. doi:10.1051/0004-6361/201936516
\bibitem[Herwig(2005)]{Herwig05} Herwig, F.\ 2005, \araa, 43, 435. doi:10.1146/annurev.astro.43.072103.150600
\bibitem[Hevelius(1671)]{Hevelius71} Hevelius, J.\ 1671, Philosophical Transactions of the Royal Society of London Series I, 6, 2197
\bibitem[Hildebrand(1983)]{Hildebrand83} Hildebrand, R.~H.\ 1983, \qjras, 24, 267
\bibitem[Hirota et al.(1998)]{Hirota98} Hirota, T., Yamamoto, S., Mikami, H., et al.\ 1998, \apj, 503, 717. doi:10.1086/306032
\bibitem[H{\"o}fner \& Olofsson(2018)]{Hofner18} H{\"o}fner, S. \& Olofsson, H.\ 2018, \aapr, 26, 1. doi:10.1007/s00159-017-0106-5
\bibitem[Imai et al.(2011)]{Imai11} Imai, H., Tafoya, D., Honma, M., et al.\ 2011, \pasj, 63, 81. doi:10.1093/pasj/63.1.81
\bibitem[Ivanova et al.(2013)]{Ivanova13} Ivanova, N., Justham, S., Chen, X., et al.\ 2013, \aapr, 21, 59. doi:10.1007/s00159-013-0059-2
\bibitem[Jin et al.(2015)]{Jin15} Jin, M., Lee, J.-E., \& Kim, K.-T.\ 2015, \apjs, 219, 2. doi:10.1088/0067-0049/219/1/2
\bibitem[Jos{\'e} \& Hernanz(1998)]{Jose98} Jos{\'e}, J. \& Hernanz, M.\ 1998, \apj, 494, 680. doi:10.1086/305244
\bibitem[J{\o}rgensen et al.(2020)]{Jorgensen20} J{\o}rgensen, J.~K., Belloche, A., \& Garrod, R.~T.\ 2020, \araa, 58, 727. doi:10.1146/annurev-astro-032620-021927
\bibitem[Kahane et al.(2000)]{Kahane00} Kahane, C., Dufour, E., Busso, M., et al.\ 2000, \aap, 357, 669
\bibitem[Kami{\'n}ski et al.(2017)]{Kaminski17} Kami{\'n}ski, T., Menten, K.~M., Tylenda, R., et al.\ 2017, \aap, 607, A78. doi:10.1051/0004-6361/201731287
\bibitem[Kami{\'n}ski et al.(2018)]{Kaminski18} Kami{\'n}ski, T., Steffen, W., Tylenda, R., et al.\ 2018, \aap, 617, A129. doi:10.1051/0004-6361/201833165
\bibitem[Kami{\'n}ski et al.(2015)]{Kaminski15} Kami{\'n}ski, T., Menten, K.~M., Tylenda, R., et al.\ 2015, \nat, 520, 322. doi:10.1038/nature14257
\bibitem[Kami{\'n}ski et al.(2020)]{Kaminski20} Kami{\'n}ski, T., Menten, K.~M., Tylenda, R., et al.\ 2020, \aap, 644, A59. doi:10.1051/0004-6361/202038648
\bibitem[Kim et al.(2010)]{Kim10} Kim, H., Wyrowski, F., Menten, K.~M., et al.\ 2010, \aap, 516, A68. doi:10.1051/0004-6361/201014094
\bibitem[Kleinmann \& Low(1967)]{Kleinmann67} Kleinmann, D.~E. \& Low, F.~J.\ 1967, \apjl, 149, L1. doi:10.1086/180039
\bibitem[Koning et al.(2011)]{Koning11}Koning, N., Kwok, S., \& Steffen, W.\ 2011, \apj, 740, 27. doi:10.1088/0004-637X/740/1/27
\bibitem[Kounkel et al.(2017)]{Kounkel17} Kounkel, M., Hartmann, L., Loinard, L., et al.\ 2017, \apj, 834, 142. doi:10.3847/1538-4357/834/2/142
\bibitem[Le Gal et al.(2020)]{Le20} Le Gal, R., {\"O}berg, K.~I., Huang, J., et al.\ 2020, \apj, 898, 131. doi:10.3847/1538-4357/ab9ebf
\bibitem[Li et al.(2016)]{Li16} Li, F., Zhu, C., L{\"u}, G., et al.\ 2016, \pasj, 68, 39. doi:10.1093/pasj/psw030
\bibitem[Lodders(2003)]{Lodders03} Lodders, K.\ 2003, \apj, 591, 1220. doi:10.1086/375492
\bibitem[Long et al.(2021)]{Long21} Long, F., Bosman, A.~D., Cazzoletti, P., et al.\ 2021, \aap, 647, A118. doi:10.1051/0004-6361/202039336
\bibitem[Lovas(2004)]{Lovas04} Lovas, F.~J.\ 2004, Journal of Physical and Chemical Reference Data, 33, 177. doi:10.1063/1.1633275
\bibitem[Lucas \& Liszt(2000)]{Lucas00} Lucas, R. \& Liszt, H.~S.\ 2000, \aap, 358, 1069
\bibitem[Mangum \& Wootten(1993)]{Mangum93} Mangum, J.~G. \& Wootten, A.\ 1993, \apjs, 89, 123. doi:10.1086/191841
\bibitem[McGuire(2018)]{McGuire18} McGuire, B.~A.\ 2018, \apjs, 239, 17. doi:10.3847/1538-4365/aae5d2
\bibitem[McGuire(2022)]{McGuire22} McGuire, B.~A.\ 2022, \apjs, 259, 30. doi:10.3847/1538-4365/ac2a48
\bibitem[M{\"u}ller et al.(2001)]{Muller01} M{\"u}ller, H.~S.~P., Thorwirth, S., Roth, D.~A., et al.\ 2001, \aap, 370, L49. doi:10.1051/0004-6361:20010367
\bibitem[M{\"u}ller et al.(2005)]{Muller05} M{\"u}ller, H.~S.~P., Schl{\"o}der, F., Stutzki, J., et al.\ 2005, Journal of Molecular Structure, 742, 215. doi:10.1016/j.molstruc.2005.01.027
\bibitem[Nakashima et al.(2008)]{Nakashima08} Nakashima, J.-I., Deguchi, S., Imai, H., et al.\ 2008, Organic Matter in Space, 251, 165. doi:10.1017/S1743921308021467
\bibitem[Nakashima et al.(2011)]{Nakashima11} Nakashima, J.-I.., Deguchi, S., Imai, H., et al.\ 2011, \apj, 728, 76. doi:10.1088/0004-637X/728/2/76
\bibitem[Nakashima \& Deguchi(2004)]{Nakashima04a} Nakashima, J.-I. \& Deguchi, S.\ 2004, \apjl, 610, L41. doi:10.1086/423240
\bibitem[Nakashima et al.(2004)]{Nakashima04b} Nakashima, J.-I., Deguchi, S., \& Kuno, N.\ 2004, \pasj, 56, 193. doi:10.1093/pasj/56.1.193
\bibitem[Nakashima \& Deguchi(2000)]{Nakashima00} Nakashima, J.-I. \& Deguchi, S.\ 2000, \pasj, 52, L43. doi:10.1093/pasj/52.5.L43
\bibitem[Nakashima \& Deguchi(2005)]{Nakashima05} Nakashima, J.-I. \& Deguchi, S.\ 2005, \apj, 633, 282. doi:10.1086/444609
\bibitem[Nakashima et al.(2015)]{Nakashima15} Nakashima, J.-I., Sobolev, A.~M., Salii, S.~V., et al.\ 2015, \pasj, 67, 95. doi:10.1093/pasj/psv064
\bibitem[Nakashima \& Deguchi(2007)]{Nakashima07} Nakashima, J.-I. \& Deguchi, S.\ 2007, \apj, 669, 446. doi:10.1086/520825
\bibitem[Nakashima et al.(2016)]{Nakashima16} Nakashima, J.-I., Ladeyschikov, D.~A., Sobolev, A.~M., et al.\ 2016, \apj, 825, 16. doi:10.3847/0004-637X/825/1/16
\bibitem[Nakashima \& Deguchi(2003)]{Nakashima03} Nakashima, J.-I. \& Deguchi, S.\ 2003, \pasj, 55, 229. doi:10.1093/pasj/55.1.229
\bibitem[Neri et al.(1998)]{Neri98} Neri, R., Kahane, C., Lucas, R., et al.\ 1998, \aaps, 130, 1. doi:10.1051/aas:1998213
\bibitem[Nishimura et al.(2015)]{Nishimura15} Nishimura, A., Tokuda, K., Kimura, K., et al.\ 2015, \apjs, 216, 18. doi:10.1088/0067-0049/216/1/18
\bibitem[Olofsson et al.(1998)]{Olofsson98} Olofsson, H., Lindqvist, M., Nyman, L.-A., et al.\ 1998, \aap, 329, 1059
\bibitem[Ortiz-Le{\'o}n et al.(2020)]{Ortiz-Leon20} Ortiz-Le{\'o}n, G.~N., Menten, K.~M., Kami{\'n}ski, T., et al.\ 2020, \aap, 638, A17. doi:10.1051/0004-6361/202037712
\bibitem[Pickett et al.(1998)]{Pickett98} Pickett, H.~M., Poynter, R.~L., Cohen, E.~A., et al.\ 1998, \jqsrt, 60, 883. doi:10.1016/S0022-4073(98)00091-0
\bibitem[Riera et al.(1995)]{Riera95} Riera, A., Garcia-Lario, P., Manchado, A., et al.\ 1995, \aap, 302, 137
\bibitem[Robitaille et al.(2006)]{Robitaille06} Robitaille, T.~P., Whitney, B.~A., Indebetouw, R., et al.\ 2006, \apjs, 167, 256. doi:10.1086/508424
\bibitem[Romano \& Matteucci(2003)]{Romano03} Romano, D. \& Matteucci, F.\ 2003, \mnras, 342, 185. doi:10.1046/j.1365-8711.2003.06526.x
\bibitem[Schilke et al.(1992)]{Schilke92} Schilke, P., Walmsley, C.~M., Pineau Des Forets, G., et al.\ 1992, \aap, 256, 595
\bibitem[Schmidt et al.(2018a)]{Schmidt18a} Schmidt, D.~R., Woolf, N.~J., Zega, T.~J., et al.\ 2018, \nat, 564, 378. doi:10.1038/s41586-018-0763-1
\bibitem[Schmidt et al.(2018b)]{Schmidt18b} Schmidt, D.~R., Zack, L.~N., \& Ziurys, L.~M.\ 2018, \apjl, 864, L31. doi:10.3847/2041-8213/aadc09
\bibitem[Schneider et al.(2010)]{Schneider10} Schneider, N., Csengeri, T., Bontemps, S., et al.\ 2010, \aap, 520, A49. doi:10.1051/0004-6361/201014481
\bibitem[Sch{\"o}ier et al.(2006)]{Schoier06} Sch{\"o}ier, F.~L., Olofsson, H., \& Lundgren, A.~A.\ 2006, \aap, 454, 247. doi:10.1051/0004-6361:20054795
\bibitem[Shang et al.(2004)]{Shang04} Shang, H., Lizano, S., Glassgold, A., et al.\ 2004, \apjl, 612, L69. doi:10.1086/424566
\bibitem[Sheehan \& Eisner(2017)]{Sheehan17} Sheehan, P.~D. \& Eisner, J.~A.\ 2017, \apj, 851, 45. doi:10.3847/1538-4357/aa9990
\bibitem[Steffen \& Koning(2017)]{Steffen17} Steffen, W. \& Koning, N.\ 2017, Astronomy and Computing, 20, 87. doi:10.1016/j.ascom.2017.06.002
\bibitem[Steffen et al.(2013)]{Steffen13} Steffen, W., Koning, N., Esquivel, A., et al.\ 2013, \mnras, 436, 470. doi:10.1093/mnras/stt1583
\bibitem[Steffen et al.(2011)]{Steffen11} Steffen, W., Koning, N., Wenger, S., et al.\ 2011, IEEE Transactions on Visualization and Computer Graphics, 17, 454. doi:10.1109/TVCG.2010.62
\bibitem[Sutton et al.(1985)]{Sutton85} Sutton, E.~C., Blake, G.~A., Masson, C.~R., et al.\ 1985, \apjs, 58, 341. doi:10.1086/191045
\bibitem[Tafoya et al.(2020)]{Tafoya20} Tafoya, D., Imai, H., G{\'o}mez, J.~F., et al.\ 2020, \apjl, 890, L14. doi:10.3847/2041-8213/ab70b8
\bibitem[Tang et al.(2018)]{Tang18} Tang, X.~D., Henkel, C., Wyrowski, F., et al.\ 2018, \aap, 611, A6. doi:10.1051/0004-6361/201732168
\bibitem[Tielens(2013)]{Tielens13} Tielens, A.~G.~G.~M.\ 2013, Reviews of Modern Physics, 85, 1021. doi:10.1103/RevModPhys.85.1021
\bibitem[Tylenda \& Soker(2006)]{Tylenda06} Tylenda, R. \& Soker, N.\ 2006, \aap, 451, 223. doi:10.1051/0004-6361:20054201
\bibitem[Ungerechts et al.(1997)]{Ungerechts97} Ungerechts, H., Bergin, E.~A., Goldsmith, P.~F., et al.\ 1997, \apj, 482, 245. doi:10.1086/304110
\bibitem[Urquhart et al.(2018)]{Urquhart18} Urquhart, J.~S., K{\"o}nig, C., Giannetti, A., et al.\ 2018, \mnras, 473, 1059. doi:10.1093/mnras/stx2258
\bibitem[Velilla Prieto et al.(2017)]{Velilla17} Velilla Prieto, L., S{\'a}nchez Contreras, C., Cernicharo, J., et al.\ 2017, \aap, 597, A25. doi:10.1051/0004-6361/201628776
\bibitem[Winters et al.(2003)]{Winters03} Winters, J.~M., Le Bertre, T., Jeong, K.~S., et al.\ 2003, \aap, 409, 715. doi:10.1051/0004-6361:20031110
\bibitem[Wu et al.(2004)]{Wu04} Wu, Y., Wei, Y., Zhao, M., et al.\ 2004, \aap, 426, 503. doi:10.1051/0004-6361:20035767
\bibitem[Xi \& Bo(1965)]{Xi65} Xi, Z.-Z. \& Bo, S.-R.\ 1965, Acta Astronomica Sinica, 13, 1
\bibitem[Xi(1955)]{Xi55} Xi, Z.-Z.\ 1955, Acta Astronomica Sinica, 3, 183
\bibitem[Yang et al.(2021)]{Yang21} Yang, Y., Jiang, Z., Chen, Z., et al.\ 2021, \apj, 922, 144. doi:10.3847/1538-4357/ac22ab
\bibitem[Yen et al.(2014)]{Yen14} Yen, H.-W., Takakuwa, S., Ohashi, N., et al.\ 2014, \apj, 793, 1. doi:10.1088/0004-637X/793/1/1
\bibitem[Yung et al.(2013)]{Yung13} Yung, B.~H.~K., Nakashima, J.-I., Imai, H., et al.\ 2013, \apj, 769, 20. doi:10.1088/0004-637X/769/1/20
\end{thebibliography}
%

\clearpage

\newgeometry{left=1.5cm,right=1.5cm,top=2.5cm,bottom=2.5cm}

\clearpage

\clearpage
\newpage

\clearpage

\clearpage
\begin{table*}
\caption{Spectral scan parameters.}
\centering
\begin{tabular}{ccccccc}
\hline\hline
Band range        & $F_{\rm eff}$   & $B_{\rm eff}$    & On-time{\boldmath$^{a}$}       & $T_{\rm sys}$    & HPBW{\boldmath} & Spatial resolution$^{b}$    \\
GHz               & \%              &  \%              & minute                         & K                & $''$            & cm                          \\  
\hline
83.999--91.781    & 95              &   81             &    54.3                        & 124               &  29.3--26.8   & 1.7--1.5 $\times$ 10$^{18}$  \\          
218.249--226.031  &  92             &   59             &    54.3                        & 256               &  11.3--10.9   & 6.4--6.2 $\times$ 10$^{17}$  \\  
\hline
\end{tabular}
\tablefoot{ 
\tablefoottext{a}{On-source time for each band.} \\
\tablefoottext{b}{Assume a distance of 3.8 kpc.} \\
}
\label{Table1}
\end{table*}

\clearpage

\begin{landscape}
\begin{longtable}{ccccccccccc}
\caption{\label{Table2} The transitions detected with the IRAM 30 m  telescope in I19312.} \\
\hline\hline
Frequency           & Molecule             & Transition                          &  Flux                   & Velocity              & $\Delta V_{1/2}$     & $T_{\rm mb}$  & RMS    &   dv             & HPBW       &  Note \\ 
MHz                 &                      & upper-lower                         &  mK\,km\,s$^{-1}$       & km\,s$^{-1}$          & km\,s$^{-1}$         & mK            & mK     &  km\,s$^{-1}$   & \arcsec    &      \\
\hline
220398.7            & $^{13}$CO            & $J=2-1$                             &  $4916.3\pm156.8$       & $29.1\pm0.1$          & $1.6\pm0.1$          & 2826.1      & 54.6     & 0.27    & 11.1    &  N       \\ 
                    &                      &                                     &  $38847.0\pm470.0$      & $36.3\pm0.1$          & $3.0\pm0.1$          & 12345.0     & 54.6     & 0.27    & 11.1    &  N       \\ 
                    &                      &                                     &  $2993.4\pm81.2$        & $44.6\pm0.1$          & $0.9\pm0.1$          & 3200.0      & 54.6     & 0.27    & 11.1    &  N       \\ 
                    &                      &                                     &  $33139.0\pm367.7$      & $35.7\pm0.1$          & $11.1\pm0.1$         & 2817.5      & 54.6     & 0.27    & 11.1    &  B       \\ 
$^{\star}$224714.4  & C$^{17}$O            & $J=2-1$                             &  $326.6\pm56.6$         & $29.4\pm0.3$          & $1.8\pm0.3$          & 167.8       & 36.4     & 0.26    & 10.9    &  N       \\ 
                    &                      &                                     &  $3711.0\pm56.6$        & $36.1\pm0.3$          & $2.6\pm0.3$          & 1362.6      & 36.4     & 0.26    & 10.9    &  N       \\ 
                    &                      &                                     &  $59.6\pm56.6$          & $44.9\pm0.3$          & $0.6\pm0.3$          & 96.7        & 36.4     & 0.26    & 10.9    &  N       \\ 
                    &                      &                                     &  $1250.2\pm56.6$        & $36.6\pm0.3$          & $13.8\pm0.3$         & 85.1        & 36.4     & 0.26    & 10.9    &  B       \\ 
$^{\star}$219560.4  & C$^{18}$O            & $J=2-1$                             &  $940.92\pm192.6$       & $29.2\pm0.3$          & $1.1\pm0.3$          & 773.9       & 50.1     & 0.27    & 11.2    &  N       \\ 
                    &                      &                                     &  $10851.0\pm192.6$      & $36.1\pm0.3$          & $2.3\pm0.3$          & 4502.1      & 50.1     & 0.27    & 11.2    &  N       \\ 
                    &                      &                                     &  $362.85\pm192.6$       & $44.6\pm0.3$          & $0.4\pm0.3$          & 892.9       & 50.1     & 0.27    & 11.2    &  N       \\ 
                    &                      &                                     &  $3074.6\pm192.6$       & $35.7\pm0.3$          & $9.9\pm0.3$          & 292.4       & 50.1     & 0.27    & 11.2    &  B       \\ 
86094.0             & SO                   & $J,K=2,2-1,1$                       &  $687.6\pm79.4$         & $36.8\pm0.3$          & $6.5\pm0.7$          & 99.4        & 11.0     & 1.36    & 28.6    &  N       \\ 
                    &                      &                                     &  $396.9\pm110.0$        & $15.5\pm4.2$          & $23.0\pm7.6$         & 16.2        & 11.0     & 1.36    & 28.6    &  B       \\ 
%
$^{\star}$219949.4  & SO                   & $J,K=6,5-5,4$                       &  $8604.9\pm244.7$       & $37.2\pm0.1$          & $5.8\pm0.1$          & 1389.2      & 46.8     & 0.27    & 11.2    &  N       \\ 
                    &                      &                                     &  $15344.0\pm262.9$      & $34.1\pm0.1$          & $25.2\pm0.6$         & 571.2       & 46.8     & 0.27    & 11.2    &  B       \\ 
86847.0             & SiO                  & $J=2-1$                             &  $1387.8\pm171.8$       & $37.4\pm0.2$          & $7.0\pm0.7$          & 187.3       & 19.6     & 1.35    & 28.3    &  N       \\ 
                    &                      &                                     &  $5067.2\pm197.2$       & $33.6\pm0.4$          & $27.6\pm1.1$         & 172.8       & 19.6     & 1.35    & 28.3    &  B       \\ 
86243.4             & SiO                  & $J=2-1, v=1$                        &  $423.6\pm132.3$        & $52.8\pm0.3$          & $4.9\pm0.9$          & 81.9        & 9.7      & 1.36    & 28.5    &  M       \\ 
                    &                      &                                     &  $456.7\pm156.1$        & $45.7\pm1.8$          & $11.2\pm3.8$         & 38.5        & 9.7      & 1.36    & 28.5    &  M       \\
$^{\star}$85640.5   & SiO                  & $J=2-1, v=2$                        &  $115.3\pm22.5$         & $58.8\pm0.1$          & $1.4\pm0.2$          & 78.7        & 16.4     & 0.68    & 28.7    &  M       \\ 
                    &                      &                                     &  $168.9\pm45.1$         & $65.5\pm0.9$          & $5.9\pm1.6$          & 26.7        & 16.4     & 0.68    & 28.7    &  M       \\
$^{\star}$85759.2   & $^{29}$SiO           & $J=2-1$                             &  $120.4\pm48.4$         & $37.4\pm0.7$          & $3.5\pm1.6$          & 32.4        & 11.3     & 1.37    & 28.7    &  N       \\ 
                    &                      &                                     &  $581.1\pm136.3$        & $35.6\pm0.4$          & $50.8\pm12.2$        & 10.8        & 11.3     & 1.37    & 28.7    &  B       \\ 
$^{\star}$84746.0   & $^{30}$SiO           & $J=2-1$                             &  $37.9\pm49.9$          & $39.3\pm2.3$          & $3.9\pm4.1$          & 9.2         & 19.6     & 1.35    & 28.3    &  N       \\ 
                    &                      &                                     &  $713.5\pm138.4$        & $30.0\pm3.5$          & $35.6\pm7.3$         & 18.9        & 19.6     & 1.35    & 28.3    &  B       \\ 
88630.4             & HCN                  & $J=1-0, F=1-1$                      &  $2041.5\pm31.7$        & $36.2\pm0.6$          & $3.1\pm0.1$          & 610.9       & 12.8     & 0.66    & 27.8    &  N       \\ 
88631.8             & HCN                  & $J=1-0, F=2-1$                      &  $1939.6\pm34.5$        & $36.0\pm0.1$          & $2.2\pm0.1$          & 815.9       & 12.8     & 0.66    & 27.8    &  N       \\ 
88633.9             & HCN                  & $J=1-0, F=0-1$                      &  $1272.4\pm36.8$        & $36.1\pm0.1$          & $2.6\pm0.1$          & 453.2       & 12.8     & 0.66    & 27.8    &  N       \\
88631.6             & HCN                  & $J=1-0$                             &  $9874.3\pm105.3$       & $31.6\pm0.2$          & $27.9\pm0.3$         & 333.8       & 12.8     & 0.66    & 27.8    &  B       \\ 
86338.7             & H$^{13}$CN           & $J=1-0, F=1-1$                      &  $297.3\pm54.1$         & $35.9\pm0.2$          & $2.8\pm0.5$          & 101.3       & 21.3     & 0.68    & 28.5    &  N       \\ 
86340.2             & H$^{13}$CN           & $J=1-0, F=2-1$                      &  $459.9\pm56.9$         & $36.4\pm0.1$          & $2.8\pm0.4$          & 151.8       & 21.3     & 0.68    & 28.5    &  N       \\ 
86342.3             & H$^{13}$CN           & $J=1-0, F=0-1$                      &  $100.1\pm47.2$         & $35.6\pm0.5$          & $2.3\pm1.1$          & 40.2        & 21.3     & 0.68    & 28.5    &  N       \\ 
86339.9             & H$^{13}$CN           & $J=1-0$                             &  $414.4\pm164.4$        & $35.6\pm3.3$          & $30.2\pm9.1$         & 12.9        & 21.3     & 0.68    & 28.5    &  B       \\ 
$^{\star}$86055.0   & HC$^{15}$N           & $J=1-0$                             &  $446.8\pm93.7$         & $38.0\pm2.4$          & $22.1\pm5.4$         & 19.0        & 9.2      & 2.72    & 28.6    &  N\&B       \\ 
90663.6             & HNC                  & $J=1-0$                             &  $492.1\pm147.3$        & $29.7\pm0.6$          & $4.5\pm0.6$          & 102.6       & 14.4     & 0.65    & 27.1    &  N       \\ 
                    &                      &                                     &  $4370.2\pm147.3$       & $36.2\pm0.6$          & $2.7\pm0.6$          & 1522.1      & 14.4     & 0.65    & 27.1    &  N       \\ 
                    &                      &                                     &  $135.9\pm147.3$        & $44.2\pm0.6$          & $4.0\pm0.6$          & 32.1        & 14.4     & 0.65    & 27.1    &  N       \\ 
$^{\star}$87090.9   & HN$^{13}$C           & $J=1-0$                             &  $251.3\pm29.8$         & $36.7\pm0.1$          & $2.1\pm0.3$          & 110.8       & 17.1     & 0.67    & 28.2    &  N       \\ 
89188.5             & HCO$^{+}$            & $J=1-0$                             &  $2129.4\pm29.5$        & $37.2\pm0.1$          & $2.5\pm0.1$          & 812.9       & 12.2     & 0.66    & 27.6    &  N       \\ 
                    &                      &                                     &  $7503.3\pm86.6$        & $32.3\pm0.2$          & $33.4\pm0.4$         & 211.3       & 12.2     & 0.66    & 27.6    &  B       \\ 
86754.3             & H$^{13}$CO$^{+}$     & $J=1-0$                             &  $1148.7\pm34.7$        & $36.1\pm0.1$          & $2.1\pm0.1$          & 512.0       & 15.8     & 0.68    & 28.4    &  N       \\ 
                    &                      &                                     &  $616.7\pm116.7$        & $33.6\pm2.5$          & $23.0\pm6.7$         & 25.2        & 15.8     & 0.68    & 27.6    &  B       \\ 
$^{\star}$85162.2   & HC$^{18}$O$^{+}$     & $J=1-0$                             &  $149.5\pm27.3$         & $36.3\pm0.2$          & $2.0\pm0.4$          & 70.0        & 16.9     & 0.69    & 27.6    &  N       \\ 
$^{\star}$221965.2  & SO$_{2}$             & $J_{K_a,K_c}=11_{1,11}-10_{0,10}$   &  $1363.8\pm168.7$       & $37.0\pm0.3$          & $4.6\pm0.6$          & 279.2       & 44.8     & 1.06    & 11.1    &  N       \\ 
87284.2             & C$_{2}$H             & $N,J,F=1,3/2,1-0,1/2,1$             &  $278.0\pm46.0$         & $36.5\pm0.2$          & $2.3\pm0.6$          & 112.7       & 19.7     & 0.67    & 28.2    &  N       \\ 
87316.9             & C$_{2}$H             & $N,J,F=1,3/2,2-0,1/2,1$             &  $1243.4\pm80.1$        & $36.4\pm0.1$          & $2.0\pm0.1$          & 579.5       & 17.8     & 0.67    & 28.2    &  N       \\ 
87328.6             & C$_{2}$H             & $N,J,F=1,3/2,1-0,1/2,0$             &  $847.3\pm40.9$         & $36.3\pm0.1$          & $2.3\pm0.1$          & 347.2       & 17.8     & 0.67    & 28.2    &  N       \\ 
87402.0             & C$_{2}$H             & $N,J,F=1,1/2,1-0,1/2,1$             &  $797.0\pm40.5$         & $36.3\pm0.1$          & $2.1\pm0.1$          & 364.7       & 21.6     & 0.67    & 28.2    &  N       \\ 
87407.2             & C$_{2}$H             & $N,J,F=1,1/2,0-1,1/2,1$             &  $349.7\pm30.0$         & $36.3\pm0.1$          & $1.9\pm0.2$          & 177.4       & 19.2     & 0.67    & 28.1    &  N       \\ 
87446.5             & C$_{2}$H             & $N,J,F=1,1/2,1-0,1/2,0$             &  $364.1\pm38.7$         & $36.4\pm0.1$          & $2.5\pm0.3$          & 134.5       & 19.0     & 0.67    & 28.1    &  N       \\ 
$^{\star}$218475.6  & $p$-H$_{2}$CO        & $J_{K_a,K_c}=3_{2,2}-2_{2,1}$       &  $525.9\pm124.8$        & $36.6\pm0.2$          & $2.6\pm0.5$          & 190.5       & 37.4     & 0.54    & 11.3    &  N       \\ 
                    &                      &                                     &  $745.0\pm176.6$        & $34.6\pm1.7$          & $13.3\pm4.0$         & 52.8        & 37.4     & 0.54    & 11.3    &  B       \\ 
$^{\star}$218760.1  & $p$-H$_{2}$CO        & $J_{K_a,K_c}=3_{2,1}-2_{2,0}$       &  $631.4\pm148.2$        & $36.6\pm0.2$          & $2.9\pm0.6$          & 202.7       & 31.7     & 0.54    & 11.2    &  N       \\ 
                    &                      &                                     &  $989.0\pm178.7$        & $37.1\pm1.4$          & $17.0\pm5.9$         & 54.6        & 31.7     & 0.54    & 11.2    &  B       \\ 
$^{\star}$225697.8  & $o$-H$_{2}$CO        & $J_{K_a,K_c}=3_{1,2}-2_{1,1}$       &  $2679.9\pm117.7$       & $36.3\pm0.1$          & $2.7\pm0.1$          & 939.8       & 33.5     & 0.54    & 10.9    &  N       \\ 
                    &                      &                                     &  $4064.7\pm125.7$       & $37.0\pm0.1$          & $10.7\pm0.5$         & 356.4       & 33.5     & 0.54    & 10.9    &  B       \\ 
$^{\star}$90979.0   & HC$_{3}$N            & $J=10-9$                            &  $805.9\pm20.2$         & $36.4\pm0.1$          & $2.4\pm0.1$          & 317.3       & 11.0     & 0.64    & 27.0    &  N       \\ 
$^{\star}$218324.7  & HC$_{3}$N            & $J=24-23$                           &  $319.79\pm79.1$        & $36.2\pm0.5$          & $3.5\pm1.0$          & 85.1        & 51.3     & 1.07    & 11.3    &  N       \\ 
$^{\star}$84521.2   & CH$_{3}$OH           & $J_{K_a,K_c}=5_{-1,0}-4_{0,0} E$    &  $263.5\pm39.0$         & $35.8\pm0.2$          & $3.1\pm0.6$          & 79.4        & 17.6     & 0.64    & 29.1    &  N       \\ 
$^{\star}$218440.1  & CH$_{3}$OH           & $J_{K_a,K_c}=4_{2,0}-3_{1,0} E$     &  $1031.9\pm94.3$        & $36.9\pm0.2$          & $4.9\pm0.6$          & 197.6       & 34.7     & 0.54    & 11.3    &  N       \\ 
$^{\star}$85338.9   & $c$-C$_{3}$H$_{2}$   & $J_{K_a,K_c}=2_{1,2}-1_{0,1}$       &  $733.2\pm29.8$         & $36.5\pm0.1$          & $2.0\pm0.1$          & 352.5       & 17.1     & 0.69    & 27.0    &  N       \\ 
\hline
\multicolumn{10}{l}{\textsuperscript{}{ Notes.  
The transition lines marked with ($^{\star}$) are new detections in I19312. N and B represent the narrow and broad components, respectively.}} \\
\end{longtable}
\end{landscape}

\clearpage

\begin{table*}
\caption{Optical depths, excitation temperatures, column densities, and molecular abundances.}
\centering
\begin{tabular}{cccccccccccc}
\hline
\hline
Species                 &  $\tau$                     & $T_{\rm ex}$(K)  & $N$(cm$^{-2}$)                     &  $f_{\rm X}$                                     & Mark     \\ 
\hline
 HC$_{3}$N              & ...                         & $^{a}$8.6 $\pm$ 1.0                  &  $^{a}$(2.89 $\pm$ 0.92) $\times$ 10$^{13}$      &      $2.91\times10^{-9}$      & N,C \\ 
                        & ...                         & $^{a}$37.0 $\pm$ 3.6                 &  $^{a}$(6.73 $\pm$ 0.74) $\times$ 10$^{12}$      &      $6.79\times10^{-10}$     & N,W \\ 
 SO                     & ...                         & $^{a}$20.3 $\pm$ 0.8                 &  $^{a}$(1.65 $\pm$ 0.09) $\times$ 10$^{14}$      &      $1.66\times10^{-8}$      & N   \\ 
                        & ...                         & $^{a}$26.5 $\pm$ 4.4                 &  $^{a}$(1.03 $\pm$ 0.22) $\times$ 10$^{15}$      &      $1.27\times10^{-7}$      & B   \\ 
 C$^{17}$O              & ...                         & $^{a}$9.8 $\pm$ 0.6                  &  $^{a}$(2.32 $\pm$ 0.22) $\times$ 10$^{16}$      &      $2.34\times10^{-6}$      & N   \\ 
                        & ...                         & 26.5                                 &  (2.51 $\pm$ 0.28) $\times$ 10$^{16}$            &      $3.10\times10^{-6}$      & B   \\ 
 CH$_{3}$OH             & ...                         & $^{a}$21.3 $\pm$ 1.8                 &  $^{a}$(3.09 $\pm$ 0.45) $\times$ 10$^{14}$      &      $3.12\times10^{-8}$      & N   \\ 
 $^{13}$CO              & 0.6                         & 9.8                                  &  $^{b}$(1.30 $\pm$ 0.13) $\times$ 10$^{17}$      &      $1.31\times10^{-5}$      & N   \\ 
                        & 0.1                         & 26.5                                 &   $^{b}$(2.61 $\pm$ 0.26) $\times$ 10$^{17}$     &     $3.23\times10^{-5}$       & B   \\ 
 C$^{18}$O              & 0.2                         & 9.8                                  &   $^{b}$(1.40 $\pm$ 0.15) $\times$ 10$^{16}$     &     $1.41\times10^{-6}$       & N   \\ 
                        & 0.01                        & 26.5                                 &   $^{b}$(4.59 $\pm$ 0.50) $\times$ 10$^{16}$     &     $5.68\times10^{-6}$       & B   \\ 
 SiO                    & ...                         & 20.3                                 &  (1.75 $\pm$ 0.22) $\times$ 10$^{13}$            &     $1.76\times10^{-9}$       & N   \\ 
                        & ...                         & 26.5                                 &  (5.62 $\pm$ 0.31) $\times$ 10$^{14}$            &     $6.95\times10^{-8}$       & B   \\ 
 $^{30}$SiO             & ...                         & 20.3                                 &  (5.19 $\pm$ 6.83) $\times$ 10$^{11}$            &     $5.23\times10^{-11}$      & N   \\ 
                        & ...                         & 26.5                                 &  (8.70 $\pm$ 1.72) $\times$ 10$^{13}$            &     $1.08\times10^{-8}$       & B   \\ 
 HCN                    & 2.6                         & 15.0                                 &  $^{b}$(1.39 $\pm$ 0.37) $\times$ 10$^{14}$      &     $1.40\times10^{-8}$       & N   \\ 
                        & ...                         & 26.5                                 &  (1.09 $\pm$ 0.03) $\times$ 10$^{15}$            &     $1.35\times10^{-7}$       & B   \\ 
 H$^{13}$CN             & 0.7                         & 15.0                                 &  $^{b}$(7.65 $\pm$ 0.21) $\times$ 10$^{13}$      &     $7.72\times10^{-9}$       & N   \\ 
                        & ...                         & 26.5                                 &  (5.07 $\pm$ 2.02) $\times$ 10$^{13}$            &     $6.27\times10^{-9}$       & B   \\ 
 HNC                    & ...                         & 15.0                                 &  (4.37 $\pm$ 0.40) $\times$ 10$^{13}$            &     $4.41\times10^{-9}$       & N   \\ 
 HN$^{13}$C             & ...                         & 15.0                                 &  (2.28 $\pm$ 0.27) $\times$ 10$^{13}$            &     $2.30\times10^{-9}$       & N   \\ 
 HCO$^{+}$              & ...                         & 15.0                                 &  (1.19 $\pm$ 0.03) $\times$ 10$^{13}$            &     $^{c}1.20\times10^{-9}$   & N   \\ 
                        & ...                         & 26.5                                 &  (4.61 $\pm$ 0.13) $\times$ 10$^{14}$            &     $^{c}5.70\times10^{-8}$   & B   \\ 
 H$^{13}$CO$^{+}$       & ...                         & 15.0                                 &  (7.00 $\pm$ 0.25) $\times$ 10$^{12}$            &     $7.06\times10^{-10}$      & N   \\ 
                        & ...                         & 26.5                                 &  (4.16 $\pm$ 0.79) $\times$ 10$^{13}$            &     $5.14\times10^{-9}$       & B   \\ 
 HC$^{18}$O$^{+}$       & ...                         & 15.0                                 &  (1.34 $\pm$ 0.25) $\times$ 10$^{12}$            &     $1.35\times10^{-10}$      & N   \\ 
 SO$_{2}$               & ...                         & 20.3                                 &  (3.09 $\pm$ 0.45) $\times$ 10$^{14}$            &     $3.12\times10^{-8}$       & N   \\ 
 C$_{2}$H               & 1.7                         & 15.0                                 &  $^{b}$(6.66 $\pm$ 0.44) $\times$ 10$^{14}$      &     $6.72\times10^{-8}$       & N   \\ 
 $p$-H$_{2}$CO          & ...                         & 15.0                                 &  $^{d}$(2.03 $\pm$ 0.78) $\times$ 10$^{14}$      &     $2.05\times10^{-8}$       & N   \\ 
                        & ...                         & 26.5                                 &  $^{d}$(3.97 $\pm$ 1.87) $\times$ 10$^{14}$      &     $4.91\times10^{-8}$       & B   \\ 
 $o$-H$_{2}$CO          & ...                         & 15.0                                 &  (1.84 $\pm$ 0.28) $\times$ 10$^{13}$            &     $1.86\times10^{-9}$       & N   \\ 
                        & ...                         & 26.5                                 &  (9.64 $\pm$ 2.03) $\times$ 10$^{13}$            &     $1.19\times10^{-8}$       & B   \\ 
 $c$-C$_{3}$H$_{2}$     & ...                         & 15.0                                 &  (2.87 $\pm$ 0.14) $\times$ 10$^{13}$            &     $2.89\times10^{-9}$       & N   \\ 
\hline
\hline
\end{tabular}
\tablefoot{ 
N and B represent the narrow and broad components, respectively.
C and W represent the cold and warm components, respectively. \\
\tablefoottext{a}{Derived from the rotation diagram analysis.} \\
\tablefoottext{b}{The optical depth effect has been corrected (see Sect.~\ref{Sect.3.4})} \\
\tablefoottext{c}{Taken from \cite{Deguchi04}.} \\
\tablefoottext{d}{The average of the values derived from the $p$-H$_{2}$CO  $J_{K_a,K_c}=3_{2,2} \rightarrow 2_{2,1}$ and  $3_{2,1} \rightarrow 2_{2,0}$} lines.\\
       }
\label{Table3}
\end{table*}

\clearpage

\begin{table*}
\caption{Isotopic abundance ratios.}
\centering
\begin{tabular}{ccccccccc}
\hline\hline
Isotopic ratio          & Species                              & Transition     & \multicolumn{2}{c}{I19312}                         & Nova\,1670$^{b}$  & IRC+10216$^{c}$      & Orion\,KL$^{d}$ & Solar$^{e}$ \\
\cline{4-5}
\cline{4-5} 
                         &                                      &               & Narrow                     & Broad                 &                  &                    &                    &          \\
\hline
$^{12}$C/$^{13}$C        & HCO$^{+}$/H$^{13}$CO$^{+}$            &  $J$=1--0    & 1.9:                       & 12.2:                 & 2.5:             & 45 $\pm$ 3         & >74.9              & 89.3    \\
                         & HCN/H$^{13}$CN                        &  $J$=1--0    & $^{a}$9.8:                 & 23.8:                 & 2.0:             &                    & >9.9               &         \\ 
                         &                                       &              & >6.1                       &                       &                  &                    &                    &         \\
                         & HNC/HN$^{13}$C                        &  $J$=1--0    & 17.4:                      & ...                   & 6.2:             &                    & >67.3              &         \\ 
\hline
$^{14}$N/$^{15}$N        & HCN/HC$^{15}$N                        &  $J$=1--0    & $^{a}$209.0:               & >35.4                 & 12.8:            & >4400              & >42.5              & 272.0   \\
                         &                                       &              & >70.0                      &                       &                  &                    &                    &         \\
\hline
$^{16}$O/$^{18}$O        & HCO$^{+}$/HC$^{18}$O$^{+}$            &  $J$=1--0    & 14.2:                      & >13.7                 & 53.1:            & 1260 $\pm$ 288     & >1613.5            & 498.8   \\
$^{16}$O/$^{17}$O        & HCO$^{+}$/HC$^{17}$O$^{+}$            &  $J$=1--0    & >66.6                      & >17.3                 & ...              & 840 $\pm$ 200      & >2039.0            & 2680.6  \\
$^{18}$O/$^{17}$O        & C$^{18}$O/C$^{17}$O                   &  $J$=2--1    & $^{a}$3.2:                 & $^{a}$2.5:            & 6.7              & 0.7 $\pm$ 0.2      & 21.5 $\pm$ 21.6    & 5.4     \\
                         &                                       &              & >2.9                       & >2.4                  &                  &                    &                    &         \\
$^{13}$CO/C$^{17}$O      & $^{13}$CO/C$^{17}$O                   &  $J$=2--1    & $^{a}$14.0:                & $^{a}$27.8:           & 286.5            & 18.7 $\pm$ 4.4     & 5.9 $\pm$ 1.7      & 30.0    \\
                         &                                       &              & >10.5                      & >26.5                 &                  &                    &                    &         \\
$^{13}$CO/C$^{18}$O      & $^{13}$CO/C$^{18}$O                   &  $J$=2--1    & $^{a}$4.3:                 & $^{a}$11.3:           & 52.1             & 28 $\pm$ 6         & 47.6 $\pm$ 13.5    & 5.6     \\
                         &                                       &              & >3.6                       & >10.8                 &                  &                    &                    &         \\
                         &  H$^{13}$CO$^{+}$/HC$^{18}$O$^{+}$    &  $J$=1--0    & 7.7 $\pm$ 1.4              & >1.5                  & 86.5             &                    & 8.1 $\pm$ 2.3      &         \\ 
\hline
$^{28}$Si/$^{30}$Si      & $^{28}$SiO/$^{30}$SiO                 &  $J$=2--1    & 36.6:                      & 7.1:                  & 8.9:             & $>$22              & 37.4 $\pm$ 1.2     & 29.9    \\
$^{29}$Si/$^{30}$Si      & $^{29}$SiO/$^{30}$SiO                 &  $J$=2--1    & 3.2 $\pm$ 4.4              & 0.8 $\pm$ 0.2         & 1.2               & 1.45               & 2.5 $\pm$ 0.1      & 1.5     \\
\hline
\hline
\end{tabular}
\tablefoot{ 
Symbol (:) means uncertain measurements due to uncertain optical depths of the main lines.\\
$^{a}$ The optical depth effects have been corrected (see Sect.~\ref{Sect.3.4}).
The row below gives the lower limits obtained by the velocity-integrated intensity ratios.\\
$^{b}$ Taken from \cite{Kaminski17}. \\
$^{c}$ Taken from \cite{Kahane00}. \\
$^{d}$ Taken from \cite{Sutton85} and \cite{Frayer15}. \\
$^{e}$ Taken from \cite{Lodders03}. \\
       }
\label{Table4}
\end{table*}

\clearpage

\restoregeometry

\clearpage

\newgeometry{left=0.5cm,right=1.5cm,top=2.5cm,bottom=2.5cm}

\begin{table*}
\caption{Velocity-integrated intensity of the emission lines for Kolmogorov-Smirnov test sample.}
\centering
\begin{small}
\begin{tabular}{ccccccccc}
\hline\hline
         &     &    Orion\,KL    &    IRAS+04016    &    IRC+10216    &    IK Tauri   &    CK\,Vul     &    IRAS+19312   & IRAS+19312      \\
         &     &    YSO          &    YSO           &    C-rich       &    O-rich     &    red nova    & narrow feature  & broad feature   \\
\hline
Frequency & Line  & $\int T_{\rm mb} {\rm dv}$   & $\int T_{\rm mb} {\rm dv}$   & $\int T_{\rm mb} {\rm dv}$   & $\int T_{\rm A}^{*} {\rm dv}$       & $\int T_{\rm A}^{*} {\rm dv}$   & $\int T_{\rm mb} {\rm dv}$   & $\int T_{\rm mb} {\rm dv}$   \\
MHz       &       & mK km\,s$^{-1}$              & mK km\,s$^{-1}$              & K km\,s$^{-1}$               & K km\,s$^{-1}$                      &  K km\,s$^{-1}$                 & mK km\,s$^{-1}$              & mK km\,s$^{-1}$   \\
\hline
84410.7  &  $^{34}$SO\,(2$_2$--1$_1$)                               & 37.3    & ...    & ...   & ...       & ...       & ...         & ...       \\
84521.2  &  CH$_{3}$OH\,(5$_{-1,5}$--4$_{0,4} E$)                   & 57.7    & ...    & ...   & ...       & ...       & ...         & ...       \\
84410.7  &  $^{29}$SiO\,(2--1, $v$=2)                               & ...     & ...    & ...   & 12.6      & ...       & ...         & ...       \\
84746.2  &  $^{30}$SiO\,(2--1)                                      & 20.8    & ...    & ...   & 3.7       & 1.2       & 37.9        & 713.5     \\
85139.1  &  OCS\,(7--6)                                             & 40.8    & ...    & ...   & ...       & ...       & ...         & ...       \\
85162.2  &  HC$^{18}$O$^{+}$\,(1--0)                                & ...     & 17.9   & ...   & ...       & ...       & 149.5       & ...       \\
85167.0  &  $^{29}$SiO\,(2--1,$v$=1)                                & ...     & ...    & ...   & 1.5       & ...       & ...         & ...       \\
85201.3  &  HC$_{5}$N\,(32--31)                                     & ...     & ...    & 4.4   & ...       & ...       & ...         & ...       \\
85338.9  &  $c$-C$_{3}$H$_{2}$\,(2--1)                              & ...     & ...    & 3.0   & ...       & ...       & 733.2       & ...       \\
85347.9  &  HCS$^{+}$\,(2--1)                                       & ...     & 7.4    & ...   & ...       & ...       & ...         & ...       \\
85568.1  &  CH$_{3}$OH\,(6$_{-2,5}$--7$_{-1,7} E$)                  & 15.1    & ...    & ...   & ...       & ...       & ...         & ...       \\
85634.0  &  C$_{4}$H\,(9--8)                                        & ...     & ...    & 13.0  & ...       & ...       & ...         & ...       \\
85640.5  &  SiO\,(2--1,$v$=2)                                       & 15.7    & ...    & ...   & 0.8       & ...       & 115.3       & 168.9     \\
85759.2  &  $^{29}$SiO\,(2--1)                                      & 41.1    & ...    & 1.6   & 5.9       & 1.2       & 120.4       & 581.1     \\
86055.0  &  HC$^{15}$N\,(1--0)                                      & 34.2    & 8.3    & ...   & ...       & 2.9       & 200.0       & 246.8     \\
86094.0  &  SO\,(2$_2$--1$_1$)                                      & 324.7   & ...    & ...   & 0.7       & ...       & 687.6       & 396.9     \\
86210.1  &  CH$_{3}$OCHO\,(7$_{4,4}$--6$_{4,3}$)                    & 3.5     & ...    & ...   & ...       & ...       & ...         & ...       \\
86243.4  &  SiO\,(2--1,$v$=1)                                       & 677.8   & ...    & ...   & 154.7     & ...       & 456.7       & 423.6     \\
86339.9  &  H$^{13}$CN\,(1--0)                                      & 157.1   & 27.2   & 60.7  & 0.9       & 18.1      & 857.3       & 414.4     \\
86615.6  &  CH$_{3}$OH\,(7$_{2,6}$--6$_{3,3} A$)                    & 8.9     & ...    & ...   & ...       & ...       & ...         & ...       \\
86639.1  &  SO$_{2}$\,(8$_{3,5}$--9$_{2,8}$)                        & 49.0    & ...    & ...   & ...       & ...       & ...         & ...       \\
86754.3  &  H$^{13}$CO$^{+}$\,(1--0)                                & 4.6     & 200.2  & ...   & ...       & 1.1       & 1148.7      & 616.7     \\
86847.0  &  SiO\,(2--1)                                             & 337.2   & ...    & 18.2  & 31.2      & 8.6       & 1387.8      & 5067.2    \\
86902.9  &  CH$_{3}$OH\,(7$_{2,5}$--6$_{3,4} E$)                    & 12.7    & ...    & ...   & ...       & ...       & ...         & ...       \\   
87090.8  &  HN$^{13}$C\,(1--0)                                      & 2.0     & 19.1   & ...   & ...       & 5.3       & 251.3       & ...       \\
87316.9  &  C$_{2}$H\,(1--0)                                        & 36.0    & 460.5  & 48.8  & ...       & 0.7       & 3879.5      & ...       \\ 
87863.6  &  HC$_{5}$N\,(33--32)                                     & 2.7     & ...    & 4.8   & ...       & ...       & ...         & ...       \\
87925.2  &  HNCO\,(4$_{0,4}$--3$_{0,3}$)                            & 28.8    & ...    & ...   & ...       & ...       & ...         & ...       \\   
88285.8  &  Si$^{34}$S\,(4$_{0,4}$--3$_{0,3}$)                      & ...     & ...    & ...   & 0.1       & ...       & ...         & ...       \\   
88323.8  &  CH$_{3}$CH$_{2}$CN\,(10$_{0,10}$--9$_{0,9}$)            & 18.0    & ...    & ...   & ...       & ...       & ...         & ...       \\   
88594.8  &  CH$_{3}$OH\,(15$_{3,13}$--14$_{4,10} A$)                & 12.3    & ...    & ...   & ...       & ...       & ...         & ...       \\   
88631.8  &  HCN\,(1--0)                                             & 1559.6  & 603.1  & 174.0 & 8.1       & 32.5      & 5253.5      & 9874.3    \\
88709.2  &  CH$_{3}$OCH$_{3}$\,(15$_{2,13}$--15$_{1,14} A$)         & 26.4    & ...    & ...   & ...       & ...       & ...         & ...       \\   
88865.7  &  H$^{15}$NC\,(1--0)                                      & 1.0     & 5.5    & ...   & ...       & 0.8       & ...         & ...       \\
88940.0  &  CH$_{3}$OH\,(15$_{3,12}$--14$_{4,11} A$)                & 13.2    & ...    & ...   & ...       & ...       & ...         & ...       \\   
89045.6  &  C$_{3}$N\,(9--8)                                        & ...     & ...    & 19.1  & ...       & ...       & ...         & ...       \\  
89188.5  &  HCO$^{+}$\,(1--0)                                       & 281.1   & 1228.1 & ...   & 0.5       & 2.6       & 2129.4      & 7503.3    \\
89297.6  &  CH$_{3}$CH$_{2}$CN\,(10$_{2,9}$--9$_{2,8}$)             & 14.0    & ...    & ...   & ...       & ...       & ...         & ...       \\   
89314.6  &  CH$_{3}$OCHO\,(8$_{1,8}$--7$_{1,7}$)                    & 13.3    & ...    & ...   & ...       & ...       & ...         & ...       \\   
89505.8  &  CH$_{3}$OH\,(8$_{-4,5}$--9$_{-3,7} E$)                  & 16.0    & ...    & ...   & ...       & ...       & ...         & ...       \\   
88323.8  &  CH$_{3}$CH$_{2}$CN\,(10--9)                             & 142.7   & ...    & ...   & ...       & ...       & ...         & ...       \\   
90145.6  &  CH$_{3}$OCHO\,(7$_{2,5}$--6$_{2,4}$)                    & 23.0    & ...    & ...   & ...       & ...       & ...         & ...       \\   
90453.4  &  CH$_{3}$CH$_{2}$CN\,(10$_{2,8}$--9$_{2,7}$)             & 15.3    & ...    & ...   & ...       & ...       & ...         & ...       \\   
90525.9  &  HC$_{5}$N\,(34--33)                                     & ...     & ...    & 3.5   & ...       & ...       & ...         & ...       \\
90548.2  &  SO$_{2}$\,(25$_{3,23}$--24$_{4,20}$)                    & 24.2    & ...    & ...   & ...       & ...       & ...         & ...       \\   
90663.6  &  HNC\,(1--0)                                             & 122.9   & 497.9  & 22.3  & 0.1       & 13.1      & 4998.2      & ...       \\
90771.6  &  SiS\,(5--4)                                             & 5.4     & ...    & 30.4  & 1.0       & 0.4       & ...         & ...       \\
90938.1  &  CH$_{3}$OCH$_{3}$\,(6$_{0,6}$--5$_{1,5}$)               & 5.9     & ...    & ...   & ...       & ...       & ...         & ...       \\   
90979.0  &  HC$_{3}$N\,(10--9)                                      & 95.4    & ...    & 56.3  & ...       & 0.9       & 805.9       & ...       \\
91550.4  &  SO$_{2}$\,(18$_{5,13}$--19$_{4,16}$)                    & 51.3    & ...    & ...   & ...       & ...       & ...         & ...       \\  
\hline
\multicolumn{2}{l}{Intensity of the faintest relative to HCN (1--0)} & 0.065\% & 0.91\% & 0.94\% & 1.1\%   & 1.2\%     & 0.72\%      & 1.7\%     \\
\multicolumn{2}{l}{Intensity of the faintest relative to the strongest} & 0.065\% & 0.45\% & 0.94\% & 0.058\% & 1.2\%  & 0.72\%      & 1.7\%     \\
\hline
\hline
\end{tabular}
\end{small}
\tablefoot{ 
The spectra of Orion\,KL, IRAS+04016, IRC+10216, IK\,Tauri, and CK\,Vul are taken from \cite{Frayer15}, \cite{Le20}, Tuo et al. (2022, in preparation), \cite{Velilla17}, and \cite{Kaminski17}, respectively.\\
       }
\label{Table5}
\end{table*}
\restoregeometry

\clearpage
\newgeometry{left=1.5cm,right=1.5cm,top=2.5cm,bottom=2.5cm}

\begin{table*}
\caption{The results of the two-sample Kolmogorov-Smirnov test.}
\centering
\begin{tabular}{ccccccccccccccc}
\hline\hline
                  &                & \multicolumn{5}{c}{Normalied to the strongest}                                &     & \multicolumn{5}{c}{Normalied to HCN\,(1--0)}         \\
                                                    \cline{3-7}                                                          \cline{9-13}                     
                  &                & \multicolumn{2}{c}{Narrow}     &    & \multicolumn{2}{c}{Broad}         &     & \multicolumn{2}{c}{Narrow}   &  & \multicolumn{2}{c}{Broad}  \\
                                  \cline{3-4}                           \cline{6-7}                            \cline{9-10}                          \cline{12-13}   

Classification    &  Objects       &  $D$        &  p-value    &    & $D$    &  p-value   &  &  $D$         &  p-value    &  &  $D$         &  p-value    \\
\hline
                  &                &  \multicolumn{11}{c}{All lines}      \\
\hline
high-mass YSO     &  Orion\,KL     &  0.18       &  0.62       &    &  0.26  &  0.22      &  &  0.16       &  0.68        &  &  0.24       &  0.25       \\
low-mass YSO      &  IRAS+04016    &  0.32       &  0.044      &    &  0.16  &  0.78      &  &  0.22       &  0.31        &  &  0.091      &  0.99       \\
C-rich  AGB       &  IRC+10216     &  0.18       &  0.62       &    &  0.19  &  0.56      &  &  0.16       &  0.68        &  &  0.18       &  0.60       \\
O-rich  AGB       &  IK\,Tauri     &  0.29       &  0.085      &    &  0.16  &  0.78      &  &  0.11       &  0.97        &  &  0.24       &  0.25       \\
red nova          &  CK\,Vul       &  0.21       &  0.42       &    &  0.13  &  0.94      &  &  0.19       &  0.48        &  &  0.12       &  0.96       \\
\hline
                  &                &  \multicolumn{11}{c}{Non-zero intensity lines}    \\
\hline
 high-mass YSO    &  Orion\,KL     &  0.50       &  0.00013    &    &  0.36  &  0.083      &  &  0.50       &  0.00013     &  &  0.36       &  0.083      \\
low-mass YSO      &  IRAS+04016    &  0.61       &  0.0012     &    &  0.38  &  0.23       &  &  0.47       &  0.018       &  &  0.20       &  0.89       \\
C-rich   AGB      &  IRC+10216     &  0.26       &  0.36       &    &  0.30  &  0.28       &  &  0.26       &  0.36        &  &  0.30       &  0.28       \\
O-rich  AGB       &  IK\,Tauri     &  0.55       &  0.0026     &    &  0.42  &  0.19       &  &  0.23       &  0.56        &  &  0.57       &  0.012      \\
red nova          &  CK\,Vul       &  0.35       &  0.13       &    &  0.25  &  0.63       &  &  0.35       &  0.13        &  &  0.25       &  0.63       \\
 \hline
\hline
\end{tabular}
\tablefoot{The $D$ is the maximum deviation of the cumulative distribution function 
(i.e. $F(x)$ -- the fractional number of the lines with normalized intensities of $\leq$x) between two spectra.
The p-value is the probability of two objects having similar spectral properties. 
The upper and lower parts list the results of using all the lines listed in Table~\ref{Table5} 
and the lines with non-zero intensity in at least one of the objects as the input data sets, respectively.
}
\label{Table6}
\end{table*}

\restoregeometry

\appendix

\section{Morphology: 3-D modeling with {\it SHAPE}}
\label{ShapeX}

\begin{figure}
\centering
        \includegraphics[width=16cm]{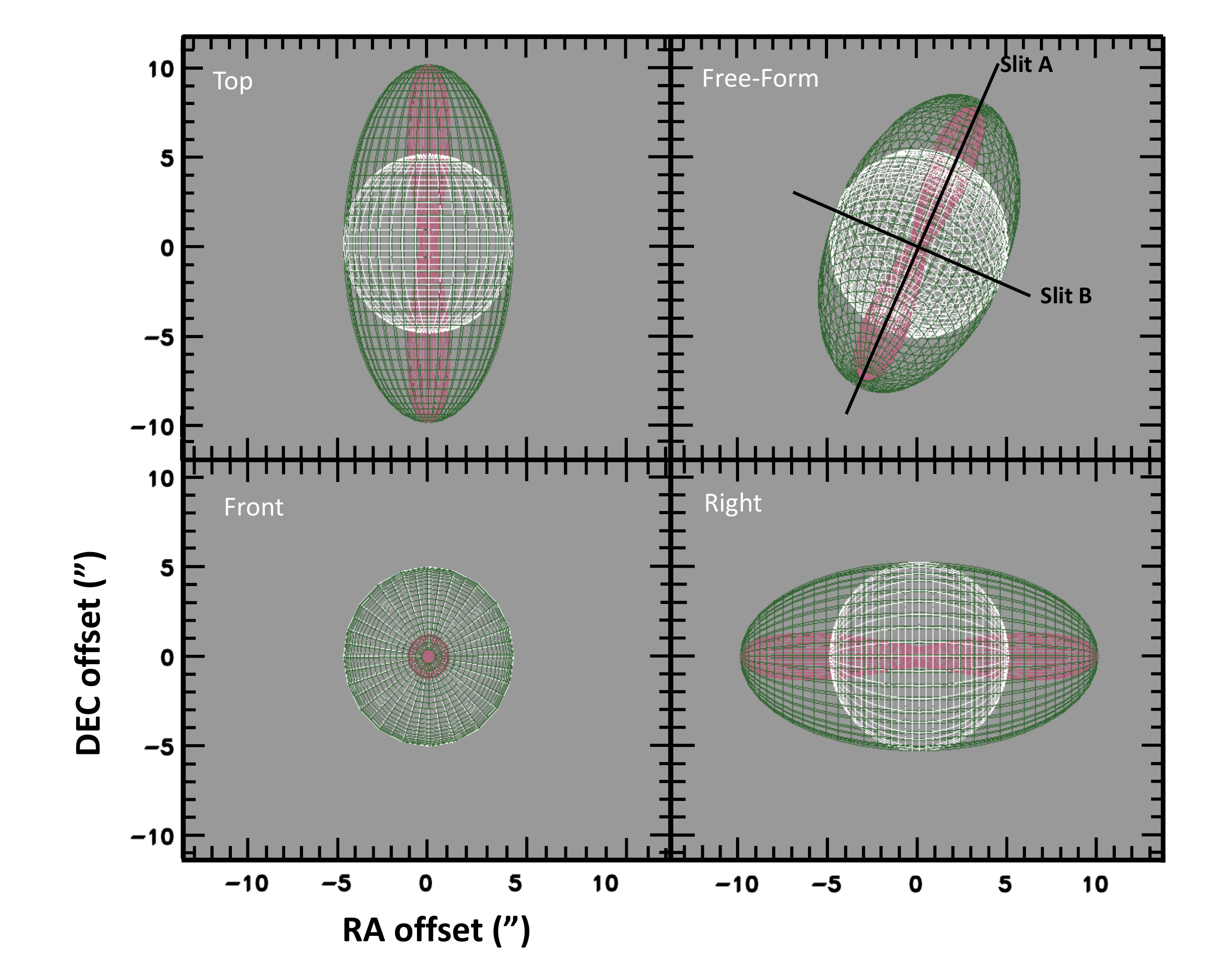}
 \caption{Four view ports of the {\it SHAPE} 3-D mesh geometry model of I19312. The Free-Form view is the observer's view with a position angle of $-37^{\circ}$, where up is north and left is east. The modeled structures are composed of an inner spherical envelope (white), an outer elliptical envelope (green), and a pair of symmetric bipolar outflows (red). The solid lines on the upper-right panel are slits used to generate the position-velocity diagram shown in Fig.~\ref{Figure_A_2}. 
}
\label{Figure_A_1}
\end{figure}

\begin{figure}
\centering
        \includegraphics[width=16cm]{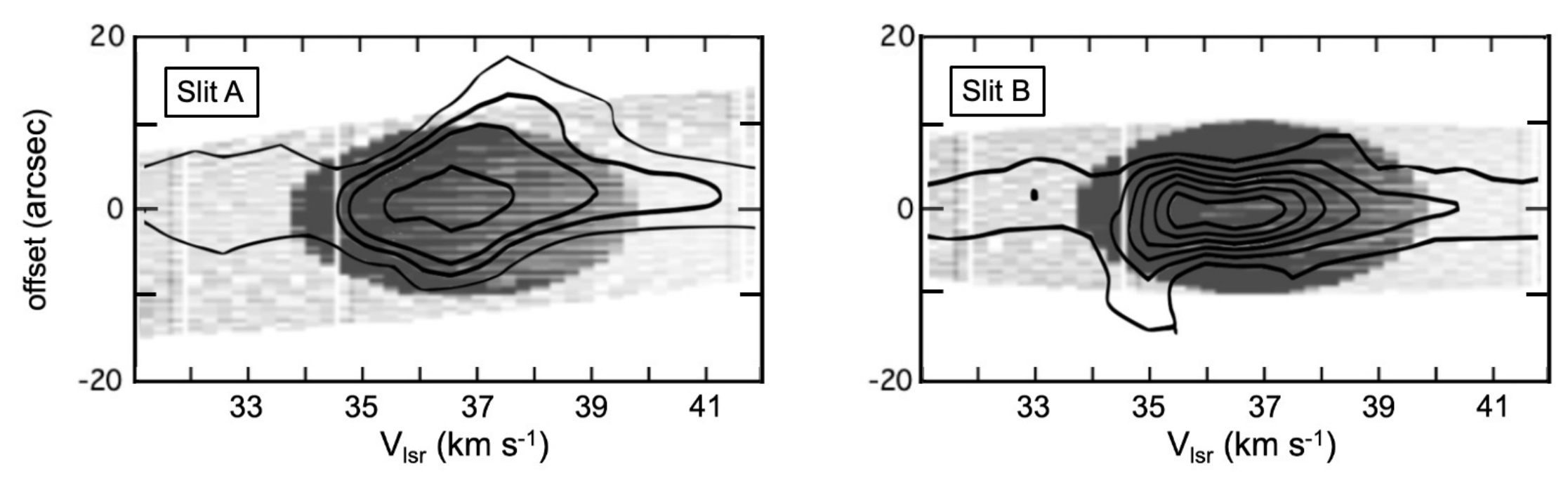}
 \caption{Simulated position-velocity diagrams (grey) of the $^{13}$CO $J=2\rightarrow1$ line in two orthogonal cuts indicated in Fig.~\ref{Figure_A_1}, 
 where $V_{\rm lsr}$ represents the velocity with respect to the Local Standard of Rest.
 The contours represent the observational results of \cite{Nakashima05}.
}
\label{Figure_A_2}
\end{figure}

\begin{figure}
\centering
        \includegraphics[width=16cm]{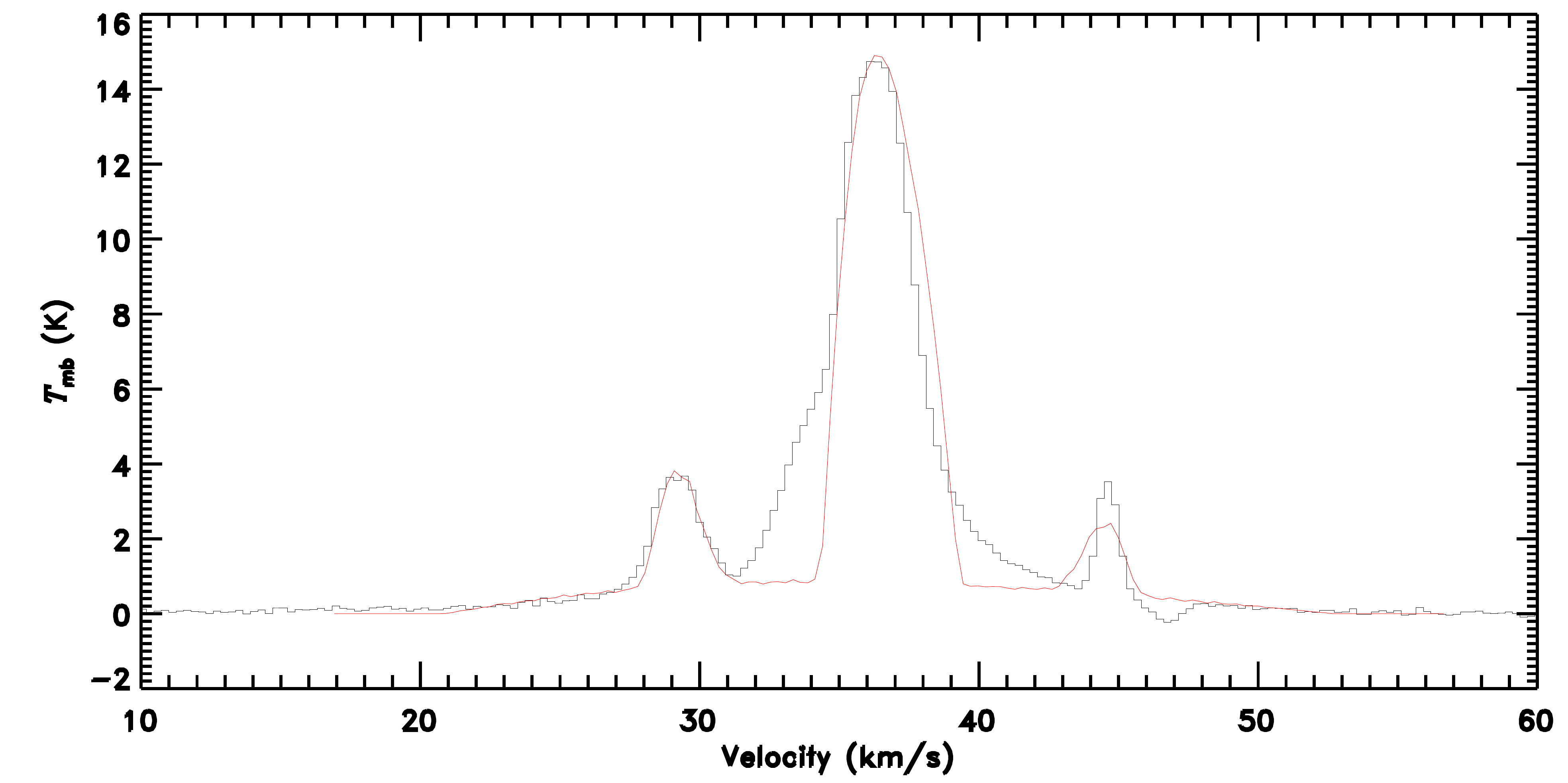}
 \caption{The {\it SHAPE}-modelled (red) and observational (black)
 $^{13}$CO $J=2\rightarrow1$ lines.
}
\label{Figure_A_3}
\end{figure}

The observed line profiles suggest that the molecular species in I19312 have complex kinematic distributions.
In order to obtain a rough idea of the geometric and kinematic structures, 
we construct a 3-D model using the software package {\it SHAPE}\footnote{http://www.astrosen.unam.mx/shape/} \citep{Steffen11,Steffen17}, 
which is a flexible interactive 3-D morpho-kinematical modeling application with volume rendering and radiation transfer for astrophysics and has been successfully used to simulate the complex structure of PNe in many studies (e.g. red rectangle PPN, \citealt{Koning11}; multi-polar PN, \citealt{Steffen13}; strigiform (owl-like) nebulae, \citealt{Garcia18}).
Firstly, we construct a 3-D structure in the $^{13}$CO $J=2\rightarrow1$ line and compare it with the previous mapping results to verify our modeling. 
The $^{13}$CO $J=2\rightarrow1$ line was selected as a guideline for the initial model setup for two reasons: the distribution of CO molecules can be representative of the distribution of the entire molecular gas, and this line is assumed to be optically thin.
Based on the previous interferometric observations with the Berkeley-Illinois-Maryland Association array \citep{Nakashima05,Nakashima11}, 
our 3-D model of I19312 in the $^{13}$CO $J=2\rightarrow1$ line consists of three major components: 
inner spherical expanding component, outer expanding elliptical component, and bipolar outflow, as displayed in Fig.~\ref{Figure_A_1}.
We fixed the physical size of the inner spherical component to 10$\arcsec$ and the outer boundary of the elliptical component in the long and short axis to 20$\arcsec$ and 10$\arcsec$, respectively.
Because the bipolar outflow has been observed only in maser lines \citep{Nakashima11}, its actual size is unknown.
Then we assume that the bipolar outflow has the same physical size as the outer envelope of 20$\arcsec$ at its jet axis. 
The densities of the inner spherical component and the outer expanding envelope are assumed to be constant.
We also fixed the position angle of the jet axis of the bipolar outflow and the long axis of the outer elliptical component to be $-37^{\circ}$. For simplicity, the symmetric axis of the 3-D geometry was assumed to be perpendicular to the line-of-sight direction.
We assumed that the velocity distributions of the inner spherical component and the outer elliptical component follow a single Hubble law with the velocity at the outer boundary of 16 and 3 km\,s$^{-1}$, respectively.
For the velocity distributions of the bipolar outflow, we assume a constant speed of 10 km\,s$^{-1}$, 
which is equal to the value indicated by the high spatial resolved observations of the HCO$^{+}$ $J=1\rightarrow0$ line \citep{Nakashima04b}. 
In order to make a qualitative comparison between the modeling results and the interferometric observations of \cite{Nakashima05}, 
we plot in Fig.~\ref{Figure_A_2} the predicted and observed position-velocity diagrams of the $^{13}$CO $J=2\rightarrow1$ line 
along the same position angles. They show a good agreement. 
Figure~\ref{Figure_A_3} shows the profile of the $^{13}$CO $J=2\rightarrow1$ line reproduced by our model overlaid on the present observation.
We find that the model and observed line profiles agree well with each other. 

\begin{figure}
\centering
        \includegraphics[width=16cm]{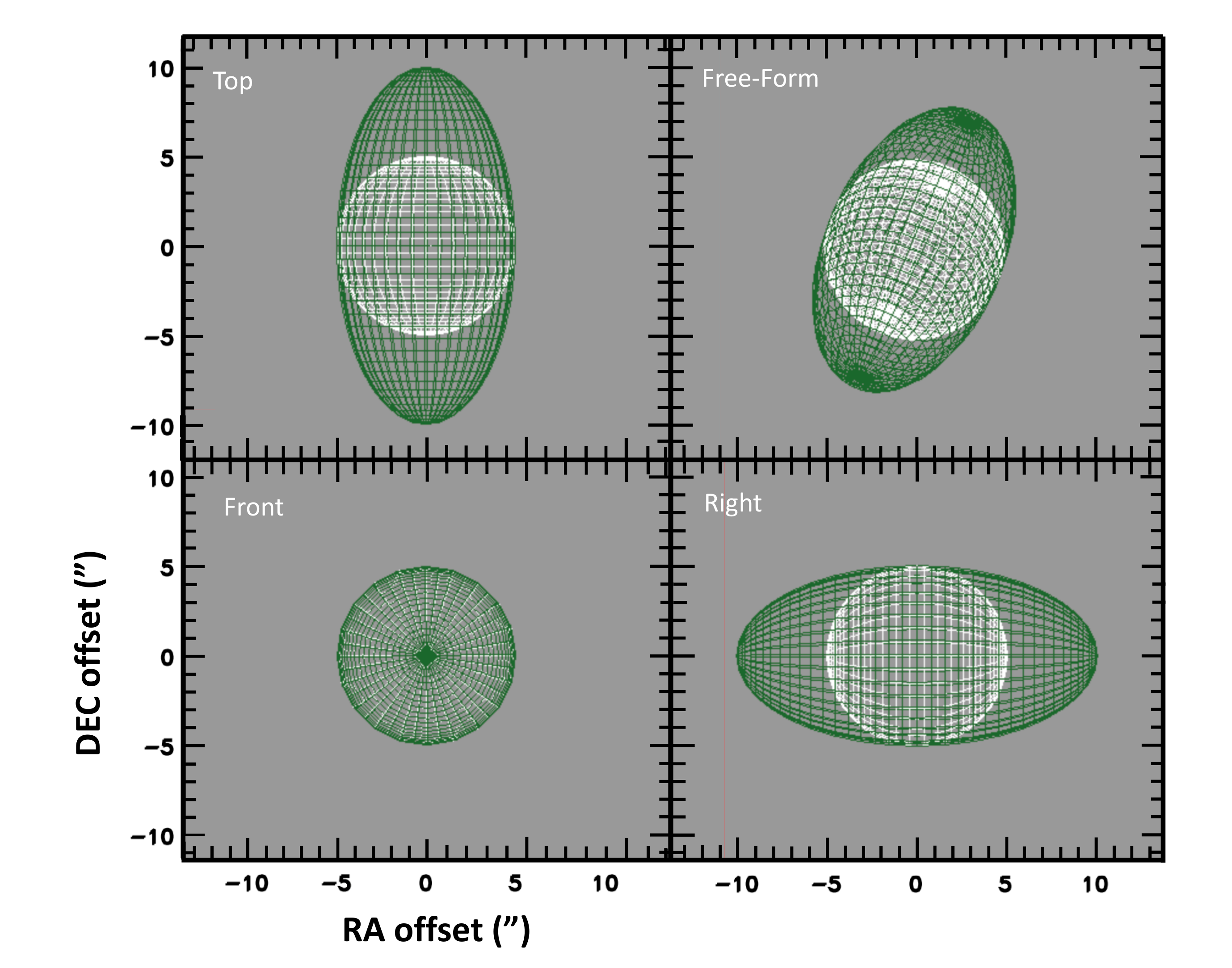}
 \caption{Same as Fig.~\ref{Figure_A_1} but for the SO $J = 6 \rightarrow 5$ line.
 }
\label{Figure_A_4}
\end{figure}

\begin{figure}
\centering
        \includegraphics[width=16cm]{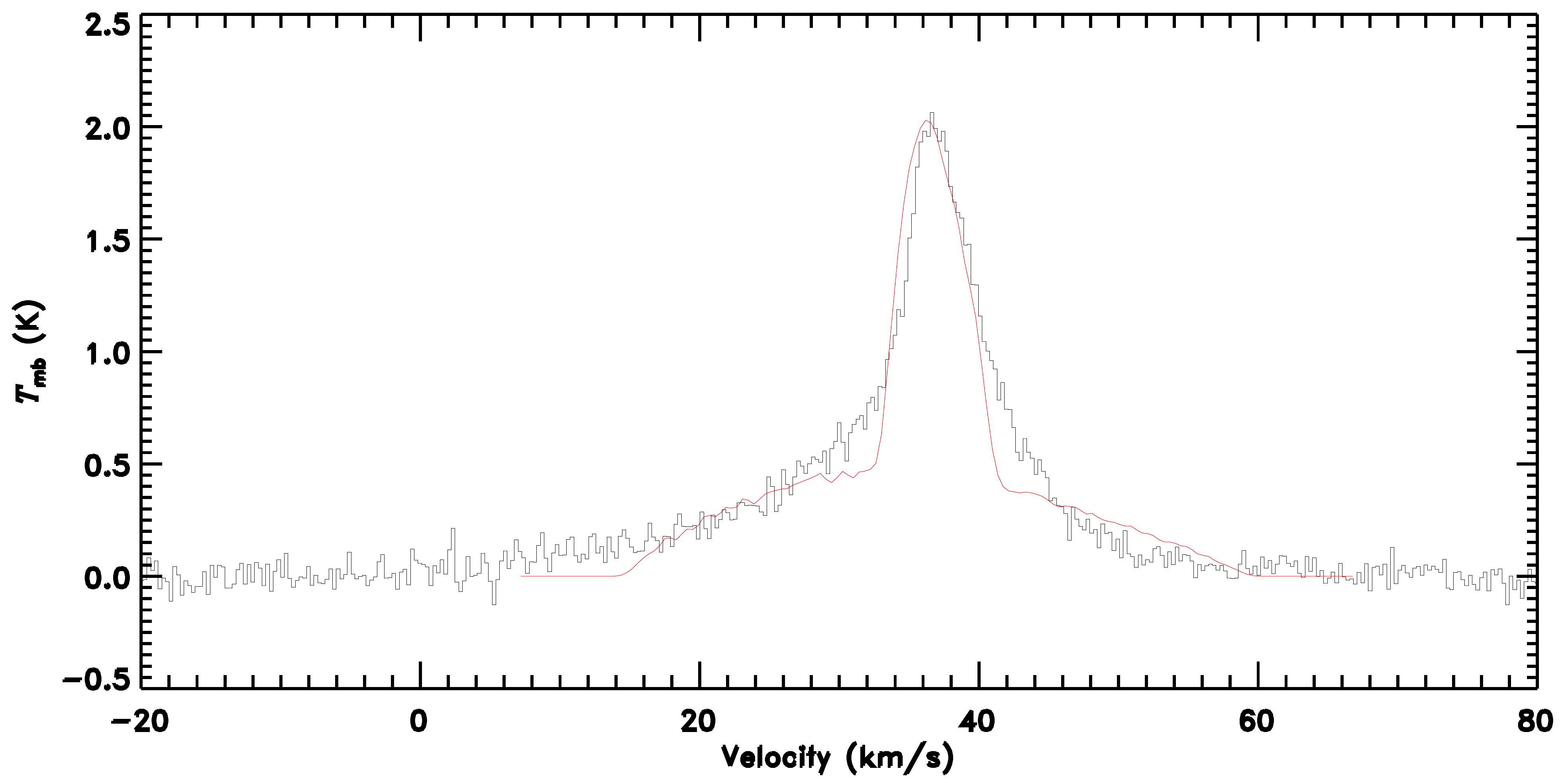}
 \caption{Same as Fig.~\ref{Figure_A_3} but for the SO $J = 6 \rightarrow 5$ line. 
}
\label{Figure_A_5}
\end{figure}

\begin{figure}
\centering
        \includegraphics[width=16cm]{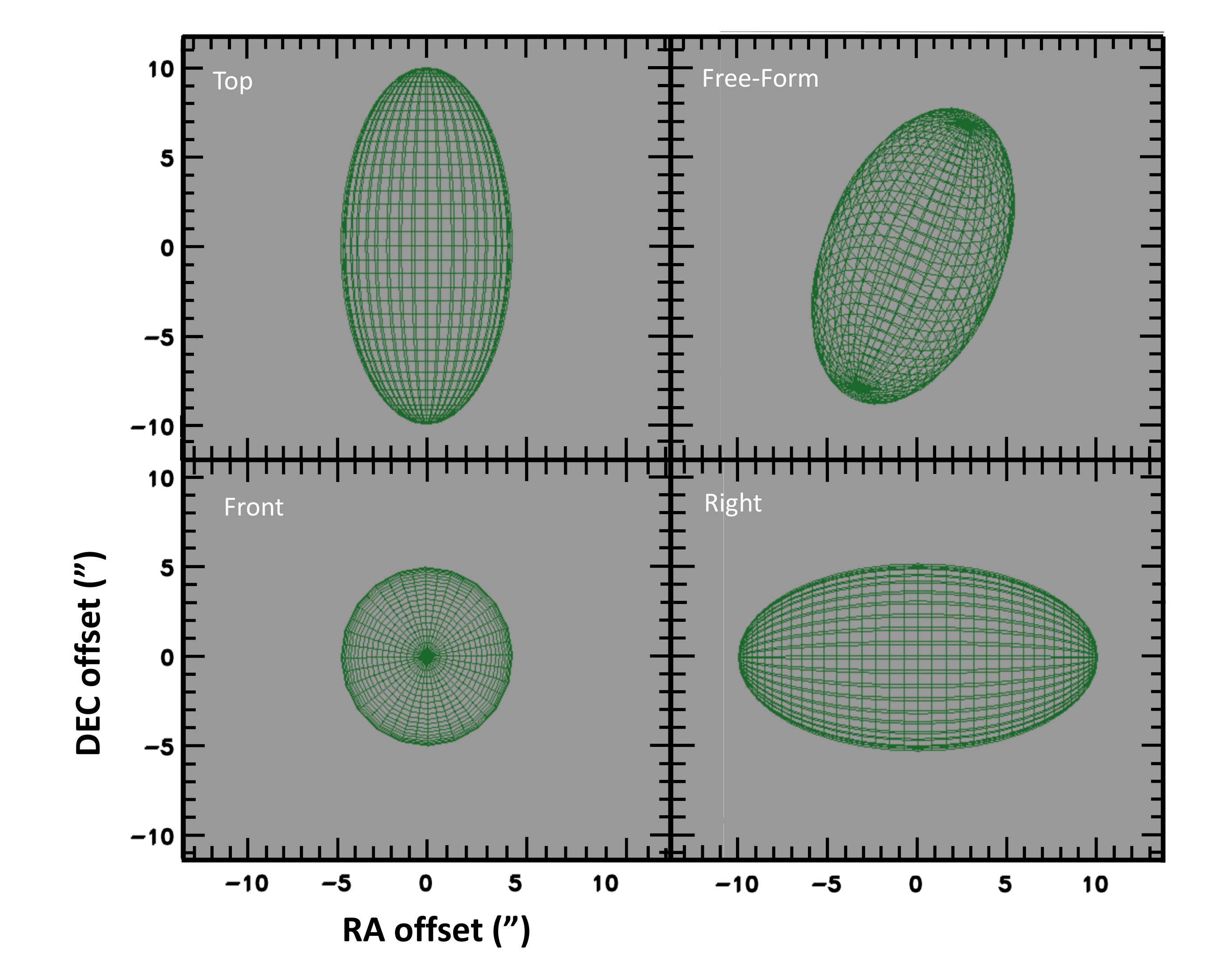}
 \caption{Same as Fig.~\ref{Figure_A_1} but for the HC$_{3}$N $J=10 \rightarrow 9$ line.
}
\label{Figure_A_6}
\end{figure}

\begin{figure}
\centering
        \includegraphics[width=16cm]{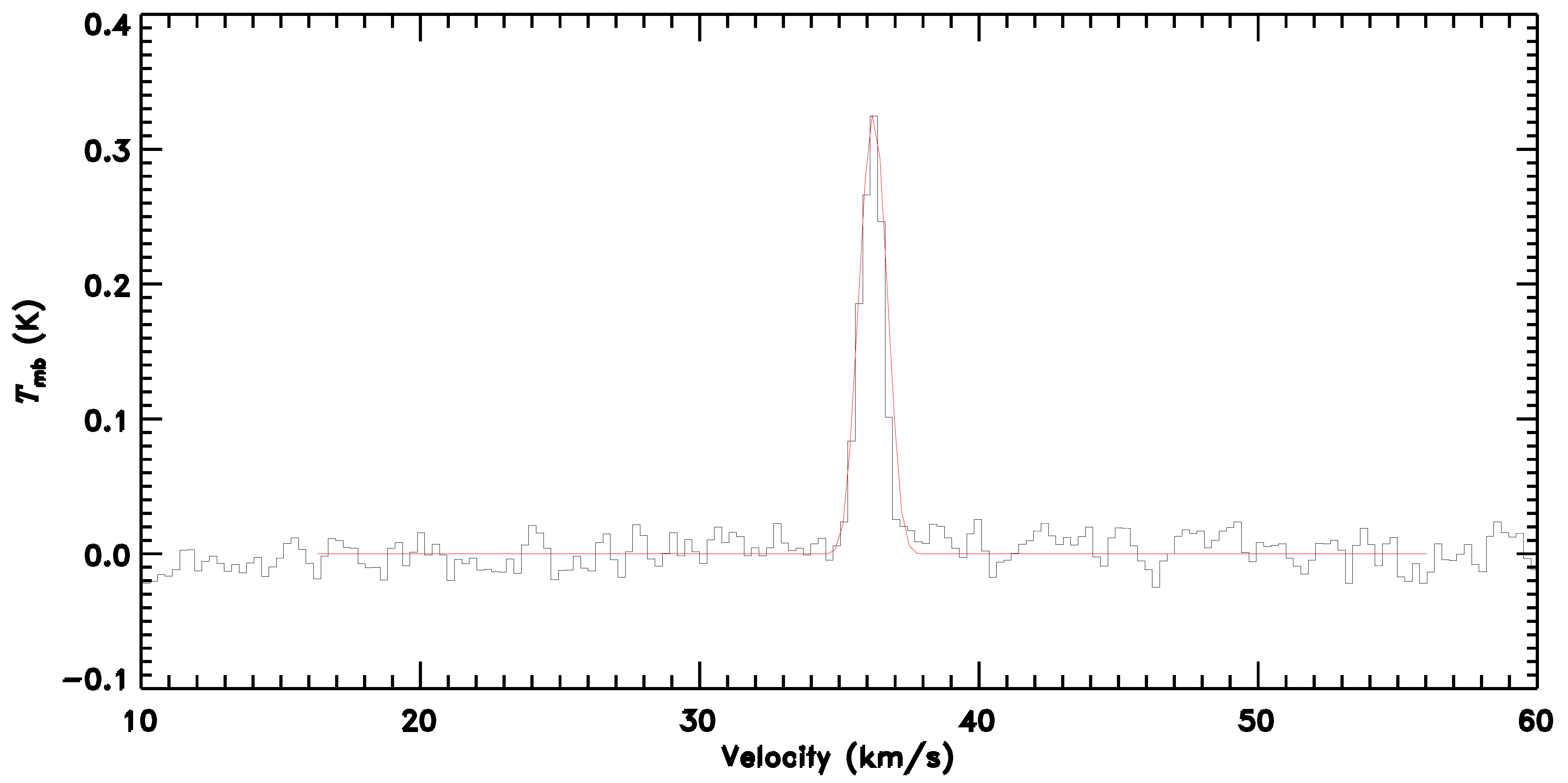}
 \caption{Same as Fig.~\ref{Figure_A_3} but for the
    HC$_{3}$N $J=10 \rightarrow 9$ line.
}
\label{Figure_A_7}
\end{figure}

Based on the same assumptions, we also constructed the 3-D model of other molecular species using the {\it SHAPE} package.
We selected the SO $J = 6 \rightarrow 5$ line as a typical representative of the emission lines with both narrow and broad features and the HC$_{3}$N $J=10 \rightarrow 9$ line for the emission lines with only a narrow feature. (Note: there are no lines with only broad components). 
The mesh geometry of the best-fit model for the SO $J = 6 \rightarrow 5$ line is shown in Fig.~\ref{Figure_A_4}.
The model and observed line profiles of the SO $J = 6 \rightarrow 5$ line are compared in Fig.~\ref{Figure_A_5}.
Similarly, for the HC$_{3}$N $J=10 \rightarrow 9$ line, the best-fit model geometry is shown in Figs.~\ref{Figure_A_6}.
The model and observed line profiles of the HC$_{3}$N $J=10 \rightarrow 9$ line are compared in Fig.~\ref{Figure_A_7}.
As we can see in the figures, the SO $J = 6 \rightarrow 5$ line is explained as a combination of the inner spherical expanding component and outer expanding elliptical component. 
In contrast, the HC$_{3}$N $J=10 \rightarrow 9$ line is reproduced by an outer expanding elliptical component. 
In other words, the profiles of the SO $J = 6 \rightarrow 5$ and HC$_{3}$N $J=10 \rightarrow 9$ lines can be explained by assuming one or two of the three kinematic components needed to explain the $^{13}$CO $J=2\rightarrow1$ line.
Thus, the modeling result suggests that the emission sources of the molecular lines detected towards I19312 can be classified into either an inner spherical expanding component, an outer expanding elliptical component, or a bipolar outflow.
Based on this result, it would be natural to infer that similar explanations are possible for other lines exhibiting similar line profiles analyzed here.

For the $^{13}$CO\,$J=2\rightarrow1$ and SO $J = 6 \rightarrow 5$ lines, there are  discrepancies between the model and observations in the velocity range of about 30--35 and 40--45 km\,s$^{-1}$. 
This discrepancy can be reduced if we add another spherically expanding shell, which has a different velocity, to the inner component.
This may suggest that the inner component has a more complex kinematic structure rather than a simple spherically expanding shell.
We would note that this situation could be explained by considering that the origin of the inner component is not a simple stellar wind, but a gas component with a complex kinematic structure that is  created by a stellar merger.
Additionally, information about which molecular species are associated with which kinematic components may also limit the chemical composition of the progenitor stars to mergers and the chemical reactions induced by stellar mergers.

It should be noted that the model rests on many assumed parameters that
may not represent the most justified configuration. Moreover, the optically thin
assumption may not be true for the $^{13}$CO line. Therefore, a more elaborated model
is required to make a quantitative comparison with the observations, which is beyond
the scope of this paper. 
Here we only demonstrate the feasibility of using such a geometry 
to account for the available observations of this source.

\newgeometry{left=1.5cm,right=1.5cm,top=2.5cm,bottom=2.5cm}

\clearpage

\restoregeometry

\end{CJK*}
\end{document}